\documentclass[a4paper,11pt]{article}
\pdfoutput=1 

\usepackage{nomencl}

\usepackage{jheppub} 

\usepackage[T1]{fontenc} 

\usepackage{color}

\usepackage{graphicx} 

\usepackage{booktabs} 
\usepackage{array} 
\usepackage{paralist} 
\usepackage{verbatim} 
\usepackage{subfig} 

\usepackage{enumerate} 

\usepackage{fancyhdr} 
\usepackage{sectsty}
\allsectionsfont{\sffamily\mdseries\upshape} 

\usepackage[nottoc,notlof,notlot]{tocbibind} 
\usepackage[titles,subfigure]{tocloft} 

\usepackage{amsfonts}

\usepackage{yfonts}
\usepackage{amsmath}
\usepackage{amssymb}

\title{\boldmath UV Divergence and Tensor Reduction}

\author[a]{Qingjun Jin}

\affiliation[a]{Graduate School of China Academy of Engineering Physics, No. 10 Xibeiwang East Road, Haidian District, Beijing, 100193, China\\ }

\emailAdd{qjin@gscaep.ac.cn}

\abstract{We present an efficient algorithm to decompose the ultraviolet (UV) divergences of Feynman integrals to local divergences and various types of sub-divergences. 
With some reasonable assumptions the local divergences of Feynman integrals can be uniquely defined in dimensional regularization scheme.
By an asymptotic expansion in the hard momenta, the computation of local and sub-divergences is reduced to the computation of local divergences of massless vacuum integrals. 
In theories with spin $\le\frac{1}{2}$, the beta functions and anomalous dimensions can be extracted directly from the local divergence of integrals.

We also propose two methods to reduce the tensor structures which can be used in the computation of local divergence. The first method is based on dimensional shift and is extremely powerful for integrals with loop number $L\le3$. The second method is based on a PV reduction in a $d_{\infty}$ dimension subspace, and it is more suited in four and more loops.
}

\makenomenclature

\begin{document} 
\maketitle
\flushbottom

\section{Introduction}

The beta functions and anomalous dimensions govern the renormalization group flow of physical quantities in quantum fields theories, and their evaluation involves the computation of ultraviolet (UV) divergences of Feynman integrals. In the MS-scheme (or $\overline{MS}$-scheme) all the UV counterterms are polynomial of momenta and in masses \cite{Collins:1974da}, so it is tempting to carry out a Taylor expansion in external momenta and masses before integration over loop momenta, and reduce the complicated Feynman integrals with multiple scales to vacuum integrals with no scale. However, besides UV divergences, these vacuum integrals also has IR(infrared) divergences, which must be regulated or subtracted. In \cite{Vladimirov:1979zm,Chetyrkin:1980pr} the 'infrared rearrangement' approach was introduced which regulate the IR divergence by adding artificial masses or external momenta in certain lines of a given Feynman diagram. IR divergences can also be removed using a more sophisticated $R^*$ operation \cite{Chetyrkin:1984xa,Larin:2002sc,Batkovich:2014rka,Baikov:2016tgj, Herzog:2017bjx, Herzog:2017jgk} technique, and the UV counterterm of $(L+1)$-loop Feynman integral can be expressed in terms  $L$-loop massless propagators. 

Another way of regulating IR divergence is achieved by introducing the same artificial mass to all propagators, which reduces the computation of complicated Feynman diagrams to relatively simple vacuum integrals \cite{Chetyrkin:1997fm}. 
Combined with IBP reduction, the fully massive vacuum integrals was used in the computation of the beta functions of $\phi^3$ theory in 6-dimension \cite{Kompaniets:2021hwg}.
Similar IR regulator was used in \cite{Bern:2010ue,Bern:2012uc,Bern:2013yya,Bern:2015ooa,Bern:2017ucb, Bern:2018jmv} to study the UV behavior of super Yang-Mills and supergravity amplitudes at the critical dimensions.

Most of the known methods are less efficient when applied to integrals with high rank tensor structures, which appear for instance during the computation of UV divergences in gravity theories, and anomalous dimensions of high dimensional operators in effective field theories. This work is part of the effort to address this problem.

We mainly follow the storyline of $R^*$ operation, but will propose multiple improvements to the algorithm by exploiting the UV structure of massless vacuum integrals.
The UV divergence of generic Feynman integrals are decomposed into local divergences and sub-divergences in different regions, and the renormalization Z-factors can be determined solely from the the local divergences.
The local divergences of generic Feynman integrals are expressed by the local divergences of massless vacuum integrals via an asymptotic expansion around hard loop momenta, and the sub-divergences can be computed from the local divergences of the corresponding lower loop sub-integrals.
The local IR divergences of massless vacuum integrals are regulated by adding an auxiliary mass to a single propagator. Then the local UV divergences are obtained by subtracting the remaining lower loop IR and UV sub-divergences from the mass regulated integral.

In order to evaluate the local divergence of integrals with high rank tensor structures, we need efficient tensor reduction methods which reduce the local divergence of tensor integrals to that of scalar integrals.
The conventional Passarino-Veltman (PV) reduction \cite{Passarino:1978jh} cannot be employed because the local divergence operator does not commute with Lorentz contraction.
We propose two new tensor reduction methods which are suited for this task. The first method is based on dimensional shift \cite{Tarasov:1996br,Tarasov:1996bz}, which relates tensor integrals to scalar integrals in higher dimensions, and it is extremely powerful at lower loops ($L\le 3$). The second method is based on the PV reduction in a $d_{\infty}$ dimensional subspace, which relates tensor integrals to scalar integrals containing $d_{\infty}$ dimensional Lorentz products, and it is more efficient in higher loops.

We will try to give a self-contained introduction to the whole program. 
In Section \ref{section:1-loop}, we demonstrated the efficiency of massless vacuum integral approach by evaluating the 1-loop UV divergences.
In Section \ref{uv-decom}, after presenting some examples and introducing conventions on integrals and divergence degrees, we discuss the UV decomposition formula which holds for a single Feynman integral. 
In Section \ref{CV-2-3}, we evaluate the local divergence of two and three loop massless vacuum integral using UV decomposition.
In Section \ref{IR-div}, we discuss the computation of IR divergences. We propose a scheme in which the total IR divergence is a simple sum of all IR divergences in different regions.
In Section \ref{tensor-reduction}, we present two approaches to tensor reduction based on dimensional shift and $d_{\infty}$ dimensional PV reduction.
We also evaluate the local divergence of some 5-loop tensor integrals.
In Section \ref{z-factors}, we extend the UV decomposition formula to correlation functions.
In our formalism, the sub-divergences automatically cancel each other, and the renormalization factors are simply given by the local divergences of the corresponding "unrenormalized" correlation functions. 
We also demonstrate the method by computing the renormalization factors in the 6-d $\phi^3$ theory  to 3-loop, and discuss its application in generic quantum field theories.



\section{The UV divergence of 1-loop integrals}
\label{section:1-loop}

We start by considering a one loop integral in Euclidean space which would appears in the two-gluon correlation function $\langle A^{\mu}A^{\nu}\rangle$:
\begin{equation}\label{1-loop-ex1}
I^{\mu\nu}=S_{\epsilon}\int \frac{d^D l}{(2\pi)^D}\frac{l^{\mu}l^{\nu}}{l^2(l+p)^2}\ ,
\end{equation}
in which $S_{\epsilon}=(4\pi)^{\frac{D}{2}}e^{\epsilon \gamma_E}$ is a prefactor which is introduced in the MSbar scheme to make the expression compact, and for $L$ loop integrals the prefactor is $S_{\epsilon}^L$. We will work in dimensional regularization scheme, and use $d$ to denote the unregularized spacetime dimension, and $D=d-2\epsilon$. In this example $d=4$, but later we will also study integrals in other dimensions.

We will mainly work in Euclidean space $\mathbb{R^D}$ in this paper. As far as UV divergences and soft IR divergences are concerned, there is no essential difference between Euclidean and Minkowski space. In Minkowski space there are collinear divergences in the presence of external momenta. However, in this work the only IR divergences we are interested in are those of vacuum integrals.



The UV divergence of $I^{\mu\nu}$ can be obtained by expanding its analytic expression,
\begin{equation}\label{1 loop direct}
I^{\mu\nu}
=\frac{e^{\epsilon \gamma_E}(2-\epsilon)\Gamma^2(1-\epsilon)\Gamma(\epsilon)}{2(3-2\epsilon)\Gamma(2-2\epsilon)(-p^2)^{\epsilon}}(p^{\mu}p^{\nu}-\frac{p^2\eta^{\mu\nu}}{4-2\epsilon})\sim \frac{1}{3\epsilon}(p^{\mu}p^{\nu}-\frac{p^2\eta^{\mu\nu}}{4})\ ,
\end{equation}
where $A\sim B$ means $A$ and $B$ have the same UV divergence\footnote{$A\sim B$ does not mean $A$ and $B$ have the same $\epsilon$-poles, because $A$ and $B$ may have different IR divergences.}.

However, it can be very difficult to find the analytic expression of integrals with more scales and/or loops, so we need alternative methods to evaluate the UV divergences. A very illuminating approach was presented in \cite{Chetyrkin:1997fm}, which we briefly review in the next subsection.

\subsection{Regulating IR divergence with an auxiliary mass}

Following \cite{Chetyrkin:1997fm} one can perform a regulated expansion to the propagators in its external momentum,
\begin{equation}\label{decompose propagator}
\frac{1}{(l+p)^2}=\frac{1}{l^2+m^2}+\frac{m^2-p^2-2l\cdot p}{(l^2+m^2)^2}+\frac{(m^2-p^2-2l\cdot p)^2}{(l^2+m^2)^3}+\frac{(m^2-p^2-2l\cdot p)^3}{(l^2+m^2)^3(l+p)^2}\ ,
\end{equation}
in which $m$ is an auxiliary mass which serves as a regulator of IR divergence.
We have chose the regulated propagator $\frac{1}{l^2+m^2}$ instead of $\frac{1}{l^2-m^2}$ as in \cite{Chetyrkin:1997fm}, because we work in Euclidean space.

Apply \eqref{decompose propagator} to both propagators in $I^{\mu\nu}$, and drop UV finite terms,
\begin{equation}
\begin{aligned}
&I^{\mu\nu}\sim
\frac{l^{\mu}l^{\nu}}{(l^2+m^2)^2}+\frac{(2m^2-p^2-2l\cdot p)l^{\mu}l^{\nu}}{(l^2+m^2)^3}
+\frac{(2l\cdot p)^2l^{\mu}l^{\nu}}{(l^2+m^2)^4}\ .\\
\end{aligned}\label{Imunu decom}
\end{equation}

The one loop vacuum integral in \eqref{Imunu decom} vanishes when there are odd number of $l^{\mu}$ in the numerator. When there are even number of $l^{\mu}$,
\begin{equation}
\begin{aligned}
&\frac{l^{\mu_1}\cdots l^{\mu_{2a}}}{(l^2+m^2)^n}=\frac{\Gamma(n-a-\frac{D}{2})}{2^a\Gamma(n)}m^{D+2a-2n} \eta_s^{\mu_1 \cdots \mu_{2a}}\ ,\\
\end{aligned}\label{int-1loop-mass-tensor}
\end{equation}
where 
\begin{equation}
\eta_s^{\mu_1 \cdots \mu_{2a}}=\eta^{\mu_1\mu_2}\cdots \eta^{\mu_{2a-1}\mu_{2a}}+\text{non-repetitive permutations of }\mu_i\ .
\end{equation}
When $n\le a+\frac{d}{2}$, the UV divergence is non-zero:
\begin{equation}
\begin{aligned}
&\frac{l^{\mu_1}\cdots l^{\mu_{2a}}}{(l^2+m^2)^n}
\sim \frac{(-m^2)^{\frac{d}{2}+a-n}}{2^a\Gamma(n)(\frac{d}{2}+a-n)!\epsilon}\eta_s^{\mu_1 \cdots \mu_{2a}}\ .\\
\end{aligned}\label{fnmassive}
\end{equation}
Plug into \eqref{Imunu decom},
\begin{equation}
\begin{aligned}
&\frac{l^{\mu}l^{\nu}}{l^2(l+p)^2}\sim
-\frac{m^2}{2\epsilon}\eta^{\mu\nu}+\frac{(2m^2-p^2)\eta^{\mu\nu}}{4\epsilon}
+\frac{p^2\eta^{\mu\nu}+2p^{\mu}p^{\nu}}{6\epsilon}
=\frac{-p^2\eta^{\mu\nu}+4p^{\mu}p^{\nu}}{12\epsilon}\ ,\\
\end{aligned}
\end{equation}
which is in agreement with \eqref{1 loop direct}. 

The UV divergence of generic one-loop integrals with more external momenta and masses can be obtained using the same technique. The key idea of this method is reducing the UV divergence of complicated integrals to that of massive vacuum integrals, which can be simply evaluated. In the next subsection we will show that the method can be further refined by considering massless vacuum integrals.

\subsection{The UV divergence of massless vacuum integrals}
An auxiliary mass $m$ was introduced in \cite{Chetyrkin:1997fm} as an IR regulator of the resulting vacuum integrals. However, the UV divergence of one-loop vacuum integrals can be easily obtained even if they are massless. The simplest way is setting $m=0$ in \eqref{fnmassive},
\begin{equation}
\begin{aligned}
&\frac{l^{\mu_1}\cdots l^{\mu_{2a}}}{(l^2)^n}\sim \frac{\delta_{n-a,\frac{d}{2}}}{2^a\Gamma(n)\epsilon}
\eta_s^{\mu_1 \cdots \mu_{2a}}\ .\\
\end{aligned}\label{fn}
\end{equation}
Using a Taylor expansion about $p=0$, the integral $I_{\mu\nu}$ in \eqref{Imunu decom} can be decomposed to some massless vacuum integrals and a UV finite remainder term $R(l)$,
\begin{equation}
\begin{aligned}
\frac{l^{\mu}l^{\nu}}{l^2(l+p)^2}=
&\frac{l^{\mu}l^{\nu}}{(l^2)^2}+\frac{(-p^2-2l\cdot p)l^{\mu}l^{\nu}}{(l^2)^3}
+\frac{(2l\cdot p)^2l^{\mu}l^{\nu}}{(l^2)^4}+R(l)\ \\
\sim &-\frac{p^2l^{\mu}l^{\nu}}{(l^2)^3}+\frac{(2l\cdot p)^2l^{\mu}l^{\nu}}{(l^2)^4}
\sim \frac{-p^2\eta^{\mu\nu}+4p^{\mu}p^{\nu}}{12\epsilon}\ ,\\
\end{aligned}\label{1 loop massless decomposition}
\end{equation}
which is again in agreement with \eqref{1 loop direct}.

There is a crucial difference between the massive and massless vacuum integrals. 
In the massive case, the integral has non-zero UV divergence if the superficial degree of divergence $\omega=2a+d-2n\ge 0$, while in the massless case,  the UV divergence is non-zero only if $\omega=2a+d-2n= 0$.
So a smaller number of integrals contribute in the massless approach compared with the massive approach. 
This simplification can be important in multiloop calculations, where a large number of vacuum integrals may appear after the decomposition. 

The UV divergence degree $\omega$ for generic integrals will be discussed in Section \ref{Degree}.
Massless vacuum integrals with $\omega= 0$ will be extensively studied and used in this work, and we will call them \textbf{critical vacuum integrals} (CV). The condition $\omega= 0$ gives a relation among the number of propagators, the rank of tensor structure and the dimension $d$, and will be called the \textbf{critical condition}.

\section{The decomposition of UV divergence}
\label{uv-decom}

\begin{figure}[htb]
\centering
\includegraphics[scale=0.4]{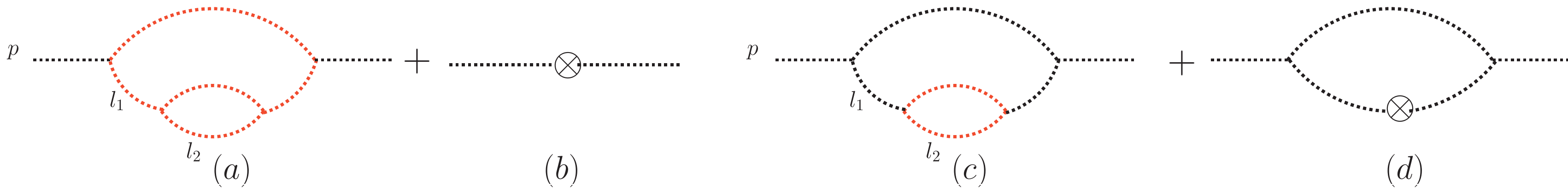}
\caption{The local and sub UV divergences and their counterterms. The red lines are the hard loop momenta.}
\label{fig:localsub}
\end{figure}

Now let us investigate the UV divergence of multiloop integrals.
A multiloop integral contains both local UV divergence and sub UV divergences (see e.g. Chapter 10.4 of \cite{Peskin:1995ev}). The local divergence is the overall divergence in the region where all loop momenta are hard compared with external momenta and masses ($|l_i|\gg |m|, |p_i|$). During renormalization, it is canceled by a local counterterm which has polynomial dependence on momentum variables. An example of local divergence and the corresponding counterterm is shown in Figure \ref{fig:localsub} (a) and (b). The sub-divergences appear in regions where only a subset of the loop momenta are hard, and they are canceled by non-local counterterms which are lower loop integral containing local counterterms as vertices. An example of sub-divergence and the corresponding counterterm is shown in Figure \ref{fig:localsub} (c) and (d).

In this section we will start with the integral in Figure \ref{fig:localsub} and show that the UV divergence of this integral can be naturally decomposed into a local divergence part and sub-divergence part. Then we will study the UV decomposition of generic multiloop integrals systematically.

\subsection{The UV decomposition of a two-loop integral}

The Feynman diagrams in Figure  \ref{fig:localsub} appear in the $\langle\phi\phi\rangle$ correlation function in 6-d $\phi^3$ theory at two-loop. The integral in Figure  \ref{fig:localsub}(a) is
\begin{equation}\label{phi3-I1}
I=\frac{1}{(l_1^2)^2(l_1+p)^2l_2^2(l_1-l_2)^2}\equiv S_{\epsilon}^2\int \frac{d^Dl_1d^Dl_2}{(2\pi)^{2D}}\frac{1}{(l_1^2)^2(l_1+p)^2l_2^2(l_1-l_2)^2}\ .
\end{equation}
From now on, for compactness we will drop $S_{\epsilon}$ factor and the integration symbol, and use the integrand to represent the integral. 

The integral has no IR divergence in Euclidean space. The 2 loop local UV divergence corresponds to the region where $|l_1|, |l_2|\ge |p|$. The integral may have 3 different one loop sub UV divergences corresponding to the following regions:
\begin{enumerate}
\item Region 1: $|l_1|\gg |l_2|, |p|$.
\item Region 2: $|l_2|\gg |l_1|, |p|$. (Shown in Figure \ref{fig:localsub}(c).)
\item Region 3: $|l_1|, |l_2|\gg |l_1-l_2|, |p|$.
\end{enumerate}

Both Region 1 and Region 3 contain four hard propagators and are free of UV divergence in 6-d, but there is a non-zero sub-divergence in Region 2. In order to isolate this sub-divergence from the local divergence, we decompose the $l_1+p$ propagators following \eqref{decompose propagator},
\begin{equation}
\begin{aligned}\label{phi3-I1-split}
I=&I_{v}+\frac{1}{(l_1^2)^2l_2^2(l_1-l_2)^2}R(l_1)\ ,\\
I_{v}=&
\frac{1}{(l_1^2)^2l_2^2(l_1-l_2)^2}\Bigl[\frac{1}{l_1^2+m^2}+\frac{m^2-p^2-2l_1\cdot p}{(l_1^2+m^2)^2}+\frac{(2l_1\cdot p)^2}{(l_1^2+m^2)^3}\Bigr]\ ,\\
\end{aligned}
\end{equation}
Then the original integral is split into several vacuum integrals $I_v$ and a remainder part which has negative superficial degree of divergence. The UV divergence of the $I_v$ can be evaluated using (see e.g. \cite{Smirnov:2006ry}),
\begin{equation}
\begin{aligned}
\frac{1}{(l_1^2)^2(l_1^2+m^2)l_2^2(l_1-l_2)^2}
\sim &m^2\Bigl(\frac{1}{24\epsilon ^{2}}+\frac{1}{\epsilon }(\frac{25}{144}-\frac{\ln m}{6})\Bigr)\ ,\\
\frac{1}{(l_1^2)^2(l_1^2+m^2)^2l_2^2(l_1-l_2)^2}
\sim &-\frac{1}{24\epsilon ^{2}}+\frac{1}{\epsilon }(\frac{-13}{144}+\frac{\ln m}{6})\ ,\\
\frac{(l_1\cdot p)^2}{(l_1^2)^2(l_1^2+m^2)^3l_2^2(l_1-l_2)^2}
\sim &p^2\Bigl(-\frac{1}{144\epsilon ^{2}}+\frac{1}{\epsilon }(\frac{-1}{96}+\frac{\ln~m}{36})\Bigr)\ ,\\
&\\
\end{aligned}\label{2 loop m00}
\end{equation}
and the result is\footnote{Besides Feynman integrals, the complete correlation functions also contain $\mu^{\epsilon}$ factors which are introduced by the bare coupling constants. The UV divergences of correlation functions may contain $\ln \frac{m^2}{\mu^2}$ and $\ln \frac{p^2}{\mu^2}$ terms. In this paper we neglect the $\mu^{\epsilon}$ factors, so we have terms like $\ln m$ and $\ln p^2$ in UV divergences.}
\begin{equation}\label{I-vacuum}
I_{v}\sim \frac{m^2}{12\epsilon }
+p^2\Bigl(\frac{1}{72\epsilon ^{2}}+\frac{1}{\epsilon }(\frac{7}{144}-\frac{\ln m}{18})\Bigr)\ .
\end{equation}


The UV divergence of the remainder term comes from the divergence of the $l_2$ integral, 
\begin{equation}
\begin{aligned}
\frac{R(l_1)}{(l_1^2)^2l_2^2(l_1-l_2)^2}
\sim&\frac{1}{6\epsilon l_1^2}\Bigl[
\frac{1}{l_1^2+m^2}+\frac{m^2-p^2-2l_1\cdot p}{(l_1^2+m^2)^2}+\frac{(2l_1\cdot p)^2}{(l_1^2+m^2)^3}
-\frac{1}{(l_1+p)^2}\Bigr]\\
\sim&-\frac{p^{2}}{36\epsilon ^{2}}+\frac{-18m^{2}-5p^{2}+12p^{2}\ln m}{216\epsilon }
-\frac{1}{6\epsilon l_1^2(l_1+p)^2}\ ,\\
\end{aligned}\label{remainder part}
\end{equation}
where in the first line we replaced $\frac{1}{l_2^2(l_1-l_2)^2}\rightarrow -\frac{l_1^2}{6\epsilon}$, and the expression of $R(l_1)$ can be obtained from \eqref{phi3-I1-split}. To derive the second line we evaluated the 1-loop vacuum integrals with the help of \eqref{int-1loop-mass-tensor}. Combining \eqref{I-vacuum} and \eqref{remainder part}, we find
\begin{equation}
\begin{aligned}
&\frac{1}{(l_1^2)^2(l_1+p)^2l_2^2(l_1-l_2)^2}
\sim p^2\Bigl(-\frac{1}{72\epsilon ^{2}}+\frac{11}{432\epsilon}\Bigr)
-\frac{1}{6\epsilon}\frac{1}{l_1^2(l_1+p)^2}\ .\\
\end{aligned}\label{2 loop decomposition example 1}
\end{equation}

One may continue to evaluate the 1 loop integral in \eqref{2 loop decomposition example 1} and find the total UV divergence, but the current form demonstrates the structure of UV divergence more clearly. The UV divergence of the integral is decomposed into two terms. The first term is the local UV divergence, and can be canceled by the local counterterm in Figure \ref{fig:localsub}(b). The second term is the sub-divergence from the region $l_2\gg l_1,p$, which matches the form of counterterm in Figure \ref{fig:localsub}(d).

\subsection{Degree of superficial UV and IR divergences}
\label{Degree}

The UV divergences of generic multiloop integrals have similar structures as the two-loop integral we studied in the last subsection. 
But before discussing these structures, in this subsection we present some formal definitions about integrals and their divergences. 

A Feynman integral is a product of several "\textbf{lines}", 
\begin{equation}\label{int-def-1}
F=\prod_{i=1}^N\mathtt{Line}(l_i)\ .
\end{equation}
The line $\mathtt{Line}(l_i)$ contains all numerators with of the form $l_i^{\mu}$ or $l_i\cdot k$, and all propagators\footnote{We will not discuss linear propagators like $l_i\cdot p$ in this work, and assume all propagators are quadratic in it loop momentum.} whose loop momentum is $l_i$, for example,
\begin{equation}\label{line}
\mathtt{Line}(l_i)=\frac{l_i^{\mu_1}\cdots l_i^{\mu_a}l_i\cdot k_1\cdots l_i\cdot k_b}
{(l_i^2)^{n_0}[(l_i+p_1)^2+M_1^2]^{n_1}[(l_i+p_2)^2+M_2^2]^{n_1}\cdots}\ .
\end{equation}

As an example, the integral $I$ in \eqref{phi3-I1} can be written as $I=\mathtt{Line}\{l_1,l_2,l_1-l_2\}$ in terms of lines, in which
\begin{equation}
\begin{aligned}\label{phi3-I2}
&\mathtt{Line}(l_1)=\frac{1}{(l_1^2)^2(l_1+p)^2},\ 
\mathtt{Line}(l_2)=\frac{1}{l_2^2},\ 
\mathtt{Line}(l_1-l_2)=\frac{1}{(l_1-l_2)^2}\ .\\
\end{aligned}
\end{equation}

The UV degree, $\omega$, of a line characterize the behavior of the line when $|l_i|\rightarrow \infty$, and the IR degree, $\omega_{ir}$, characterize the behavior of the line when $|l_i|\rightarrow 0$.
\begin{equation}
\mathtt{Line}(l_i)|_{l_i\rightarrow \infty}\rightarrow |l_i|^{\omega},\ 
\mathtt{Line}(l_i)|_{l_i\rightarrow 0}\rightarrow |l_i|^{-\omega_{ir}}.\ 
\end{equation}
For example, for \eqref{line}, $\omega=a+b-2\sum_i n_i$ and $\omega_{ir}=2n_0-a-b$.

An integral $\gamma$ is called a \textbf{sub-integral} of $F$, if the lines of $\gamma$ is a subset of the lines of $F$. Since sub-integrals will be used extensively all through this paper, we find it convenient to regard a integral as a set composed of all its lines:
\begin{equation}\label{int-def-2}
F=\{\mathtt{Line}(l_i)|i=1,2,\cdots,N\}\ .
\end{equation}
In this formalism, $\gamma$ is a sub-integral of $F$ can be simply denoted by $\gamma\subset F$. We will use \eqref{int-def-1} and \eqref{int-def-2} in different scenarios, and regard them as two representations of the same quantity.

For compactness, sometimes we will also use the following notation,
\begin{equation}
\mathtt{Line}\{l_1,\cdots,l_n\}\equiv \{\mathtt{Line}(l_1),\cdots,\mathtt{Line}(l_n)\}\ .
\end{equation}

$\gamma$ will be called an \textbf{IR sub-integral}, if $\forall \mathtt{Line}(l_i)\in F\setminus \gamma$, $l_i$ cannot be written as a linear combination of loop momenta of $\gamma$. The set of all IR sub-integrals of $F$ is denoted by $\Upsilon(F)$. The complement of an IR sub-integral is called a \textbf{UV sub-integral}.  The set of all UV sub-integral is denoted by $\Theta(F)$.  

The loop number of a sub-integral $\gamma$ will be denoted by $\mathbb{L}(\gamma)$. The loop number of an IR sub-integral equals the number of independent line momenta. The loop number of a UV sub-integral $\theta\subset F$ is given by
\begin{equation}
\mathbb{L}(\theta)=\mathbb{L}(F)-\mathbb{L}(F\setminus\theta)\ .
\end{equation}

\begin{figure}[htb]
\centering
\includegraphics[scale=0.5]{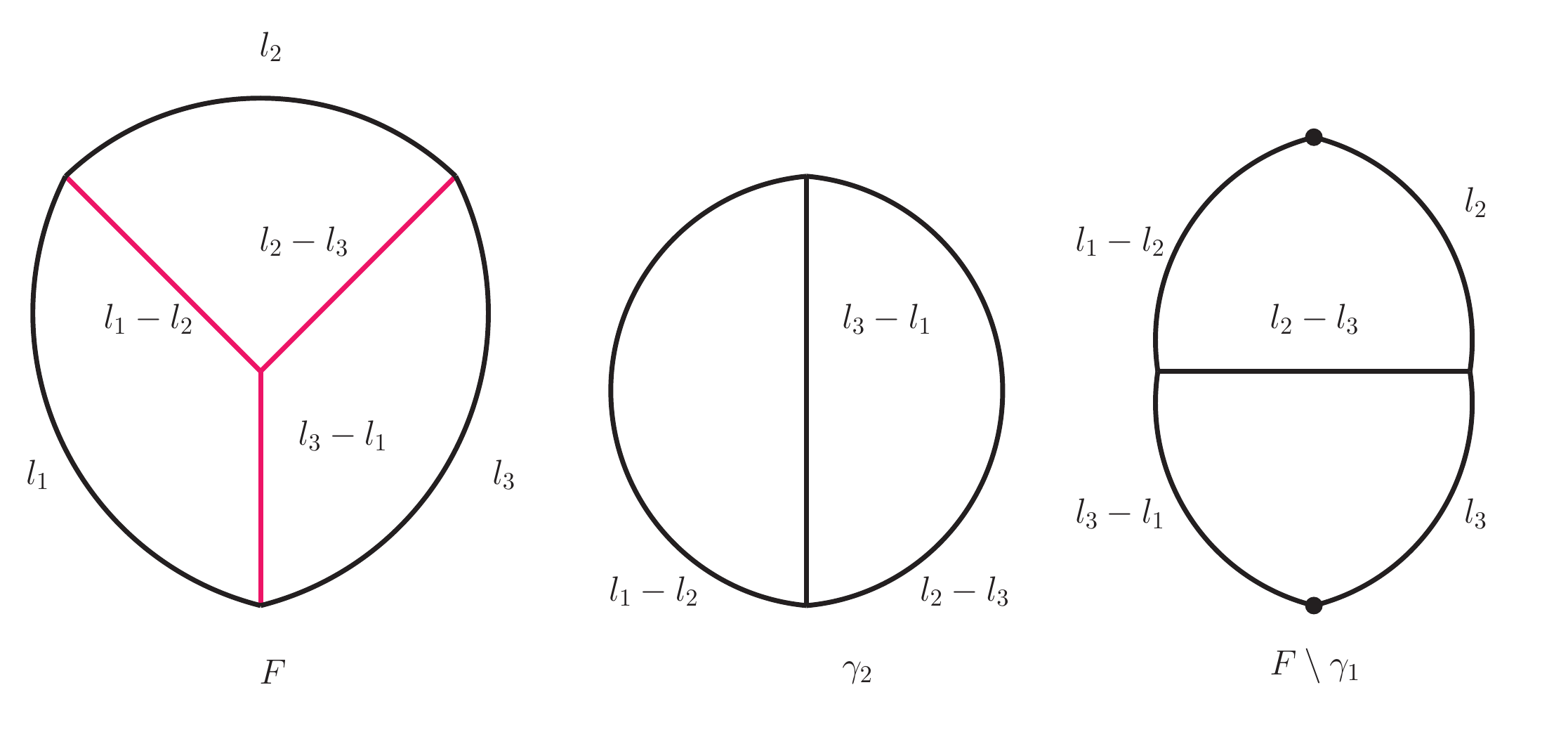}
\caption{An IR sub-integrals $\gamma_2$, and a UV sub-integral $F\setminus\gamma_1$.}
\label{fig:minimal maximal}
\end{figure}

For example, in Figure \ref{fig:minimal maximal}, $F=\mathtt{Line}\{l_1,l_2,l_3,l_2-l_3,l_1-l_3,l_1-l_2\}$.  $\gamma_1=\{\mathtt{Line}(l_1)\}$ is a 1-loop IR sub-integral, $\gamma_2=\mathtt{Line}\{l_1,l_2,l_{1}-l_2\}$ is a 2-loop IR sub-integral. $F\setminus \gamma_2=\mathtt{Line}\{l_3,l_{2}-l_3,l_1-l_{3}\}$ is a 1-loop UV sub-integral, and $F\setminus \gamma_1=\mathtt{Line}\{l_2,l_3,l_{2}-l_3,l_1-l_{3},l_{1}-l_2\}$ is a 2-loop UV sub-integral.

We can see that although the IR sub-integral $\gamma_2$ is a vacuum integral, it is not a "closed" diagram when it is embedded in the original integral (the lines in red color in Figure \ref{fig:minimal maximal}). In order to obtain the correct topology for $\gamma_2$, we should shrink $F\setminus\gamma_2$ to a point in $F$.

In a UV sub-integral, not all independent line momenta can be regraded as loop momenta of the sub-integral. For example, in the UV sub-integral $F\setminus \gamma_1=\mathtt{Line}\{l_3,l_{2}-l_3,l_1-l_{3}\}$, only $l_3$ is the loop momenta. $l_1$ and $l_2$ should be regarded as as external momenta of $F\setminus \gamma_1$.

Some useful properties of IR and UV sub-integrals are:
\begin{enumerate}
\item The empty set and the integral itself are both IR and UV sub-integrals.

\item The intersection of two IR sub-integrals is also an IR sub-integral.

\item The union of two UV sub-integrals is also a UV sub-integral.  

\item However, the union of two IR sub-integrals is not always an IR sub-integral.

\item If $\gamma\notin\{\emptyset,F\}$ is both a UV sub-integral and an IR sub-integral of $F$, then $F$ is called a \textbf{disconnected integral}, and $\gamma$ and $F\setminus\gamma$ are components of $F$. A disconnected Feynman diagram corresponds to a disconnected integral.

\end{enumerate}

 An $L$-loop (soft\footnote{We will only consider the IR divergence of vacuum integrals in this paper, so IR divergence always mean soft IR divergence.}) IR divergence is defined in the region where $L$ independent loop momenta 
$\{l^s_1,\cdots, l^s_ L\}$ become soft, i.e. $l^s_i$ is much smaller than external momenta, masses and other  (hard) loop momenta $l^h_i$. In this region, a propagator depending on both soft and hard loop momenta behaves like $\frac{1}{(l_h+l_s)^2+M^2}\rightarrow \frac{1}{l_h^2+M^2}$, and does not change the degree of IR divergence. So each IR divergence corresponds to an IR sub-integral, and the (superficial) degree of IR divergence is
\begin{equation}
\omega_{ir}(\gamma)=-d L+\sum_{L_i\in\gamma}\omega_{ir}(L_i).
\end{equation}

Similarly, each UV sub-divergence corresponds to a UV sub-integral. An $L$-loop UV divergence appears in the region where $L$ independent loop momenta $\{l^h_1,\cdots, l^h_ L\}$ become hard, i.e. $l^h_i$ is much larger than external momenta, masses and other (soft) loop momenta $l^s_i$. In this region, the propagator $\frac{1}{(l_h+l_s)^2+M^2}\rightarrow \frac{1}{l_h^2}$, and alters the degree of UV divergence.
\begin{equation}
\omega(\theta)=d L+\sum_{L_i\in\theta}\omega(L_i).
\end{equation}

\subsection{The BPHZ R-operation}
\label{subsection:BPHZ}
A milestone in the development of renormalization in quantum field theory is the famous BPHZ renormalization scheme \cite{Bogoliubov:1957gp, Hepp:1966eg, Zimmermann:1967fj, Zimmermann:1969jj}, which provides a standard approach to the systematic subtraction of divergences in Feynman diagrams.
A Feynman diagram $\Gamma$ can be rendered finite by the BPHZ R-operation:
\begin{equation}\label{BPHZ-R-1}
R(\Gamma)=\sum_{\theta\subset \Gamma}(\Gamma\setminus \theta)Z(\theta)\ ,
\end{equation}
in which $R(\Gamma)$ is the finite renormalized integral, $Z$ is the counterterms operator, and $\theta$ runs over all "spinneys" contained in $\Gamma$. We will not discuss the definition and properties of "spinneys" here, but would like to point out that they are equivalent to UV sub-integrals defined in the last subsection, so in our notation \eqref{BPHZ-R-1} becomes
\begin{equation}\label{BPHZ-R-2}
\sum_{\theta\in \Theta(\Gamma)}(\Gamma\setminus \theta)Z(\theta)\sim 0\ .
\end{equation}

An artifact in the definition of $Z$ is that $Z(\emptyset)=1$, and \eqref{BPHZ-R-2} can be written as
\begin{equation}\label{BPHZ-R-3}
\Gamma+\sum_{\theta\in \Theta'(\Gamma)}(\Gamma\setminus \theta)Z(\theta)\sim 0\ ,
\end{equation}
in which $\Theta'$ means the set of non-empty UV sub-integrals. 
The physical meaning of Eq. \eqref{BPHZ-R-3} is very clear: the UV divergence of $\Gamma$ can be canceled by the counterterms in all different regions. We would like to use a different form of \eqref{BPHZ-R-3} which we believe to be more natural when discussing the UV divergence of a single Feynman integral $F$,
\begin{equation}\label{uvdecom-2}
F\sim\sum_{\theta\in \Theta(F)}\mathcal{V}_{\theta}F, \ 
 \mathcal{V}_{\theta}F\equiv (F\setminus \theta)\mathbf{L}\theta\ ,
\end{equation}
in which $\mathbf{L}$ is the \textbf{local divergence operator}, and $\mathcal{V}_{\theta}$ is the \textbf{sub-divergence operator} corresponding to $\theta$.
$\mathbf{L}\theta=-Z(\theta)$ if $\theta$ is non-empty, but we define $\mathbf{L}\emptyset=0$ because $\emptyset$ corresponds to a constant.

The local divergence is a special type of sub-divergence which corresponds to the integral itself, and \eqref{uvdecom-2} can also be written as
\begin{equation}\label{uvdecom-3}
F\sim \mathbf{L}F+\sum_{\theta\in \Theta(F)}^{\theta\ne F}\mathcal{V}_{\theta}F\ ,
\end{equation}
which states that the UV divergence of an integral $F$ can be decomposed into a local divergence and various types of sub-divergences. Therefore, \eqref{uvdecom-2} and \eqref{uvdecom-3} will be called the \textbf{UV decomposition formula}.
The formula seems to be only a simple paraphrase of the BPHZ R-operation, but as will be discussed in Section \ref{IR-div}, it can be conveniently extended to integrals with IR divergences.

Since $\mathbf{L}F$ can be determined once the total UV divergence and all sub-divergences are known, and sub-divergences can be computed from lower loop local divergences, the UV decomposition formula actually gives a recursive definition of $\mathbf{L}$ operator. 
In this section, we will show that in the MSbar scheme, $\mathbf{L}$ can be uniquely determined for arbitrary Feynman integrals, given the following two requirements:
\begin{enumerate}[\textbf{R1}]
\item The local divergence has polynomial dependence on mass and external momenta.
\item An integral with $\omega<0$ has no local divergence.
\end{enumerate}
For completeness, in Section \ref{subsection:decom-proof} we will also directly prove that the UV decomposition formula holds with our definition of $\mathbf{L}$.

In the next subsection we would like to study the properties of $\mathbf{L}$ in the case of massless vacuum integrals, which will useful in order to prove \eqref{uvdecom-2} for generic Feynman integrals.

\subsection{Massless vacuum integrals}
\label{subsection:massless}

Suppose $V$ is a $L$-loop massless vacuum integral with $\omega\ge0$. The IR divergence of $V$ can be regulated by adding mass $m$ to all the propagators, and the resulting massive integral will be denoted by $V(m)$. In order for $\mathbf{L}V(m)$ to have the correct mass dimension, it must be of the form
\begin{equation}\label{lvm1}
\mathbf{L}V(m)=f(\epsilon)m^{\omega}
\equiv m^{\omega}\sum_{i=1}^k \frac{f_i}{\epsilon^i} \ ,
\end{equation}
in which $k$ is a positive integer and $f_i$ are constants. 

Using $V=V(0)$, the $m\rightarrow 0$ limit of \eqref{lvm1} gives
\begin{equation}
\begin{aligned}
&\mathbf{L}V=f(\epsilon),&\text{ if }\omega=0,\ \\
&\mathbf{L}V=0, &\text{ if }\omega>0.\ \\
\end{aligned}
\end{equation}
This means a massless vacuum integral has no local divergence unless it is a CV, and the local divergence of a CV is unaltered after IR regulation.

Let $V$ be a CV, and $\theta$ be a $L_1$-loop UV sub-integral of $V$ with $\omega(\theta)\ge0$, then the local divergence of $\theta$ has the form
\begin{equation}
\mathbf{L}\theta=f_0(\epsilon)(l^s_i)^{\omega(\theta)}+f_1(\epsilon)(l^s_i)^{\omega(\theta)-1}m+\cdots
+f_{\omega(\theta)}(\epsilon)m^{\omega(\theta)}\ ,
\end{equation}
therefore the sub-divergence corresponding to $\theta$ can be written as
\begin{equation}\label{sub-vm-1}
\mathcal{V}_{\theta}V(m)=(V(m)\setminus \theta)\mathbf{L}\theta
=f_{\theta}(\epsilon)m^{-2(L-L_1)\epsilon}\ .
\end{equation}

Using these expressions, we can prove the following property of CV:
\begin{enumerate}[\textbf{CV1}]
\item If \eqref{uvdecom-2} holds for any integral with at most $L_0$-loop sub-divergence, then it also holds for $(L_0+1)$-loop CV.
\end{enumerate}
Let $V$ be a $(L_0+1)$-loop CV, and $V(m)$ be the IR regulated integral. Then using relations like
\begin{equation}
\frac{1}{(l^2)^n}-\frac{1}{(l^2+m^2)^n}
=\frac{m^2}{(l^2)^n(l^2+m^2)}+\frac{m^2}{(l^2)^{n-1}(l^2+m^2)^2}+\cdots+\frac{m^2}{l^2(l^2+m^2)^n}\ ,
\end{equation}
the integral $V-V(m)$ can be rearranged into a form which has $\omega<0$, and so it has at most $L_0$-loop sub-divergence. Since we assumed \eqref{uvdecom-2} holds for any integral with at most $L_0$-loop sub-divergence, we have
\begin{equation}
V-V(m)\sim\sum_{\theta\in\Theta,\theta\ne V}\mathcal{V}_{\theta}\Bigl[V-V(m)\Bigr]\ .
\end{equation}
Using \eqref{sub-vm-1}, The UV divergence of $V$ can be expressed by
\begin{equation}
\begin{aligned}
V\sim &\sum_{\theta\in\Theta,\theta\ne V}\mathcal{V}_{\theta}V+V(m)
-\sum_{\theta\in\Theta,\theta\ne V}\mathcal{V}_{\theta}V(m)\\
\sim &\sum_{\theta\in\Theta,\theta\ne V}\mathcal{V}_{\theta}V+v(\epsilon)m^{-2(L_0+1)\epsilon}
-\sum_{\theta\in\Theta,\theta\ne V}f_{\theta}(\epsilon)m^{-2[L_0+1-\mathbb{L}(\theta)]\epsilon}\ ,\\
\end{aligned}\label{sub-vm-2}
\end{equation}
in which we used \eqref{sub-vm-1}, and expressed $V_m$ by $v(\epsilon)m^{-2(L_0+1)\epsilon}$.

Since $V$ does not depend on $m$, the $\epsilon$-pole part of the quantity
\begin{equation}
f(\epsilon,m)\equiv v(\epsilon)m^{-2(L_0+1)\epsilon}
-\sum_{\theta\in\Theta,\theta\ne V}f_{\theta}(\epsilon)m^{-2[L_0+1-\mathbb{L}(\theta)]\epsilon}\ ,
\end{equation}
must be free of $m$-dependence:
\begin{equation}
f(\epsilon,m)=\sum_{i=1}^k\frac{f_k}{\epsilon^k}+\mathcal{O}(\epsilon^0)\ .
\end{equation}
We will define $\mathbf{L}V\equiv \sum_{i=1}^k\frac{f_k}{\epsilon^k}$, then
\begin{equation}
V\sim \sum_{\theta\in\Theta,\theta\ne V}\mathcal{V}_{\theta}V+\mathbf{L}V
=\sum_{\theta\in\Theta}\mathcal{V}_{\theta}V\ ,
\end{equation}
which finishes the proof of \textbf{CV1}.

Before ending this subsection, let us point out that \eqref{sub-vm-2} provides a method to compute the local divergence of CV:
\begin{equation}\label{cv-local}
\mathbf{L}V\sim V(m)
-\sum_{\theta\in\Theta,\theta\ne V}\mathcal{V}_{\theta}V(m)\ .
\end{equation}
The local divergence is obtained by adding masses to the CV, and subtract the sub-divergences from the the total divergence of the massive integral.

\subsection{The asymptotic expansion}
\label{subsection:asymptotic}

As shown in \eqref{1 loop massless decomposition}, the local divergence of 1-loop integrals can be computed from the local divergence of 1-loop CV. In the following we will show that this is also true for generic multiloop integrals.

An integral $F$ depends on loop momenta $l_i$, external momenta\footnote{In a UV sub-integral only $l^h_i$ are regarded as loop momenta, while $l^s_i$ are regarded as external momenta.} $p_i$ and mass $M_i$,
\begin{equation}
F=F(l_i,p_i,M_i)\ .
\end{equation}

Since $|l_i|\ge |p_i|, M_i$, we can define a large number $\Lambda$, so that $\hat{l}_i\equiv\frac{l_i}{\Lambda}$ are in the same order as $|p_i|$ and $M_i$. Now let us expand the integral into asymptotic series around large $\Lambda$,
\begin{equation}\label{asym-expand-1}
F=F(\Lambda \hat{l}_i,p_i,M_i)=\sum_{i=0}^{\omega(F)}\frac{1}{\Lambda^i}F_k(\hat{l}_i,p_i,M_i)+\mathcal{O}(\frac{1}{\Lambda})\ ,
\end{equation}
in which we have included the $\Lambda^{d\mathbb{L}(F)}$ term from the loop integration measure.
The $\frac{1}{\Lambda^i}$ factors can be absorbed if we replace $\hat{l}_i\rightarrow \frac{l_i}{\Lambda}$, and \eqref{asym-expand-1} can be written as
\begin{equation}\label{asym-expand-2}
F=\mathcal{A}F+R(F),\ 
\mathcal{A}F=\sum_{i=0}^{\omega(F)}\mathcal{A}^iF
\equiv\sum_{i=0}^{\omega(F)}F_k(l_i,p_i,M_i)\ ,
\end{equation}
in which $\mathcal{A}$ will be called the \textbf{asymptotic expansion operator}, and $R(F)$ is remainder term with $\omega<0$.

Obviously $\mathcal{A}^iF$ has polynomial dependence on $p_i$ and $M_i$, and it is a massless vacuum integrals with $\omega=i$. The local divergence of $\mathcal{A}^iF$ vanishes unless $i=0$, so we have 
\begin{equation}\label{local-from-cv}
\mathbf{L}F=\mathbf{L}\mathcal{A}^0F\ .
\end{equation}

Since $\mathcal{A}^0F$ is a CV, \eqref{local-from-cv} implies that the local divergence of general integrals can be determined by the local divergence of CV. But as discussed in Section \ref{subsection:massless}, the local divergence of a $(L_0+1)$-loop CV can be determined by the local divergence of some $L\le L_0$ loop integrals. Therefore, as long as the UV decomposition formula \eqref{uvdecom-2} holds, the local divergence of a CV can be uniquely defined, which is determined recursively from the local divergence of lower loop CV.

Another implication of \eqref{local-from-cv} is that if $d$ is not an even integer, UV divergence only appear at certain loops. This is will be discussed in Appendix \ref{appendix:fractional}.

Using \eqref{local-from-cv}, the sub-divergence operator can be written as
\begin{equation}
\mathcal{V}_{\theta}F=(F\setminus \theta)\mathbf{L}\mathcal{A}^0\theta\ .
\end{equation}
The following property of $\mathcal{A}^0$ will be useful later:
the $\mathcal{A}^0$ operator does not increase the divergence degree of $F$ if it acts on $\theta\in\Theta(F)$:
\begin{equation}\label{A0-degree}
\omega\Bigl[(F\setminus \theta)\mathcal{A}^0\theta\Bigr]\le \omega(F)\ .
\end{equation}
To prove \eqref{A0-degree}, let us denote the loop momenta of $F\setminus \theta$ and $\theta$ by $l^s_i$ and $l^h_i$, respectively, then $\theta$ can be written as $\theta(l^h_i,l^s_i,p_i, M_i)$. $\mathcal{A}^0\theta$ has polynomial dependence on $l^s_i,p_i, M_i$, and it has the form
\begin{equation}
\mathcal{A}^0\theta=\sum_{a,b\ge 0}(p_i)^a(M_i)^b(l^s_i)^{\omega(\theta)-a-b}A_{ab}(l^h_i)\ ,
\end{equation}
in which $\omega(A_{ab})=0$. 
Combined with the $F\setminus \theta$ term,
\begin{equation}
\omega\Bigl[(F\setminus \theta)(l^s_i)^{\omega(\theta)-a-b}A_{ab}(l^h_i)\Bigr]
=\omega(F)-\omega(\theta)+\omega(\theta)-a-b=\omega(F)-a-b\le \omega(F).
\end{equation}
So all terms in $(F\setminus \theta)\mathcal{A}^0\theta$ have $\omega\le \omega(F)$.

Let us demonstrate the computation of $\mathcal{A}$ by revisiting the sub UV divergence of the integral $I$ in \eqref{phi3-I1} and \eqref{phi3-I2}. 
The sub-divergence of the integral corresponds to the UV sub-integral $\theta=\mathtt{Line}\{l_2,l_1-l_2\}$, the only hard momenta is $l_2$. Replacing $l_2\rightarrow\frac{\hat{l}_2}{\Lambda}$, we obtain
\begin{equation}
\begin{aligned}
\theta=&\frac{1}{(l_1-l_2)^2l_2^2}\rightarrow \frac{\Lambda^6}{(l_1-\Lambda^2\hat{l}_2)^2\Lambda^2\hat{l}_2^2}
=\frac{\Lambda^2}{(\hat{l}_2^2-\frac{2l_1\cdot \hat{l}_2}{\Lambda}+\frac{l_1^2}{\Lambda^2})\hat{l}_2^2}\ \\
=&\frac{\Lambda^2}{(\hat{l}_2^2)^2}
+\frac{\Lambda^2(\frac{2l_1\cdot \hat{l}_2}{\Lambda}-\frac{l_1^2}{\Lambda^2})}{(\hat{l}_2^2)^3}
+\frac{\Lambda^2(\frac{2l_1\cdot \hat{l}_2}{\Lambda})^2}{(\hat{l}_2^2)^4}+\cdots\ ,\\
\end{aligned}
\end{equation}
where in the first line we have included a $\Lambda^6$ factor from integration measure.

Therefore the corresponding asymptotic expansion is given by:
\begin{equation}
\begin{aligned}
\mathcal{A}\theta=&\mathcal{A}^2\theta+\mathcal{A}^1\theta+\mathcal{A}^0\theta\ ,\\
\mathcal{A}^2\theta=&\frac{1}{(l_2^2)^2},\ 
\mathcal{A}^1\theta=\frac{2l_1\cdot l_2}{(l_2^2)^3},\ 
\mathcal{A}^0\theta=-\frac{l_1^2}{(l_2^2)^3}+\frac{(2l_1\cdot l_2)^2}{(l_2^2)^4}.\ \\
\end{aligned}\label{2-loop-asymptotic}
\end{equation}

The local divergences of one-loop integrals are equal to the total UV divergences, so from \eqref{fn} one obtains:
\begin{equation}
\begin{aligned}
&\mathbf{L}\frac{l^{\mu_1}\cdots l^{\mu_{2a}}}{(l^2)^n}= \frac{\delta_{n-a,\frac{d}{2}}}{2^a\Gamma(n)\epsilon}
\eta_s^{\mu_1 \cdots \mu_{2a}}\ .\\
\end{aligned}\label{local-1-loop}
\end{equation}
The sub-divergence in of $I$ is given by
\begin{equation}
\begin{aligned}
\mathcal{V}_{\theta}I=&\frac{1}{(l_1^2)^2(l_1+p)^2}\mathbf{L}\Bigl[-\frac{l_1^2}{(l_2^2)^3}+\frac{(2l_1\cdot l_2)^2}{(l_2^2)^4}\Bigr]
=-\frac{1}{6\epsilon}\frac{1}{l_1^2(l_1+p)^2}\ ,\\
\end{aligned}
\end{equation}
which is in agreement with the sub-divergence in \eqref{2 loop decomposition example 1}.

\subsection{The proof of UV decomposition}
\label{subsection:decom-proof}

Now we are ready to prove the UV decomposition formula \eqref{uvdecom-2} for generic Feynman integrals. Still we will prove it by induction: let us assume \eqref{uvdecom-2} holds if the integral has at most $L_0$ loop sub-divergences, and prove that it also holds if the integral has at most $L_0+1$ loop sub-divergences.

Suppose $F$ has at most $L_0+1$ loop sub-divergences. Let $\Theta_{1}$ be the set of $L_0+1$ loop UV sub-integrals, and let $\Theta_0$ be the set of $L\le L_0$ loop UV sub-integrals. We can split $F$ into $F_0$ and $F_1$, 
\begin{equation}\label{uv-prove-1}
F_0\equiv F-\sum_{\theta\in \Theta_{1}}(F\setminus \theta)\mathcal{A}^0\theta,\ 
F_1\equiv \sum_{\theta\in \Theta_{1}}(F\setminus \theta)\mathcal{A}^0\theta\ .
\end{equation}

First we observe that $F_0$ has no $L_0+1$ loop sub-divergences. To see this, let $\eta$ be an arbitrary $L_0+1$ loop UV sub-integral of $F$, then
\begin{equation}
\begin{aligned}
&\mathcal{V}_{\eta}\Bigl[F-\sum_{\theta\in \Theta_{1}}(F\setminus \theta)\mathcal{A}^0\theta\Bigr]
=\mathcal{V}_{\eta}F-\sum_{\theta\in \Theta_{1}}\mathcal{V}_{\eta}\Bigl[(F\setminus \theta)\mathcal{A}^0\theta\Bigr]
=-\sum_{\theta\in \Theta_{1}}^{\theta\ne \eta}\mathcal{V}_{\eta}\Bigl[(F\setminus \theta)\mathcal{A}^0\theta\Bigr]\ ,\\
\end{aligned}
\end{equation}
where we used $\mathcal{V}_{\eta}\Bigl[(F\setminus \eta)\mathcal{A}^0\eta\Bigr]=(F\setminus \eta)\mathbf{L}\mathcal{A}^0\eta=\mathcal{V}_{\eta}F$ in the last step.

Notice that $(F\setminus \theta)\mathcal{A}^0\theta$ is a disconnected integral with two components $F\setminus \theta$ and $\mathcal{A}^0\theta$. 
In Appendix \ref{appendix:disconnected} the local divergences of disconnected integrals are studied.
Let $\eta_1=\eta\cap (F\setminus \theta)$, and $\eta_2=\eta\cap \mathcal{A}^0\theta$, then using \eqref{disconnected-1} we find
\begin{equation}
\mathcal{V}_{\eta}\Bigl[(F\setminus \theta)\mathcal{A}^0\theta\Bigr]
=\mathcal{V}_{\eta_1}(F\setminus \theta)\mathcal{V}_{\eta_2}\mathcal{A}^0\theta\ .
\end{equation}
We will denote the loop momenta of $\eta_1$ and $\mathcal{A}^0\theta$ by $l_{\eta,1}$ and $l_{\theta}$, respectively. $\{l_{\eta,1},l_{\theta}\}$ are also the loop momenta of  $\eta\cup \theta$.  Since $\mathbb{L}(\eta\cup \theta)>L_0+1$, using the induction assumption we have $\omega(\eta\cup \theta)<0$. Using \eqref{A0-degree}, the action of $\mathcal{A}^0$ on $\theta$ will not increase the UV divergence degree of $\eta\cup \theta$, so we have
\begin{equation}\label{g-eta1}
\omega(\mathcal{A}^0\theta)+\omega(\eta_1)\le\mathtt{g}(\eta\cup \theta)<0.
\end{equation}
Since $\omega(\mathcal{A}^0\theta)=0$, \eqref{g-eta1} implies $\omega(\eta_1)<0$, and consequently $\mathcal{V}_{\eta_1}(F\setminus \theta)=0$. So we have found that $\mathcal{V}_{\eta}\Bigl[(F\setminus \theta)\mathcal{A}^0\theta\Bigr]=0$, and $F_0$ has no $L_0+1$ loop divergence.

By the induction assumption $F_0$ has the following UV decomposition,
\begin{equation}
\begin{aligned}
&F_0\sim\sum_{\eta\in \Theta_0}\mathcal{V}_{\eta}
\Bigl[F-\sum_{\theta\in \Theta_{1}}(F\setminus \theta)\mathcal{A}^0\theta\Bigr]\ .\\
\end{aligned}\label{UVprove2}
\end{equation}
In \eqref{UVprove2}, if $\eta\nsubseteq\theta$, it can be shown the sub-divergence vanishes for the same reason as the $\mathbb{L}(\eta)=L_0+1$ case above. If $\eta\subset \theta$,
\begin{equation}
\begin{aligned}
&\mathcal{V}_{\eta}\Bigl[(F\setminus \theta)\mathcal{A}^0\theta\Bigr]
=(F\setminus \theta)\mathcal{V}_{\eta}\mathcal{A}^0\theta ,\\
\end{aligned}
\end{equation}
and \eqref{UVprove2} becomes
\begin{equation}
\begin{aligned}
&F_0
\sim\sum_{\eta\in \Theta_0}\mathcal{V}_{\eta}(F)
-\sum_{\theta\in \Theta_{1}}\sum_{\eta\in \Theta(\theta)}^{\eta\ne \theta}
(F\setminus \theta)\mathcal{V}_{\eta}\mathcal{A}^0\theta\ .\\
\end{aligned}\label{UVprove3}
\end{equation}

The second term in the r.h.s. of \eqref{UVprove3} can be combined with $F_1$,
\begin{equation}
\begin{aligned}
&F_1
-\sum_{\theta\in \Theta_{1}}\sum_{\eta\in \Theta(\theta)}^{\eta\ne\theta}(F\setminus \theta)\mathcal{V}_{\eta}\mathcal{A}^0\theta
=\sum_{\theta\in \Theta_{1}}(F\setminus \theta)
\Bigl[\mathcal{A}^0\theta-\sum_{\eta\in \Theta(\theta)}^{\eta\ne\theta}\mathcal{V}_{\eta}\mathcal{A}^0\theta\Bigr]\  .\\
\end{aligned}\label{UVprove4}
\end{equation}
Since $\mathcal{A}^0\theta$ is a $L_0+1$ loop CV, using \textbf{CV1} in Section \ref{subsection:massless}, it satisfies the UV decomposition,
\begin{equation}
\mathcal{A}^0\theta
=\mathbf{L}\mathcal{A}^0\theta
+\sum_{\eta\in \Theta(\theta)}^{\eta\ne\theta}\mathcal{V}_{\eta}\mathcal{A}^0\theta
+\text{UV finite terms}\ .
\end{equation}
The $F\setminus \theta$ term in \eqref{UVprove4} is UV finite, because $F$ has no $\mathbb{L}>L_0+1$ loop sub-divergences, therefore $\mathcal{A}^0\theta-\sum_{\eta\in \Theta(\theta)}^{\eta\ne\theta}\mathcal{V}_{\eta}\mathcal{A}^0\theta$ can be replaced by $\mathbf{L}\mathcal{A}^0$,
\begin{equation}
\begin{aligned}
&F_1
-\sum_{\theta\in \Theta_{1}}\sum_{\eta\in \Theta(\theta)}^{\eta\ne\theta}(F\setminus \theta)\mathcal{V}_{\eta}\mathcal{A}^0\theta
\sim\sum_{\theta\in \Theta_{1}}(F\setminus \theta)\mathbf{L}\mathcal{A}^0\theta
=\sum_{\theta\in \Theta_{1}}\mathcal{V}_{\theta}F\ .\\
\end{aligned}\label{UVprove5}
\end{equation}
Combined with the first term on the r.h.s. of \eqref{UVprove3}, $F$ has the following decomposition:
\begin{equation}
F\sim \sum_{\eta\in \Theta_0}\mathcal{V}_{\eta}F
+\sum_{\theta\in \Theta_{1}}\mathcal{V}_{\theta}F
=\sum_{\theta\in \Theta(F)}\mathcal{V}_{\theta}F\ .
\end{equation}
which completes the proof of \eqref{uvdecom-2}.

\section{The local divergence of 2 and 3 loop scalar CV}
\label{CV-2-3}
The evaluation of local divergence of CV is the kernel problem in UV decomposition, because the local divergences of generic Feynman integrals can be obtained from that of CV. 
Eq. \eqref{cv-local} provides an efficient approach to this problem.
We will demonstrate this approach by evaluating the local divergence of two and three loop scalar CV, which are CV without tensor structures like $l_1^{\mu_1}\cdots l_n^{\mu_n}$ in the numerators.

The basis of 2 loop scalar CV can be chosen as $\{I_{n_1n_2n_3}\}$, with
\begin{equation}\label{2-loop-cv}
I_{n_1n_2n_3}=\frac{1}{(l_1^2)^{n_1}(l_2^2)^{n_2}[(l_1-l_2)^2]^{n_3}}\ .
\end{equation}
The critical condition requires $n_1+n_2+n_3=d$. In Section \ref{subsection:local-div-2-loop} we evaluate the local divergence for arbitrary choice of $n_i$.

The basis of 3 loop massless scalar vacuum integrals can be chosen as $\{I_{n_1n_2n_3n_4n_5n_6}\}$, with
\begin{equation}
I_{n_1n_2n_3n_4n_5n_6}=\frac{1}{(l_1^2)^{n_1}(l_2^2)^{n_2}(l_3^2)^{n_3}[(l_2-l_3)^2]^{n_4}
[(l_3-l_1)^2]^{n_5}[(l_1-l_2)^2]^{n_6}}\ .
\end{equation}
Critical condition requires $\sum_{i=1}^6 n_i=\frac{3d}{2}$. We will demonstrate the computation by 3 examples: $I_{311211}$, $I_{441111}$ and $I_{411411}$.

\subsection{The local divergence of 2 loop scalar CV}
\label{subsection:local-div-2-loop}


Without loss of generality, we can assume $n_1\ge n_2\ge n_3$ in \eqref{2-loop-cv}. As discussed in Section \ref{subsection:massless}, 
we can regulate the IR divergence of the integral by adding masses to the propagators without changing its local divergence. The evaluation of a vacuum integral usually becomes more and more difficult as more propagators become massive, so it would be preferable to add mass to as less propagators as possible.

The integral $I_{n_1n_2n_3}$ has 1 loop IR divergences when $n_i\ge \frac{d}{2}$. Suppose $n_2<\frac{d}{2}$, then the IR divergence can be regulated by a single mass, $I(m)\equiv\frac{1}{(l_1^2+m^2)^{n_1}(l_2^2)^{n_2}(l_3^2)^{n_3}}$. The integral is recursively one loop, and can be easily evaluated \cite{Smirnov:2006ry},
\begin{equation}
\begin{aligned}
&I(m)
=\frac{e^{2\gamma\epsilon}\Gamma(2\epsilon)\Gamma(\frac{d}{2}-n_1+\epsilon)\Gamma(\frac{d}{2}-n_2-\epsilon)\Gamma(\frac{d}{2}-n_3-\epsilon)}
{\Gamma(n_1)\Gamma(n_2)\Gamma(n_3)\Gamma(\frac{d}{2}-\epsilon)(m^2)^{2\epsilon}}\ .\\
\end{aligned}\label{totaluv2loop1}
\end{equation}

The gamma functions behave as
\begin{equation}
\begin{aligned}
e^{\gamma\epsilon}\Gamma(n+\epsilon)
=&\Gamma(n)\Bigl[1+H_{n-1}\epsilon\Bigr]+\mathcal{O}(\epsilon^2),\ &n>0,\ \ \\
e^{\gamma\epsilon}\Gamma(-n+\epsilon)
=&\frac{(-1)^n}{n!}\Bigl[\frac{1}{\epsilon}+H_{n}\Bigr]+\mathcal{O}(\epsilon),\ &n\ge0,\ \\
\end{aligned}
\end{equation}
where $H_n=1+\frac{1}{2}+\cdots +\frac{1}{n}$ is the harmonic number.

If $n_1<\frac{d}{2}$, the integral has no sub-divergence, and the local divergence is
\begin{equation}
\mathbf{L}I_{n_1n_2n_3}=\frac{\Gamma(\frac{d}{2}-n_1)\Gamma(\frac{d}{2}-n_2)\Gamma(\frac{d}{2}-n_3)}
{2\epsilon\Gamma(n_1)\Gamma(n_2)\Gamma(n_3)\Gamma(\frac{d}{2})}\ .
\end{equation}

If $n_1\ge \frac{d}{2}$, the integral has a divergent sub-integral $\gamma_1=\mathtt{line}\{l_2,l_1-l_2\}$,
\begin{equation}
\begin{aligned}
\mathcal{V}_{\gamma_1}I(m)
=&\frac{1}{(l_1^2+m^2)^{n_1}}
\mathbf{L}\frac{1}{(l_2^2)^{n_2}[(l_1-l_2)^2]^{n_3}}\\
=&\frac{(l_1^2)^{n_1-\frac{d}{2}}}{(l_1^2+m^2)^{n_1}}
\frac{(-1)^{n_1-\frac{d}{2}}\Gamma(\frac{d}{2}-n_2)\Gamma(\frac{d}{2}-n_3)}
{\Gamma(n_1)\Gamma(n_2)\Gamma(n_3)(n_1-\frac{d}{2})!\epsilon}\\
= &
\frac{(-1)^{n_1-\frac{d}{2}}e^{\gamma\epsilon}\Gamma(\frac{d}{2}-n_2)\Gamma(\frac{d}{2}-n_3)\Gamma(\epsilon)\Gamma(n_1-\epsilon)}
{\epsilon\Gamma^2(n_1)\Gamma(n_2)\Gamma(n_3)\Gamma(\frac{D}{2})(n_1-\frac{d}{2})!(m^2)^{\epsilon}}\\
\end{aligned}\label{subuv2loop1}
\end{equation}
In the first line, the local divergence equals the total divergence of bubble diagram because the diagram has no IR divergence ($n_2,n_3<\frac{d}{2}$).

Subtract the sub-divergence \eqref{subuv2loop1} from the total divergence \eqref{totaluv2loop1}, we find
\begin{equation}
\begin{aligned}
\mathbf{L}I_{n_1n_2n_3}=& \frac{(-1)^{n_1-\frac{d}{2}}\Gamma(\frac{d}{2}-n_2)\Gamma(\frac{d}{2}-n_3)}{2(n_1-\frac{d}{2})!\Gamma(n_1)\Gamma(n_2)\Gamma(n_3)\Gamma(\frac{d}{2})}
\Bigl[-\frac{1}{\epsilon^2}+\frac{1}{\epsilon}Z_{n_1n_2n_3}\Bigr]\ ,\\
Z_{n_1n_2n_3}\equiv &2H_{n_1-1}+H_{n_1-\frac{d}{2}}-H_{\frac{d}{2}-1}-H_{\frac{d}{2}-n_2-1}-H_{\frac{d}{2}-n_3-1}\ .\\
\end{aligned}\label{local n1}
\end{equation}

Now we consider the case $n_2\ge \frac{d}{2}$. Using $n_1+n_2+n_3=d$ and $n_1\ge n_2\ge n_3$, we find $n_3\le 0$, and
\begin{equation}
\begin{aligned}
\mathbf{L}I_{n_1n_2n_3}=&\sum_{0\le i,j\le -n_3}\frac{(-n_3)!}{i!j!(-n_3-i-j)!}
\mathbf{L}\frac{(-2l_1\cdot l_2)^{-n_3-i-j}}{(l_1^2)^{n_1-i}(l_2^2)^{n_2-j}}\\
=&\sum_{i=\frac{d}{2}}^{n_2}\frac{(-n_3)!}{(n_1-i)!(n_2-i)!(2i-d)!}\mathbf{L}\frac{(-2l_1\cdot l_2)^{2i-d}}{(l_1^2)^{i}(l_2^2)^{i}}\ \\
=&\sum_{i=\frac{d}{2}}^{n_2}\frac{(-n_3)!}{(n_1-i)!(n_2-i)!(i-\frac{d}{2})!\Gamma(i)\Gamma(\frac{d}{2})}
\Bigl[-\frac{1}{\epsilon^2}+\frac{1}{\epsilon}(H_{i-1}-H_{\frac{d}{2}-1})\Bigr]\ ,\\
\end{aligned}\label{n3le0-1}
\end{equation}
where we used \eqref{n3le0-3} to derive the last line.

In summary, the local divergences of 2 loop scalar CV are:
\begin{equation}
\begin{aligned}
\mathbf{L}I_{n_1n_2n_3}=&
\left\{
\begin{aligned}
&\frac{\Gamma(\frac{d}{2}-n_1)\Gamma(\frac{d}{2}-n_2)\Gamma(\frac{d}{2}-n_3)}
{2\epsilon\Gamma(n_1)\Gamma(n_2)\Gamma(n_3)\Gamma(\frac{d}{2})},\ 
&n_i<\frac{d}{2}\\
& \frac{(-1)^{n_1-\frac{d}{2}}\Gamma(\frac{d}{2}-n_2)\Gamma(\frac{d}{2}-n_3)}{2(n_1-\frac{d}{2})!\Gamma(\frac{d}{2})\prod_{i=1}^3\Gamma(n_i)}
\Bigl(-\frac{1}{\epsilon^2}+\frac{Z_{n_1n_2n_3}}{\epsilon}\Bigr),
&n_2,n_3<\frac{d}{2}\le n_1\\
&\sum_{i=\frac{d}{2}}^{\min(n_1,n_2)}\frac{(-n_3)!\Bigl[-\frac{1}{\epsilon^2}+\frac{1}{\epsilon}(H_{i-1}-H_{\frac{d}{2}-1})\Bigr]
}{(n_1-i)!(n_2-i)!(i-\frac{d}{2})!\Gamma(i)\Gamma(\frac{d}{2})},\ 
&n_3<\frac{d}{2}\le n_1,n_2\\
\end{aligned}
\right. \\
\end{aligned}
\end{equation}
in which
\begin{equation}
\begin{aligned}
Z_{n_1n_2n_3}\equiv &2H_{n_1-1}+H_{n_1-\frac{d}{2}}-H_{\frac{d}{2}-1}-H_{\frac{d}{2}-n_2-1}-H_{\frac{d}{2}-n_3-1}\ .\\
\end{aligned}
\end{equation}

\subsection{IR regulation by adding a single mass}
\label{subsection:single-mass}

In the most simple case, for example $I_{311211}$ in 6-d, a single massive propagator is suffice to regulate all IR divergences,
\begin{equation}
I^{m}_{311211}\equiv\frac{1}{(l_1^2+m^2)^3l_2^2l_3^2[(l_2-l_3)^2]^2(l_3-l_1)^2(l_1-l_2)^2}\ .
\end{equation}
To evaluate the integral, first we perform an IBP reduction using FIRE \cite{Smirnov:2014hma},
\begin{equation}
\begin{aligned}
I^{m}_{311211}
=&\frac{4(D-5)(D-2)(\frac{3D}{2}-8)_5}{(D-6)^{2}(D-4)^2m^{10}}(I^m_{101101}+I^m_{110110})\\
&+\frac{(D-3)(3D-16)(3D-14)}{2(D-6)m^{8}}I^m_{111011}\ .\\
\end{aligned}
\end{equation}
The analytic expression of each of these 3 master integrals can be easily obtained by evaluating several 1-loop simple integrals successively (these integrals are called 1-loop-reducible integrals), 
\begin{equation}
\begin{aligned}
&I^m_{101101}=I^m_{110110}=
-3e^{3\epsilon \gamma}m^{3D-8}\Gamma(3-\frac{3D}{2})\Gamma(3-D)\Gamma^2(-1+\frac{D}{2})\ ,\\
&I^m_{111011}=\frac{e^{3\epsilon \gamma}\Gamma(5-\frac{3D}{2})\Gamma(2-\frac{D}{2})^{2}\Gamma(-1+\frac{D}{2})^{4}\Gamma(-4+\frac{3D}{2})}{m^{10-3D}\Gamma(-2+D)^{2}\Gamma(\frac{D}{2})}\ ,\\
\end{aligned}
\end{equation}
and the total divergence of the integral is
\begin{equation}
\begin{aligned}
I^{m}_{311211}
\sim&\frac{1}{24\epsilon ^{2}}+\frac{29-24\zeta_3-36\ln m}{144\epsilon } .\\
\end{aligned}
\end{equation}
The integral has a 2 loop sub-divergence, corresponding to the UV sub-integral $\theta=F\setminus\{\mathtt{line}(l_1)\}$,
\begin{equation}
\begin{aligned}
&\mathcal{V}_{\theta}I^{m}_{311211}
=\frac{1}{(l_1^2+m^2)^3}\mathbf{L}\frac{1}{(l_2^2)^2(l_3^2)^2[(l_2-l_3)^2]^2}
\sim \frac{1}{8\epsilon ^{2}}-\frac{\ln m}{4\epsilon }\ ,\\
\end{aligned}
\end{equation}
and the local divergence is found to be
\begin{equation}
\begin{aligned}
&\mathbf{L}I_{311211}
=-\frac{1}{12\epsilon ^{2}}+\frac{29-24\zeta_3}{144\epsilon }\ .\\
\end{aligned}
\end{equation}

\subsection{IR regulation by adding two masses}
\label{IR regulation by adding two masses}

\begin{figure}[htb]
\centering
\includegraphics[scale=0.5]{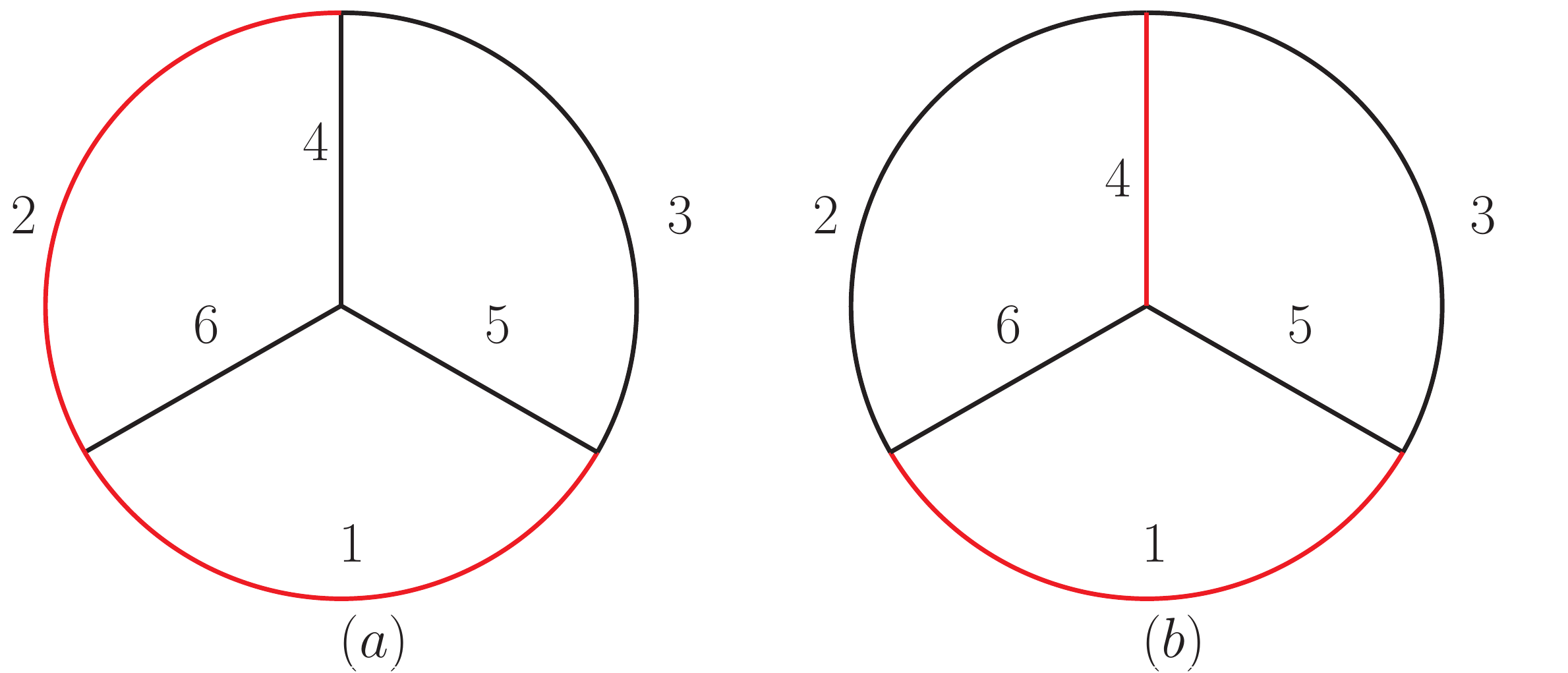}
\caption{3 loop vacuum diagrams with two $n_i\ge \frac{d}{2}$ (the red lines). The IR divergences can be regulated by adding masses to the red lines.}
\label{fig:3 loop vacuum}
\end{figure}

Some three-loop CV have IR divergences in multiple soft regions, and adding mass to a single propagator is not enough to regulate all these IR divergences. In this subsection we consider two integrals $I_{441111}$ and $I_{411411}$ in 8-d, for which both propagators with $n_i=4$ require regulation, as shown in Figure \ref{fig:3 loop vacuum}.

First we consider the integral with two adjacent massive propagators,
\begin{equation}
F=I^{adj}_{441111}\equiv\frac{1}{(l_1^2+m^2)^4(l_2^2+m^2)^4l_3^2(l_2-l_3)^2(l_3-l_1)^2(l_1-l_2)^2}\ .
\end{equation}
After integral reduction, we end up with 5 master integrals, $I^{adj}_{111010},\ I^{adj}_{111100},\ I^{adj}_{110110},\ I^{adj}_{011011}$, $I^{adj}_{101101}$, which can be evaluated with the help of the following formula \cite{Smirnov:2006ry},
\begin{equation}
\begin{aligned}\label{2-mass-int}
&\frac{1}{(l_1^2+m^2)^{n_1}(l_2^2+m^2)^{n_2}[(l_1-l_2)^2]^{n_3}}\\
=&\frac{e^{2\epsilon \gamma}\Gamma(n_1+n_3-\frac{D}{2})\Gamma(n_2+n_3-\frac{D}{2})\Gamma(\frac{D}{2}-n_3)\Gamma(n_1+n_2+n_3-D)}
{\Gamma(n_1)\Gamma(n_2)\Gamma(n_1+n_2+2n_3-D)\Gamma(\frac{D}{2})(m^2)^{n_1+n_2+n_3-D}}\ .\\
\end{aligned}
\end{equation}
The total UV divergence of $F$ is found to be
\begin{equation}
F\sim -\frac{1}{1296\epsilon ^{3}}+\frac{-\frac{7}{7776}+\frac{\ln m}{216}}{\epsilon ^{2}}+\frac{5-9\pi^{2}+252\ln m-648\ln m^{2}}{46656\epsilon }\ .
\end{equation}
$F$ has a 1-loop sub-divergence, corresponding to $\theta_1=\mathtt{line}\{l_3,l_2-l_3,l_3-l_1\}$,
\begin{equation}
\begin{aligned}
\mathcal{V}_{\theta_1}F
=&
\frac{1}{(l_1^2+m^2)^4(l_2^2+m^2)^4(l_1-l_2)^2}
\mathbf{L}\Bigl[-\frac{l_1^2+l_2^2}{(l_3^2)^4}
+\frac{4l_{3\mu}l_{3\nu}}{(l_3^2)^5}(l_1^{\mu}l_1^{\nu}+l_1^{\mu}l_2^{\nu}+l_2^{\mu}l_2^{\nu})
\Bigr]\\
=&
\frac{1}{24\epsilon}\frac{-l_1^2-l_2^2-(l_1-l_2)^2}{(l_1^2+m^2)^4(l_2^2+m^2)^4(l_1-l_2)^2}\\
\sim&-\frac{1}{432\epsilon ^{3}}+\frac{-\frac{1}{2592}+\frac{\ln m}{108}}{\epsilon ^{2}}+\frac{5-3\pi^{2}+12\ln m-144\ln^2m}{7776\epsilon }\ ,\\
\end{aligned}
\end{equation}
where we used \eqref{2-mass-int} in the last step.

There are two identical 2-loop sub-divergences, corresponding to $\theta_2=F\setminus \{\mathtt{line}(l_1)\}$ and $\theta_3=F\setminus \{\mathtt{line}(l_2)\}$, respectively,
\begin{equation}
\begin{aligned}
\mathcal{V}_{\theta_2}F
=&\frac{1}{(l_1^2+m^2)^4}\mathbf{L}\frac{1}{(l_2^2)^5(l_3^2)^2(l_2-l_3)^2}\\
\sim &\frac{1}{864\epsilon ^{3}}+\frac{-\frac{5}{5184}-\frac{\ln m}{432}}{\epsilon ^{2}}+\frac{\pi^{2}+20\ln m+24\ln^2m}{10368\epsilon }\ .\\
\end{aligned}
\end{equation}
The local divergence is
\begin{equation}
\mathbf{L}I_{441111}=-\frac{1}{1296\epsilon ^{3}}+\frac{11}{7776\epsilon ^{2}}
-\frac{25}{46656\epsilon }\ .
\end{equation}

Next we compute the local divergence of $I_{411411}$,
\begin{equation}
\begin{aligned}
F=&I^{non-adj}_{411411}\equiv\frac{1}{(l_1^2+m^2)^4l_2^2l_3^2[(l_2-l_3)^2+m^2]^4(l_3-l_1)^2(l_1-l_2)^2}\ .\\
\end{aligned}
\end{equation}
After integral reduction, we end up with 9 master integrals. 8 of them are 1 loop reducible, $I^{non-adj}_{101110}$, $I^{non-adj}_{111100}$, $I^{non-adj}_{011111}$, $I^{non-adj}_{100111}$, $I^{non-adj}_{110101}$, $I^{non-adj}_{111011}$, $I^{non-adj}_{110110}$ and $I^{non-adj}_{101101}$. The last master integral, $I^{non-adj}_{111111}$, is not 1 loop reducible. In Appendix \ref{appendix-non-adj} we compute this master integral using DRA method \cite{Lee:2009dh, Lee:2010hs, Lee:2012hp}. Combining \eqref{f211111}, \eqref{f211111shift} and \eqref{f111111}, we find at $D=8-2\epsilon$,
\begin{equation}
I^{non-adj}_{111111}\sim m^{12-6\epsilon}\Bigl[\frac{23}{12960\epsilon ^{3}}+\frac{841}{97200\epsilon ^{2}}+\frac{119027+7875\pi^{2}+129600\zeta_3}{11664000\epsilon }\Bigr]\ .
\end{equation}
The total UV divergence of $F$ is
\begin{equation}
\begin{aligned}
F\sim &\frac{1}{648\epsilon^{3}}-\frac{\frac{1}{3888}+\frac{\ln m}{108}}{\epsilon^{2}}+\frac{-901+45\pi^{2}+180\ln m+3240\ln^2m+648\zeta_3}{116640\epsilon}\ .\\
\end{aligned}
\end{equation}
$F$ has a 1-loop UV divergence corresponding to the UV sub-integral $\theta_1=\mathtt{line}\{l_2,l_3,l_3-l_1,l_1-l_2\}$,
\begin{equation}
\begin{aligned}
&\mathcal{V}_{\theta_1}F
=\frac{1}{6\epsilon}\frac{1}{(l_1^2+m^2)^4(l_2^2+m^2)^4}
\sim \frac{1}{216\epsilon^{3}}-\frac{\ln m}{54\epsilon^{2}}+\frac{\pi^{2}+48\ln^2m}{1296\epsilon}\ .\\
\end{aligned}
\end{equation}
$F$ has two identical 2-loop UV divergences, corresponding to $\theta_2=F\setminus\{\mathtt{line}(l_1)\}$ and $\theta_3=F\setminus\{\mathtt{line}(l_2-l_3)\}$, respectively,
\begin{equation}
\begin{aligned}
\mathcal{V}_{\theta_2}F
=&\frac{1}{(l_1^2+m^2)^4}\mathbf{L}\frac{1}{(l_2^2)^2(l_3^2)^2[(l_2-l_3)^2]^4}\\
\sim &-\frac{1}{432\epsilon^{3}}+\frac{-\frac{1}{2592}+\frac{\ln m}{216}}{\epsilon^{2}}+\frac{-\pi^{2}+4\ln m-24\ln^2m}{5184\epsilon}\ .\\
\end{aligned}
\end{equation}
The local divergence is
\begin{equation}\label{ld411411}
\mathbf{L}I_{411411}=\frac{1}{648\epsilon^{3}}+\frac{1}{1944\epsilon^{2}}
+\frac{1}{\epsilon}\Bigl(-\frac{901}{116640}+\frac{\zeta_3}{180}\Bigr)\ .
\end{equation}

\section{The subtraction of IR divergence}
\label{IR-div}

In the last section in order to regulate the IR divergence of 3-loop vacuum integrals, we added masses to one or two propagators. The integral reduction takes more time with more massive propagators. For example, on my laptop using FIRE5, the 3-loop vacuum integral $I_{222222}$ takes 54 seconds when there is a single massive propagator, 103 seconds when there are two adjacent massive propagators, and 122 seconds when there are two non-adjacent massive propagators. In addition, as can be seen in Section \ref{IR regulation by adding two masses}, the expressions of master integrals also become more complicated, and more effort needs be paid on the evaluation of master integrals. In order to regulate more complicated IR divergence, for example in $I_{2212(-2)1}$, even more masses must be added to the integral.

Alternatively, we may make less propagators massive, leaving some IR divergences in the integral, and  subtract these IR divergence afterwards.
It is preferable to make only a single propagator massive. Because when $L\ge 3$, the integral reduction is usually the most time consuming part in the computation, and by adding a single mass we can minimize this time. Although the regulated integral may have multiple IR divergences, adding a single mass always remove the most difficult IR divergence: the local IR divergence. The lower loop sub IR divergences take much less time to evaluate compared with IBP. Moreover, the analytic expressions of $L$-loop vacuum integrals with a single massive propagators can be easily obtained from $(L-1)$-loop massless propagator integrals, for which the analytic expressions are known to 5 loops \cite{Baikov:2010hf,Lee:2011jf,Lee:2011jt,Georgoudis:2018olj,Georgoudis:2021onj}. 

In Section \ref{subsection:uvdecom-ir}, we propose a modified version of UV decomposition formula which can be used to compute the local divergence of vacuum integrals in the presence of IR divergences. In Section \ref{subsection:sub-ir} we study the IR divergence of multiloop integrals, then in Section \ref{subsection:IR-CV} we compute the total IR divergence of 2-loop scalar CV. Last, in Section \ref{subsection:multi-IR}, we evaluate the local divergence of a vacuum integral with multiple IR divergences.

\subsection{The UV decomposition with IR divergence}
\label{subsection:uvdecom-ir}

We will use $\mathcal{I}$ to denote the \textbf{IR divergence operator}, and $\mathcal{U}=1-\mathcal{I}$ is the \textbf{IR subtraction operator}.
Start with the UV decomposition formula \eqref{uvdecom-2}, we can subtract IR divergence on both sides of the formula, and obtain the following "UV decomposition with IR subtraction" formula,
\begin{equation}
\begin{aligned}
&\mathcal{U}F\sim\sum_{\theta\in \Theta(F)}\mathcal{UV}_{\theta}F\ .\\
\end{aligned}\label{uvdecom-ir-1}
\end{equation}
The l.h.s. and r.h.s. of \eqref{uvdecom-ir-1} not only have the same UV divergence, but also have the same $\epsilon$-poles.

The local divergence of $F$ is not affected by the $\mathcal{U}$ operator, so \eqref{uvdecom-ir-1} can be rewritten as
\begin{equation}
\begin{aligned}
&\mathbf{L}F\sim F-\mathcal{I}F-\sum_{\theta \in \Theta(F)}^{\theta\ne F}\mathcal{UV}_{\theta}F \ .\\
\end{aligned}\label{uvdecom-ir-2}
\end{equation}
Eq. \eqref{uvdecom-ir-2} states that the local divergence of an IR-divergent integral can be obtained by subtracting the IR divergence and various sub-divergences from the integral.

Let $V$ be a CV, and $V(m)$ be the massive integral obtained by adding mass to a single propagator of $V$. Then
\begin{equation}
\begin{aligned}
&\mathbf{L}V(m)\sim V(m)-\mathcal{I}V(m)-\sum_{\theta \in \Theta(V)}^{\theta\ne V}\mathcal{UV}_{\theta}V(m) \ .\\
\end{aligned}\label{uvdecom-ir-3}
\end{equation}
provides a new approach to the local divergence of CV.

Similar as UV divergence, IR divergence can also be split into local IR divergence and sub IR divergences. Each sub IR divergence corresponds to an IR sub-integral, and the local IR divergence is a special type of sub IR divergence corresponding to the integral itself.

First let us consider the simplest example, the IR divergence of the 1-loop scalar vacuum integral $\frac{1}{(l^2)^{n}}$. Since the integral vanishes, its UV divergence and IR divergence must cancel each other, so we have
\begin{equation}\label{IR 1 loop scalar}
\mathcal{I}\frac{1}{(l^2)^{n}}=-\mathbf{L}\frac{1}{(l^2)^{n}}=-\frac{\delta_{n,\frac{d}{2}}}{\Gamma(\frac{d}{2})\epsilon}\ .
\end{equation}

A key feature of $\mathcal{I}$ is that it commutes with the contraction of Lorentz indices\footnote{The total UV divergence also commutes with the contraction of Lorentz indices. However, the local divergence operator $\mathbf{L}$ does not commute with the contraction of Lorentz indices, and more details can be found  in Section \ref{tensor-reduction}.}. Let $F^{\mu_1\mu_2\cdots \mu_n}$ be a tensor integral,
\begin{equation}\label{IR-lorentz-commute-1}
\eta_{\mu_1\mu_2}\mathcal{I} F^{\mu_1\mu_2\cdots \mu_n}
=\mathcal{I}\Bigl( \eta_{\mu_1\mu_2}F^{\mu_1\mu_2\cdots \mu_n }\Bigr)+\text{IR finite terms}\ .
\end{equation}
This allows us to relate the IR divergence of tensor integrals to that of scalar integrals using PV reduction. The IR finite terms in \eqref{IR-lorentz-commute-1} can be different in different schemes, and we will choose the most simple scheme in which these IR finite terms vanish:
\begin{equation}\label{IR-lorentz-commute-2}
\eta_{\mu_1\mu_2}\mathcal{I} F^{\mu_1\mu_2\cdots \mu_n }
=\mathcal{I}\Bigl( \eta_{\mu_1\mu_2}F^{\mu_1\mu_2\cdots \mu_n}\Bigr)\ .
\end{equation}
With this choice, the IR divergence of one loop tensor vacuum integral is found to be,
\begin{equation}
\begin{aligned}
&\mathcal{I}\frac{l^{\mu_1}\cdots l^{\mu_{2a}}}{(l^2)^n}
=-\frac{\delta_{n-\frac{d}{2},a}\eta_s^{\mu_1\cdots \mu_{2a}}}{2^a (\frac{D}{2})_a\Gamma(\frac{d}{2})\epsilon}\ .\\
\end{aligned}
\end{equation}

The one-loop sub IR divergence of a multiloop integral $F$ can be computed in two steps. Suppose $\gamma$ is the IR sub-integral corresponding to the sub IR divergence, and $l_s$ is the loop momenta of $\gamma$. First, one can expand the integral into asymptotic series around $l_s=0$. And second, one evaluates the IR divergence in the $l_s$ integral.

As an example, consider the following 2-loop integral in 6-d:
\begin{equation}
F=\frac{1}{(l_1^2)^4(l_1-l_2)^2(l_2^2+m^2)}\ .
\end{equation}
Let $\gamma=\mathtt{line}(l_1)$,
and the expansion around the soft momenta of $\gamma$ will be denoted by $\mathcal{R}_{\gamma}$,
\begin{equation}
\begin{aligned}
\mathcal{R}_{\gamma}F
=&\frac{1}{l_2^2(l_2^2+m^2)}\frac{1}{(l_1^2)^4}
+\frac{2l_2^{\mu}}{(l_2^2)^2(l_2^2+m^2)}\frac{l_{1\mu}}{(l_1^2)^4}\\
&-\frac{1}{(l_2^2)^2(l_2^2+m^2)}\frac{1}{(l_1^2)^3}
+\frac{4l_2^{\mu}l_2^{\nu}}{(l_2^2)^3(l_2^2+m^2)}\frac{l_{1\mu}l_{1\nu}}{(l_1^2)^4}+R(F)\ .\\
\end{aligned}
\end{equation}
In which $R(F)$ is reminder term with negative IR divergence degree.
The sub-divergence corresponding to $\gamma$ will be denoted by $\mathcal{I}_{\gamma}F$:
\begin{equation}
\begin{aligned}
\mathcal{I}_{\gamma}F
=&-\frac{1}{(l_2^2)^2(l_2^2+m^2)}\mathcal{I}\frac{1}{(l_1^2)^3}
+\frac{4l_2^{\mu}l_2^{\nu}}{(l_2^2)^3(l_2^2+m^2)}\mathcal{I}\frac{l_{1\mu}l_{1\nu}}{(l_1^2)^4}+\cdots\\
=&\frac{D-4}{2D\epsilon}\frac{1}{(l_2^2)^2(l_2^2+m^2)}\ .\\
\end{aligned}
\end{equation}

A more complicated example is given in Appendix \ref{appendix:example IR subtraction}, in which the local divergence of $I_{411411}$ in Section \ref{IR regulation by adding two masses} is reproduced using IR subtraction.

\subsection{The sub IR divergence}
\label{subsection:sub-ir}

The IR divergence corresponding to a multiloop IR sub-integral can be defined similarly as the one-loop case. Suppose $\gamma$ is an IR sub-integral of $F$, and let $l^s_i$ and $l^h_i$ be the loop momenta of $\gamma$ and $F\setminus\gamma$, respectively.
Let us define a large number $\Lambda$, so that $\hat{l}^s_i\equiv \Lambda l^s_i$ are in the same order as $l^h_i, p_i, M_i$, and expand the integral $F(\frac{\hat{l}^s_i}{\Lambda},l^h_i, p_i, M_i)$ into asymptotic series around large $\Lambda$. Following similar procedures as in Section \ref{subsection:asymptotic}, $F$ can be written as
\begin{equation}\label{ir-expand-1}
F=\mathcal{R}_{\gamma}F+R(F)
\equiv \sum_{i=0}^{\omega_{ir}(\gamma)}\mathcal{R}^i_{\gamma}F+R(F),\ 
\end{equation}
in which $R(F)$ is a remainder term without negative IR divergence degree in the region. $\mathcal{R}^i_{\gamma}F$ is a disconnected integral with a soft component $\mathcal{H}^i_{\gamma}(F)$ and a hard component $\mathcal{S}^i_{\gamma}(F)$:
\begin{equation}
\mathcal{R}^i_{\gamma}F=\mathcal{H}^i_{\gamma}(F)\mathcal{S}^i_{\gamma}(F)\ ,
\end{equation}
in which $\mathcal{S}^i_{\gamma}(F)$ has $\omega_{ir}=i$.
The sub IR divergence corresponding to $\gamma$ is given by
\begin{equation}\label{sub-ir-1}
\mathcal{I}_{\gamma}F\equiv\mathcal{H}^0_{\gamma}(F)
\mathcal{I}\mathcal{S}^0_{\gamma}(F)\ .
\end{equation}

There is an important different between the sub UV divergence and the sub IR divergence. In \eqref{uvdecom-ir-2}, the sub UV divergence $\mathcal{V}_{\gamma}F$ is preceded by the UV subtraction operator, so if a CV appears in $\mathcal{V}_{\gamma}F$, it cannot be dropped. However, massless vacuum integrals in $\mathcal{I}V(m)$ can be set to zero using dimensional regularization. This implies that in \eqref{sub-ir-1}, massless vacuum integrals in $\mathcal{H}^0_{\gamma}(F)$ can also be dropped.
Using this property it can be shown that if $\gamma$ and $\rho$ are IR sub-integrals of $F$,
\begin{equation}\label{repeated R}
\mathcal{I}_{\rho}\mathcal{R}_{\gamma}F=0, \text{unless}\ \gamma\subset\rho\ .
\end{equation}

The total IR divergence is given by the sum of sub IR divergences corresponding to all IR sub-integrals,
\begin{equation}\label{total IR}
\mathcal{I}F=\sum_{\gamma\in \Upsilon(F)}\mathcal{I}_{\gamma}F.
\end{equation}

Eq. \eqref{total IR} can be proved by induction. First we assume it holds for vacuum integrals with at most $L_0$ loop IR divergences. If $F$ has at most $L_0+1$ loop IR divergences, we can split $F$ into $F_0$ and $F_1$,
\begin{equation}
\begin{aligned}
&F_0\equiv F-\sum_{\gamma\in \Upsilon_{1}(F)}\mathcal{R}_{\gamma}F,\ 
F_1\equiv \sum_{\gamma\in \Upsilon_{1}(F)}\mathcal{R}_{\gamma}F ,\\
\end{aligned}\label{total IR1}
\end{equation}
where $\Upsilon_{1}(F)$ is the set of $L_0+1$ loop IR sub-integrals.

Suppose $\rho\in\Upsilon_{1}(F)$, and by using \eqref{repeated R} we find
\begin{equation}
\begin{aligned}
&\mathcal{I}_{\rho}F_0
=-\sum_{\gamma\in \Upsilon_{1}(F)}^{\gamma\ne\rho}\mathcal{I}_{\rho}\mathcal{R}_{\gamma}F=0\ .\\
\end{aligned}
\end{equation}
So $F_0$ has no $L_0+1$ loop IR divergences, and by induction assumption its IR divergence is given by
\begin{equation}
\begin{aligned}
&\mathcal{I}F_0
=\sum_{\rho\in \Upsilon_0(F)}\mathcal{I}_{\rho}F
-\sum_{\rho\in \Upsilon_0(F)}\sum_{\gamma\in \Upsilon_{1}F}
\mathcal{I}_{\rho}\mathcal{R}_{\gamma}F\ ,\\
\end{aligned}\label{total IR2}
\end{equation}
where $\Upsilon_0(F)$ is the set of $L\le L_0$ loop IR sub-integrals. The second term on the r.h.s. of \eqref{total IR2} vanishes because $\gamma$ cannot be a sub-integral of $\rho$. The IR divergence of $F$ is given by
\begin{equation}
\begin{aligned}
&\mathcal{I}F
=\sum_{\rho\in \Upsilon_0(F)}\mathcal{I}_{\rho}F
+\sum_{\gamma\in \Upsilon_{1}(F)}\mathcal{I}\mathcal{R}_{\gamma}F
=\sum_{\rho\in \Upsilon_0(F)}\mathcal{I}_{\rho}F
+\sum_{\gamma\in \Upsilon_{1}(F)}\mathcal{I}\Bigl[\mathcal{S}^0_{\gamma}(F)\mathcal{H}^0_{\gamma}(F)\Bigr]\ .\\
\end{aligned}\label{total IR3}
\end{equation}
The $\mathcal{H}_{\gamma}(F)$ term has no IR divergence, otherwise $F$ would have an IR divergence with $L>L_0+1$. Therefore we have
\begin{equation}
\mathcal{I}\Bigl[\mathcal{S}^0_{\gamma}(F)\mathcal{H}^0_{\gamma}(F)\Bigr]
=\mathcal{H}^0_{\gamma}(F)\mathcal{IS}^0_{\gamma}(F)=\mathcal{I}_{\gamma}F\ ,
\end{equation}
and
\begin{equation}
\begin{aligned}
&\mathcal{I}F
=\sum_{\gamma\in \Upsilon_{1}(F)}\mathcal{I}_{\gamma}F
+\sum_{\rho\in \Upsilon_0(F)}\mathcal{I}_{\rho}F
=\sum_{\gamma\in \Upsilon(F)}\mathcal{I}_{\gamma}F\ .\\
\end{aligned}\label{total IR4}
\end{equation}
This means \eqref{total IR} holds for $F$, and completes the proof of \eqref{total IR}.

Using \eqref{total IR}, the IR divergence of generic (Euclidean) integrals are reduced to that of CV. In the next subsection, we will discuss the IR divergence of 2-loop scalar CV.

Different approaches to IR subtraction have been proposed in e.g. \cite{Chetyrkin:1982nn,Chetyrkin:1984xa,Larin:2002sc,Chetyrkin:2017ppe} (see section 7 of \cite{Herzog:2017bjx} for a detailed discussion of the literature). One of the advantages of our definition is that the total IR divergence is a simple sum of the IR divergences in all regions, therefore we do not need to worry about nested or overlapped divergences. Another nice feature is that the IR subtraction operator commutes with the Lorentz contraction, and the IR divergence of tensor integrals can be easily reduced to that of scalar integrals through PV reduction.

\subsection{IR divergence of 2-loop scalar CV}
\label{subsection:IR-CV}

In the last section we expressed the IR divergence of vacuum integrals using the IR divergence of CV. The IR divergence of tensor CV can be reduced to that of scalar CV using PV reduction.
The IR divergence of a scalar vacuum integral $V$ can be obtained by
\begin{equation}\label{IR-CV-1}
\mathcal{I}(V)=V-\mathcal{U}(V)=-\mathcal{U}(V)\ ,
\end{equation}
which reduces the problem to the computation UV divergences.

As an example, let us consider the 2-loop CV $I_{n_1n_2n_3}$ in \eqref{2-loop-cv}. In the case $n_1\ge \frac{d}{2}$ and $n_2<\frac{d}{2}$, there is a sub UV divergence
\begin{equation}
\begin{aligned}
&\mathcal{U}\frac{1}{(l_1^2)^{n_1}}\mathbf{L}\frac{1}{(l_2^2)^{n_2}[(l_1-l_2)^2]^{n_3}}
=\mathcal{U}\frac{1}{(l_1^2)^{\frac{d}{2}}}
\frac{(-1)^{n_1-\frac{d}{2}}\Gamma(\frac{d}{2}-n_2)\Gamma(\frac{d}{2}-n_3)}
{\Gamma(n_1)\Gamma(n_2)\Gamma(n_3)(n_1-\frac{d}{2})!\epsilon}\\
= &
\frac{(-1)^{n_1-\frac{d}{2}}\Gamma(\frac{d}{2}-n_2)\Gamma(\frac{d}{2}-n_3)}
{\Gamma(n_1)\Gamma(n_2)\Gamma(n_3)\Gamma(\frac{d}{2})(n_1-\frac{d}{2})!\epsilon^2}\ .\\
\end{aligned}
\end{equation}
The local divergence is given in \eqref{local n1}. Using \eqref{IR-CV-1}, one obtains
\begin{equation}
\begin{aligned}
&\mathcal{I}I_{n_1n_2n_3}=\frac{(-1)^{n_1-\frac{d}{2}}\Gamma(\frac{d}{2}-n_2)
\Gamma(\frac{d}{2}-n_3)}{(n_1-\frac{d}{2})!\Gamma(n_1)\Gamma(n_2)\Gamma(n_3)\Gamma(\frac{d}{2})}
\Bigl[-\frac{1}{2\epsilon^2}-\frac{1}{2\epsilon}Z_{n_1n_2n_3}\Bigr]\ .\\
\end{aligned}
\end{equation}
Comparing with \eqref{local n1}, we find that if we write the local divergence as $\frac{a}{\epsilon^2}+\frac{b}{\epsilon}$, then the IR divergence is $\frac{a}{\epsilon^2}-\frac{b}{\epsilon}$. This relation holds for generic 2 loop scalar CV, and can be proved as follows.

The IR divergence of $V=I_{n_1n_2n_3}$ can be regulated by adding mass $m$ to 1 or 2 of its propagators. We denote the regulated integral by $V(m)$, and the UV decomposition of $V(m)$ has the following form
\begin{equation}
\begin{aligned}\label{cv2-ir-1}
V(m)\sim \frac{a}{\epsilon^2}+\frac{b}{\epsilon}+\frac{1}{\epsilon}I_1(m)\ ,
\end{aligned}
\end{equation}
in which the last term is the sub UV divergence and $I_1(m)$ is some 1 loop integral.

An important feature of \eqref{cv2-ir-1} is that the expression is formally "smooth" in the $m\rightarrow 0$ limit. The analytic expression of $I_1(m)$ has the form
\begin{equation}
I_1(m)=m^{-2\epsilon}(\frac{\alpha}{\epsilon}+\beta)+\mathcal{O}(\epsilon^2)\ ,
\end{equation}
and it is non-analytical in the $m\rightarrow 0$ limit. However, the integrand of $I_1(m)$ must be smooth, since neither the original integrand of $V(m)$ nor the $\mathbf{L}\mathcal{A}^0\theta$ operation introduce singularity to the expression. Therefore the UV decomposition of $V=V(0)$ is given by
\begin{equation}
\begin{aligned}
\mathcal{U}V\sim \frac{a}{\epsilon^2}+\frac{b}{\epsilon}+\frac{1}{\epsilon}\mathcal{U}I_1(0)\ .
\end{aligned}
\end{equation}
Since $I_1(0)$ is a 1-loop scalar CV, $\mathcal{U}I_1(0)$ is equal to $\frac{\alpha_1}{\epsilon}$ with some constant $\alpha_1$. By using
\begin{equation}
\mathcal{U}I(m)\sim \mathbf{L}I(m)\sim \mathbf{L}I(0)=\mathcal{U}I(0)\ ,
\end{equation}
we find 
\begin{equation}
\mathcal{U}I(0)=\frac{\alpha}{\epsilon}\ .
\end{equation}

On the other hand, $V(m)$ itself is proportional to $m^{-4\epsilon}$, and has the form
\begin{equation}
V(m)\sim m^{-4\epsilon}(\frac{x}{\epsilon^2}+\frac{y}{\epsilon})\ .
\end{equation}
Compare this expression with \eqref{cv2-ir-1} we find $\alpha=-2a$, then the IR divergence of $V$ is given by
\begin{equation}
\mathcal{I}V=-\mathcal{U}V=\frac{a}{\epsilon^2}-\frac{b}{\epsilon}\ .
\end{equation}

Similar discussion can also be applied to higher loop CV which gives relations between local UV divergence and total IR divergence, but the exact form of total IR divergence cannot be determined by these relations at $\mathbb{L}\ge 3$.

\subsection{A vacuum integral with multiple IR divergences}
\label{subsection:multi-IR}
Combining \eqref{uvdecom-ir-3} and \eqref{total IR}, the local divergence of mass regulated CV can be computed using
\begin{equation}
\begin{aligned}
&\mathbf{L}V(m)\sim V(m)-\sum_{\gamma\in \Upsilon(V)}^{\gamma\ne V}\mathcal{I}_{\gamma}V(m)-\sum_{\theta \in \Theta(V)}^{\theta\ne V}\mathcal{UV}_{\theta}V(m) \ .\\
\end{aligned}\label{uvdecom-ir-4}
\end{equation}
Since $V(m)$ has no local IR divergence, its IR divergence can be determined by the IR divergence of lower loop CV.

As an example of \eqref{uvdecom-ir-4}, let us compute the local divergence of $V=I_{2212(-2)1}$ in 4-d using IR subtraction. The integral has multiple IR divergences even after adding mass to the $l_1$ propagator. The total divergence is
\begin{equation}
\begin{aligned}
V(m)\equiv&\frac{[(l_3-l_1)^2]^2}{(l_1^2+m^2)^2(l_2^2)^2l_3^2[(l_2-l_3)^2]^2(l_1-l_2)^2}\\
\sim& -\frac{1}{3\epsilon^{3}}+\frac{-1+24\ln m}{12\epsilon^{2}}+\frac{-25-10\pi^{2}+12\ln m-144\ln^2m}{24\epsilon}\ .\\
\end{aligned}
\end{equation}

The integral has three 1-loop UV divergences, corresponding to $\theta_1=\{L(l_1),L(l_3-l_1),L(l_1-l_2)\}$, $\theta_2=\{L(l_2),L(l_3),L(l_3-l_1),L(l_1-l_2)\}$ and $\theta_3=\{L(l_3),L(l_2-l_3),L(l_3-l_1)\}$ respectively.
\begin{equation}
\begin{aligned}
&\mathcal{UV}_{\theta_1}F
=\mathcal{U}\frac{3l_3^2-2l_2\cdot l_3-2m^2}{(l_2^2)^2l_3^2[(l_2-l_3)^2]^2}
\sim \frac{2}{\epsilon ^{3}}\ ,\\
&\mathcal{UV}_{\theta_2}F
=\frac{1}{\epsilon}\mathcal{U}\frac{1}{(l_1^2+m^2)^2[(l_2-l_3)^2]^2}
\sim \frac{1}{\epsilon ^{3}}-\frac{2\ln m}{\epsilon ^{2}}+\frac{\frac{\pi^{2}}{12}+2\ln^2m}{\epsilon }\ ,\\
&\mathcal{UV}_{\theta_3}F
=\frac{1}{\epsilon}\mathcal{U}\frac{3l_1^2+l_2^2-4l_1\cdot l_2}{(l_1^2+m^2)^2(l_2^2)^2(l_1-l_2)^2}
\sim \frac{2}{\epsilon ^{3}}-\frac{2\ln m}{\epsilon ^{2}}+\frac{-\frac{\pi^{2}}{4}-2\ln^2m}{\epsilon }\ .\\
\end{aligned}
\end{equation}

The integral also has three 2-loop UV divergences, corresponding to $\theta_4=F\setminus\{L(l_1)\}$, $\theta_5=F\setminus\{L(l_2)\}$ and $\theta_6=F\setminus\{L(l_2-l_3)\}$ respectively.
\begin{equation}
\begin{aligned}
&\mathcal{UV}_{\theta_4}F
=\mathcal{U}\frac{1}{(l_1^2+m^2)^2}\mathbf{L}\frac{l_3^2}{(l_2^2)^3[(l_2-l_3)^2]^2}
\sim -\frac{1}{\epsilon ^{3}}+\frac{2\ln m}{\epsilon ^{2}}+\frac{-\frac{\pi^{2}}{12}-2\ln^2m}{\epsilon }\ ,\\
&\mathcal{UV}_{\theta_5}F
=\mathcal{U}\frac{1}{(l_2^2)^2}\mathbf{L}\frac{[(l_3-l_1)^2]^2}{(l_1^2)^3(l_3^2)^3}
\sim -\frac{3}{\epsilon ^{3}}+\frac{1}{2\epsilon ^{2}}\ ,\\
&\mathcal{UV}_{\theta_6}F
=\mathcal{U}\frac{1}{(l_3^2)^2}\mathbf{L}\frac{(l_1-l_2)^2}{(l_1^2)^2(l_2^2)^3}
\sim -\frac{1}{\epsilon ^{3}}\ .\\
\end{aligned}
\end{equation}

The integral has three IR divergences, corresponding to $\gamma_1=\{L(l_2)\}$, $\gamma_2=\{L(l_2-l_3)\}$ and $\gamma_3=\{L(l_2),L(l_3),L(l_2-l_3)\}$ respectively.
\begin{equation}
\begin{aligned}
\mathcal{IR}_{\gamma_1}F
=&-\frac{1}{\epsilon}\frac{[(l_3-l_1)^2]^2}{(l_1^2+m^2)^2l_1^2(l_3^2)^3}
=0\ ,\\
\mathcal{IR}_{\gamma_2}F
=&-\frac{1}{\epsilon}\frac{(l_1-l_2)^2}{(l_1^2+m^2)^2(l_2^2)^3}
=0\ ,\\
\mathcal{IR}_{\gamma_3}F
=&\frac{1}{(l_1^2+m^2)^2l_1^2}
\mathcal{I}\frac{l_1^2(-l_2^2+2l_3^2)+(2l_1\cdot l_2)^2
-2(2l_1\cdot l_2)(2l_1\cdot l_3)+(2l_1\cdot l_3)^2}{(l_2^2)^2l_3^2[(l_2-l_3)^2]^2}\\
\sim&-\frac{2}{\epsilon ^{3}}+\frac{-\frac{1}{4}+4\ln m}{\epsilon ^{2}}+\frac{-\frac{3}{8}-\frac{\pi^{2}}{6}+\frac{\ln m}{2}-4\ln^2m}{\epsilon }\ .\\
\end{aligned}
\end{equation}
After subtracting all sub-divergences, one obtains
\begin{equation}
\begin{aligned}
&\mathbf{L}I_{2212(-2)1}
=\frac{5}{3\epsilon^{3}}-\frac{1}{3\epsilon^{2}}-\frac{2}{3\epsilon}\ .\\
\end{aligned}
\end{equation}

\section{The tensor reduction}
\label{tensor-reduction}

We discussed the local divergences of scalar vacuum integrals in Section \ref{CV-2-3}, but generic vacuum integrals contain tensor structures of the form $l_1^{\mu_1}\cdots l_n^{\mu_n}$ in the numerator.
Tensor structures may come from the original integral (for example in scattering amplitudes of spinning particles), or from the asymptotic expansion \eqref{asym-expand-2} during the computation of sub-divergence.


The local divergence of tensor vacuum integrals can of course be computed by directly subtracting or regulating the IR and sub-UV divergences, similar as how we treated scalar vacuum integrals in the last section. As an example, consider the integral $\frac{l_2^{\mu}l_2^{\nu}}{(l_1^2)^5(l_2^2)^3(l_1-l_2)^2}$ in 8-d. We regulate the IR divergence by adding mass to the $l_1$-propagator,
\begin{equation}
\begin{aligned}
&F\equiv \frac{l_2^{\mu}l_2^{\nu}}{(l_1^2+m^2)^5(l_2^2)^3(l_1-l_2)^2}\sim
\eta^{\mu\nu}\Bigl[-\frac{1}{1152\epsilon ^{2}}+\frac{1}{\epsilon }\Bigl(\frac{\ln m}{288}-\frac{7}{13824}\Bigr)\Bigr]\ .\\
\end{aligned}\label{tensor.example.1}
\end{equation}
The sub-divergence is
\begin{equation}
\begin{aligned}
&\mathcal{V}_{l_2}F
=\frac{1}{\epsilon}\frac{1}{(l_1^2+m^2)^5}(\frac{l_1^{\mu}l_1^{\nu}}{60}-\frac{l_1^2\eta^{\mu\nu}}{80})
=\eta^{\mu\nu}\Bigl[-\frac{1}{576\epsilon ^{2}}+\frac{1}{\epsilon }\Bigl(\frac{\ln m}{288}+\frac{1}{1920}\Bigr)\Bigr]\ .\\
\end{aligned}
\end{equation}
The local divergence is 
\begin{equation}\label{local div direct}
\mathbf{L}\frac{l_2^{\mu}l_2^{\nu}}{(l_1^2)^5(l_2^2)^3(l_1-l_2)^2}
=\eta^{\mu\nu}(\frac{1}{1152\epsilon^2}-\frac{71}{69120\epsilon})\ .
\end{equation}
However, this method becomes less efficient because extra efforts must be paid to treat tensor structures in each step: in the computation of the total divergence, the sub UV divergences and the IR divergences. The problem become more severe as the tensor rank increases. It would be desirable if a tensor reduction can be performed before all these steps, and reduce the problem to the computation of local divergence for scalar integrals.

As will be shown later, the traditional PV reduction is not applicable in this case because the contraction of Lorentz indices does not commute with the $\mathbf{L}$ operator. In this section, we introduce two types of tensor reduction which commute with $\mathbf{L}$. The first type is the dimensional shift, which reduces $D$ dimensional tensor integrals to $D+2k$ dimensional scalar integrals, and it is extremely efficient at lower loops ($L\le 3$). The second type is the $d_{\infty}$ dimensional PV reduction, which reduces the tensor integrals to scalar integrals containing $d_{\infty}$ dimensional Lorentz products. The second approach is more efficient at higher loops when combined with the method of large $d_{\infty}$ expansion.

\subsection{Tensor reduction using dimensional shift}
\label{original-dimensional-shift}
The "naive" PV reduction is not applicable in the computation of local divergence, because the local divergence operator $\mathbf{L}$ does not commute with the Lorentz contraction. 
To see this, let us go back to the previous example in \eqref{tensor.example.1}, but first perform a PV reduction: 
\begin{equation}
\frac{l_2^{\mu}l_2^{\nu}}{(l_1^2)^5(l_2^2)^3(l_1-l_2)^2}
\rightarrow\frac{\eta^{\mu\nu}}{D }\frac{1}{(l_1^2)^5(l_2^2)^2(l_1-l_2)^2}
\end{equation}
Then a wrong local divergence will be produced,
\begin{equation}
\begin{aligned}
&\mathbf{L}\frac{\eta^{\mu\nu}}{D }\frac{1}{(l_1^2)^5(l_2^2)^2(l_1-l_2)^2}
=\eta^{\mu\nu}(\frac{1}{1152\epsilon^2}-\frac{7}{13824\epsilon})\ .\\
\end{aligned}\label{naive.tensor.example}
\end{equation}
Apparently, the local and sub-divergence structures are disrupted by the explicit $\epsilon$-dependence in $D$, which makes it impossible to extract the local divergence using the $\mathbf{L}$ operator. 
The dimensional shift \cite{Tarasov:1996br,Tarasov:1996bz} does not explicitly depend on $\epsilon$, and provides an alternative way to perform tensor reduction: rank-$2a$ tensor integrals in $D$ dimension are reduced to scalar integrals in $D+2a$ dimension. 

Let us start with a two loop critical vacuum integral in $D=d-2\epsilon$ dimension,
\begin{equation}
I_{n_1n_2n_3}=\frac{1}{(l_1^2)^{n_1}(l_2^2)^{n_2}[(l_1-l_2)^2]^{n_3}}\ .
\end{equation}
Using the alpha-parameterization, the integral can be rewritten as
\begin{equation}
\begin{aligned}
&I_{n_1n_2n_3}
=\int_0^{\infty} \prod_{i=1}^3 \frac{ dx_ix_i^{n_i-1}}{\Gamma(n_i)}
e^{-\left[x_1 (l_1^2)+x_2(l_2^2)+x_3(l_1-l_2)^2\right]}\ .\\
\end{aligned}
\end{equation}
After integrating over the loop momenta $l_1, l_2$, one obtain
\begin{equation}
\begin{aligned}
&I_{n_1n_2n_3}
= \int_0^{\infty} \prod_{i=1}^3 \frac{ dx_ix_i^{n_i-1}}{\Gamma(n_i)}
U(x)^{-\frac{D}{2}}\ ,\\
&U(x)=x_1x_2+x_2x_3+x_3x_1. \\
\end{aligned}\label{alpha2loop}
\end{equation}

The generating function of tensor integrals is defined by:
\begin{equation}
\begin{aligned}
Z(x,v)=&\int_0^{\infty} d^3x \frac{ x_1^{n_1-1}x_2^{n_2-1}x_3^{n_3-1}}{\Gamma(n_1)\Gamma(n_2)\Gamma(n_3)}e^{-[x_1 l_1^2+x_2l_2^2+x_3(l_1-l_2)^2]+v_1\cdot l_1
+v_2\cdot l_2}\\
= &\int_0^{\infty} \prod_{i=1}^3\frac{dx_i x_i^{n_i-1}}{\Gamma(n_i)}
U^{-\frac{D}{2}}e^{\frac{K(v)}{4U}}\ ,\\
\end{aligned}\label{zxv}
\end{equation}
where $K(v)=x_1v_2^2+x_2v_1^2+x_3(v_1+v_2)^2$.
Using the generating function, the tensor integrals can be expressed as
\begin{equation}
\begin{aligned}
&\frac{l_1^{\mu_1}\cdots l_1^{\mu_a}l_2^{\nu_1}\cdots l_2^{\nu_b}}
{(l_1^2)^{n_1}(l_2^2)^{n_2}[(l_1-l_2)^2]^{n_3}}
=\frac{\partial}{\partial v_{1\mu_1}}\cdots \frac{\partial}{\partial v_{1\mu_a}}
\frac{\partial}{\partial v_{2\nu_1}}\cdots \frac{\partial}{\partial v_{2\nu_b}}
Z(x, v)\Bigr|_{v=0}\ .\\
\end{aligned}
\end{equation}
The $v$-derivative produces $x_i$ and $U^{-1}$ factors. Each $x_i$ increase the $n_i$ index to $n_i+1$, and each $U^{-1}$ increases $D$ to $D+2$.

As an example, let us reduce the integral in \eqref{local div direct} using dimensional shift.
\begin{equation}
\begin{aligned}
&\frac{l_2^{\mu}l_2^{\nu}}{(l_1^2)^5(l_2^2)^3(l_1-l_2)^2}
=\frac{\partial}{\partial v_{2\mu}}\frac{\partial}{\partial v_{2\nu}}
\int_0^{\infty} d^3x\frac{ x_1^{4}x_2^{2}}{48}
U^{-\frac{D}{2}}e^{\frac{K(v)}{4U}}\Bigr|_{v=0}\\
=&\int_0^{\infty} d^3x\frac{ x_1^{4}x_2^{2}}{96}
U^{-\frac{D+2}{2}}(x_1+x_3)\eta^{\mu\nu}
=\left(\frac{5}{2}I_{631}+\frac{1}{2}I_{532}\right)\eta^{\mu\nu}.\ \\
\end{aligned}\label{tensor2ex11}
\end{equation}
The local divergences of the scalar integrals can be obtained using \eqref{local n1}, and we have
\begin{equation}
\begin{aligned}
&\mathbf{L}\frac{l_2^{\mu}l_2^{\nu}}{(l_1^2)^5(l_2^2)^3(l_1-l_2)^2}
=\mathbf{L}\left(\frac{5}{2}I_{631}+\frac{1}{2}I_{532}\right)\eta^{\mu\nu}
=\eta^{\mu\nu}(\frac{1}{1152\epsilon^2}-\frac{71}{69120\epsilon})\ ,\\
\end{aligned}\label{tensor2ex12}
\end{equation}
which is consistent with the direct computation in \eqref{local div direct}.

A possible problem with \eqref{zxv} is that $Z(x,v)$ becomes singular when some $n_i\le 0$, because of the $\Gamma(n_i)$ in the numerator. To regulate this singularity, we can shift $n_i$ by an infinitesimal number $\delta$, $n_i\rightarrow n_i+\delta$. Suppose the $y$-derivative produces a $x_i^{\alpha}$ term, then
\begin{equation}\label{ni-negative1}
\frac{x_i^{n_i+\delta-1}}{\Gamma(n_i+\delta)}x_i^{\alpha}
=\frac{x_i^{n_i+\alpha+\delta-1}}{\Gamma(n_i+\alpha+\delta)}\frac{\Gamma(n_i+\alpha+\delta)}{\Gamma(n_i+\delta)}
= \frac{x_i^{n_i+\alpha+\delta-1}}{\Gamma(n_i+\alpha+\delta)}(n_i+\delta)_{\alpha}\ .
\end{equation}
Notice that $(n_i+\delta)_{\alpha}=(n_i+\delta)(n_i+\delta+1)\cdots(n_i+\delta+\alpha-1)$ is regular when $\delta\rightarrow 0$, so \eqref{ni-negative1} can be formally written as
\begin{equation}\label{ni-negative2}
\frac{x_i^{n_i-1}}{\Gamma(n_i)}x_i^{\alpha}
= \frac{x_i^{n_i+\alpha-1}}{\Gamma(n_i+\alpha)}(n_i)_{\alpha}\ ,
\end{equation}
which means replacing $n_i\rightarrow n_i+\alpha$ in the integral, and multiplying the integral by $(n_i)_{\alpha}$. We also notice that $(n_i)_{\alpha}=0$ if $n_i\le 0$ and $\alpha> |n_i|$, which means a negative $n_i$ never become positive, therefore dimensional shift never convert a numerator into a propagator.

The dimensional shift method is extremely efficient for $L\le 3$ integrals, this allows us to handle the high rank tensor integrals in gravity and effective field theories. For example, the local divergence of the following 3-loop rank-8 tensor integral can be evaluated within seconds:
\begin{equation}
\begin{aligned}
&\mathbf{L}\frac{l_1^{\mu_1}l_1^{\mu_2}l_1^{\mu_3}l_2^{\nu_1}l_2^{\nu_2}l_2^{\nu_3}
l_3^{\rho_1}l_3^{\rho_2}}{(l_1^2)^2(l_2^2)^2(l_3^2)^2(l_2-l_3)^2(l_1-l_3)^2[(l_1-l_2)^2]^2}\\
=&\Bigl[\frac{1}{11520\epsilon ^{3}}-\frac{47}{345600\epsilon ^{2}}+\frac{-14141+2880\zeta_{3}}{41472000\epsilon }\Bigl]
\Bigl(\eta^{\mu_1\mu_2}\eta^{\mu_3\nu_3}\eta^{\nu_1\rho_1}\eta^{\nu_2\rho_2}
+(\mu_i\leftrightarrow \nu_i)\Bigr)\\
&+\Bigl[\frac{1}{11520\epsilon ^{3}}-\frac{47}{345600\epsilon ^{2}}+\frac{31459-25920\zeta_{3}}{41472000\epsilon }\Bigl]\eta^{\mu_1\nu_1}\eta^{\mu_2\nu_2}\eta^{\mu_3\rho_1}\eta^{\nu_3\rho_2}\\
&+\Bigl[\frac{1}{11520\epsilon ^{3}}-\frac{47}{345600\epsilon ^{2}}+\frac{-23741+31680\zeta_{3}}{41472000\epsilon }\Bigl]
\eta^{\mu_1\mu_2}\eta^{\mu_3\rho_1}\eta^{\nu_1\nu_2}\eta^{\nu_3\rho_2}\\
&+\Bigl[\frac{1}{11520\epsilon ^{3}}-\frac{347}{345600\epsilon ^{2}}+\frac{97759-112320\zeta_{3}}{41472000\epsilon }\Bigl]
\eta^{\mu_1\nu_1}\eta^{\mu_2\nu_2}\eta^{\mu_3\nu_3}\eta^{\rho_1\rho_2}\\
&+\Bigl[\frac{1}{11520\epsilon ^{3}}-\frac{347}{345600\epsilon ^{2}}+\frac{68959-25920\zeta_{3}}{41472000\epsilon }\Bigl]
\eta^{\mu_1\mu_2}\eta^{\nu_1\nu_2}\eta^{\mu_3\nu_3}\eta^{\rho_1\rho_2}\\
&+\text{permutations}\ ,\\
\end{aligned}
\end{equation}
in which "permutations" means the non-repetitive permutations of $(\mu_1,\mu_2,\mu_3)$, $(\nu_1,\nu_2,\nu_3)$, and $(\rho_1,\rho_2)$.

The dimensional shift method can also be applied to higher loops, but it is less efficient if $L\ge4$. This is because dimensional shift produces scalar integrals with larger $n_i$, and it can be very difficult to evaluate their local divergences.
In the next subsection, we propose a new approach based on $d_{\infty}$ dimensional PV reduction which is more effective at higher loops.

\subsection{The $d_{\infty}$ dimensional PV reduction}
As has been discussed in the previous subsection, the "naive" PV reduction does not commute with $\mathbf{L}$ because the reduction formula contains explicit $\epsilon$-dependence. To avoid this problem, we choose a $d_{\infty}=d+2k$ dimensional subspace of $\mathbb{R}^D$,  in which $k$ is an integer. 
Although it seems to make more sense to require $0<d_{\infty}\le d$, 
as discussed in e.g. \cite{Wilson:1972cf, Collins:1984xc,Stockinger:2005gx}, $R^D$ should be regarded as a infinite dimensional space, so $d_{\infty}$ can be an arbitrarily large number. We will assume that $d_{\infty}$ is sufficiently large so that all external momenta $p_a$ and loop momenta $\tilde{l}_a$ are all independent.

The metric in the $d_{\infty}$ dimensional subspace will be denoted by $\tilde{\eta}_{\mu\nu}$, which satisfies
\begin{equation}
\tilde{\eta}_{\mu\nu}\tilde{\eta}^{\nu\rho}
=\tilde{\eta}_{\mu\nu}\eta^{\nu\rho}=\tilde{\delta}_{\mu}^{\rho},\ 
\tilde{\delta}_{\mu}^{\mu}=d_{\infty}\ .
\end{equation}
The $d_{\infty}$ dimensional component of a vector $l$ will be denoted by $\tilde{l}_{\mu}\equiv\tilde{\eta}_{\mu\nu}l^{\nu}$. 

This contraction of $\tilde{\eta}^{\mu\nu}$ produces no explicit $\epsilon$-dependence, so the $d_{\infty}$ dimensional PV reduction commutes with $\mathbf{L}$. As an example, the local divergence of the integral in \eqref{tensor.example.1} can be written as
\begin{equation}\label{tensor.example.3}
\mathbf{L}\frac{l_2^{\mu}l_2^{\nu}}{(l_1^2)^5(l_2^2)^3(l_1-l_2)^2}
=c\eta^{\mu\nu}\ .
\end{equation}
Contracting both sides of \eqref{tensor.example.3} with $\tilde{\eta}_{\mu\nu}$ renders
\begin{equation}\label{tensor.example.4}
\mathbf{L}\frac{\tilde{l}_2^2}{(l_1^2)^5(l_2^2)^3(l_1-l_2)^2}
=cd_{\infty}\ .
\end{equation}
The value of $c$ can be fixed by \eqref{tensor.example.4} and we find
\begin{equation}\label{tensor.example.5}
\mathbf{L}\frac{l_2^{\mu}l_2^{\nu}}{(l_1^2)^5(l_2^2)^3(l_1-l_2)^2}
=\frac{\eta^{\mu\nu}}{d_{\infty}}\mathbf{L}\frac{\tilde{l}_2^2}{(l_1^2)^5(l_2^2)^3(l_1-l_2)^2}\ .
\end{equation}

Using the $d_{\infty}$ dimensional PV reduction, the local divergences of generic tensor integrals can be expressed by that  of scalar integrals containing $\tilde{l}_a\cdot \tilde{l}_b$ terms. These scalar integrals will be called $(D,d_{\infty})$ integrals. We will present and discuss several approaches to the local divergences of $(D,d_{\infty})$ integrals in the following subsections.

\subsection{Reducing $(D,d_{\infty})$ integrals using dimensional shift}

A variant of dimensional shift (see e.g. \cite{Bern:1995db}) reduces integrals with $\mu^2$ (products of loop momenta in $-2\epsilon$ dimension) to scalar integrals in $D+2a$ dimension.
In this subsection, we will briefly review this variant of dimensional shift, and then we will use the method to evaluate the local divergence of $(D,d_{\infty})$ integrals.

First let us consider a 2-loop vacuum integral which contains Lorentz products in $-2\epsilon$ dimension, $\mu_i\cdot \mu_j$, in the numerator:
\begin{equation}
(I_{\mu})^{a_1a_2a_3}_{n_1n_2n_3}
\equiv\frac{(\mu_1^2)^{a_1}(\mu_2^2)^{a_2}[(\mu_1-\mu_2)^2]^{a_3}}{(l_1^2+m_1^2)^{n_1}(l_2^2+m_2^2)^{n_2}[(l_1-l_2)^2+m_3^2]^{n_3}}\ .
\end{equation}

This integral can be regarded as the coefficient of $y_1^{a_1}y_2^{a_2}y_3^{a_3}$ in a generating function $G_{\mu}(x,y)$:
\begin{equation}
\begin{aligned}
G_{\mu}(x,y)\equiv &\int_0^{\infty} \prod_{i=1}^3 \frac{ dx_ix_i^{n_i-1}}{\Gamma(n_i)}
e^{-F(x)-\left[x_1 l_1^2+x_2l_2^2+x_3(l_1-l_2)^2\right]-\left[y_1 \mu_1^2+y_2\mu_2^2+y_3(\mu_1-\mu_2)^2\right]}\ ,\\
(I_{\mu})^{a_1a_2a_3}_{n_1n_2n_3}
=&(-1)^{a_1+a_2+a_3}\frac{\partial^{a_1+a_2+a_3}}{\partial y_1^{a_1}\partial y_2^{a_2}\partial y_3^{a_3}}
G_{\mu}(x,y)\Bigr|_{y_i=0}\ .\\
\end{aligned}\label{dim-shift-mu1}
\end{equation}
In order to evaluate $G_{\mu}(x,y)$, we split the loop integration into a $d$ dimension part and a $-2\epsilon$ dimension part, and we find 
\begin{equation}
\begin{aligned}
G_{\mu}(x,y)\equiv &\int_0^{\infty} \prod_{i=1}^3 \frac{ dx_ix_i^{n_i-1}}{\Gamma(n_i)}
U(x)^{-\frac{d}{2}}U(x+y)^{-\frac{D-d}{2}}e^{-F}\  .\\
\end{aligned}\label{dim-shift-mu2}
\end{equation}
Then $(I_{\mu}^d)^{a_1a_2a_3}_{n_1n_2n_3}$ can be written as scalar integrals in $D+2(a_1+a_2+a_3)$ dimension.

As an example, consider the following integral in 8-d:
\begin{equation}
\begin{aligned}
&\frac{\mu_2^2}{(l_1^2)^5(l_2^2)^3(l_1-l_2)^2}
=(I_{\mu})^{010}_{531}\\
=&-\frac{\partial}{\partial y_2}
\int_0^{\infty} \prod_{i=1}^3 \frac{ dx_ix_i^{n_i-1}}{\Gamma(n_i)}
U(x)^{-\frac{d}{2}}U(x+y)^{-\frac{D-d}{2}}\Bigr|_{y_i=0}\\
=&\frac{D-d}{2}
\int_0^{\infty} \prod_{i=1}^3 d^3x \frac{x_1^4x_2^2}{\Gamma(5)\Gamma(3)}U(x)^{-\frac{D+2}{2}}(x_1+x_3)\\
=&\frac{D-d}{2}(5I_{631}+I_{532})\ .\\
\end{aligned}\label{tensor.example.6}
\end{equation}

Again, this tensor reduction produced explicit $\epsilon$-dependence and it does not commute with $\mathbf{L}$. In order to avoid the $\epsilon$-dependence, we replace the $\mu_i\cdot\mu_j$ terms by $\tilde{l}_i\cdot \tilde{l}_j$, and define
\begin{equation}
\begin{aligned}
I^{a_1a_2a_3}_{n_1n_2n_3}
\equiv&\frac{(\tilde{l}_1^2)^{a_1}(\tilde{l}_2^2)^{a_2}[(\tilde{l}_1-\tilde{l}_2)^2]^{a_3}}{(l_1^2)^{n_1}(l_2^2)^{n_2}[(l_1-l_2)^2]^{n_3}}
=(-1)^{a_1+a_2+a_3}\frac{\partial^{a_1+a_2+a_3}}{\partial y_1^{a_1}\partial y_2^{a_2}\partial y_3^{a_3}}G(x,y)\Bigr|_{y_i=0}\ ,\\
G(x,y)\equiv &\int_0^{\infty} \prod_{i=1}^3 \frac{ dx_ix_i^{n_i-1}}{\Gamma(n_i)}
e^{-\left[x_1 l_1^2+x_2l_2^2+x_3(l_1-l_2)^2\right]-\left[y_1 \tilde{l}_1^2+y_2\tilde{l}_2^2+y_3(\tilde{l}_1-\tilde{l}_2)^2\right]}\ \\
=&\int_0^{\infty} \prod_{i=1}^3 \frac{ dx_ix_i^{n_i-1}}{\Gamma(n_i)}
U(x)^{-\frac{D-d_{\infty}}{2}}U(x+y)^{-\frac{d_{\infty}}{2}}\ .\\
\end{aligned}\label{dim-shift-tilde1}
\end{equation}

As an example, let us go back to the $(D,d_{\infty})$ integral in \eqref{tensor.example.5}, the reduction is similar to \eqref{tensor.example.6} but $D-d$ should be replaced by $d_{\infty}$:
\begin{equation}
\begin{aligned}
&\mathbf{L}\frac{l_2^{\mu}l_2^{\nu}}{(l_1^2)^5(l_2^2)^3(l_1-l_2)^2}
=\frac{\eta^{\mu\nu}}{d_{\infty}}\mathbf{L}I_{532}^{010}
=\mathbf{L}\left(\frac{5}{2}I_{631}+\frac{1}{2}I_{532}\right)\eta^{\mu\nu}\ ,\\
\end{aligned}\label{dim-shift-tilde2}
\end{equation}
and the result is consistent with \eqref{tensor2ex11}.

This "$d_{\infty}$ dimension PV reduction plus dimensional shift" approach is equivalent to the direct dimensional shift approach, but it seems to be more complicated. In the next subsection, we will see that its efficiency can be enhanced by taking the large $d_{\infty}$ limit.

\subsection{The large $d_{\infty}$ limit}

In the previous subsections, we see that although the intermediate result from $d_{\infty}$ dimensional PV reduction depends on $d_{\infty}$, after dimensional shift the final result 
is free of $d_{\infty}$-dependence.
In this section we show that he computation can be simplified using this property.

First let us consider a slightly more complicated integral in 4-d,
\begin{equation}
F=\frac{l_1^{\mu}l_1^{\nu}l_2^{\rho}l_2^{\sigma}}{(l_1^2)^3(l_2^2)^2(l_1-l_2)^2}\ .
\end{equation}
The $d_{\infty}$ dimensional PV reduction gives
\begin{equation}
\begin{aligned}
F
\rightarrow &\frac{d_{\infty} (I_{321}^{200}+I_{321}^{020}+I_{321}^{002}-2I_{321}^{101}-2I_{321}^{011})
+  2 ( d_{\infty}-2) I_{321}^{110}}{4d_{\infty}(d_{\infty}-1)(d_{\infty}+2)}(\eta^{\mu\rho}\eta^{\nu\sigma}+\eta^{\mu\sigma}\eta^{\nu\rho})\\
&+\frac{2 d_{\infty} I_{321}^{110}-I_{321}^{200}-I_{321}^{020}-I_{321}^{002} + 2 I_{321}^{101}+ 2 I_{321}^{011} }{2d_{\infty}(d_{\infty}-1)(d_{\infty}+2)}\eta^{\mu\nu}\eta^{\rho\sigma}\ .\\
\end{aligned}\label{tensor-ex-2}
\end{equation}
The reduction of $(D,d_{\infty})$ integrals using dimensional shift is given by,
\begin{equation}
\begin{aligned}
&I_{321}^{200}=\frac{d_{\infty}(d_{\infty}+2)}{2}(I_{323}+2I_{332}+3I_{341}),\ \\
&I_{321}^{020}=\frac{d_{\infty}(d_{\infty}+2)}{2}(I_{323}+3I_{422}+6I_{521}),\ \\
&I_{321}^{002}=\frac{d_{\infty}(d_{\infty}+2)}{2}(I_{321}-2I_{332}+3I_{341}-3I_{422}
+6I_{521}),\ \\
&I_{321}^{110}=\frac{d_{\infty}^2}{4}I_{321}+\frac{d_{\infty}(d_{\infty}+2)}{2}I_{323},\ \\
&I_{321}^{011}=\frac{d_{\infty}^2}{4}I_{321}+3d_{\infty}(d_{\infty}+2)I_{521},\ \\
&I_{321}^{101}=\frac{d_{\infty}^2}{4}I_{321}+\frac{3d_{\infty}(d_{\infty}+2)}{2}I_{341},\ \\
\end{aligned}\label{tensor-ex-3}
\end{equation}
in which we have reduced some 8-d integrals to the 6-d integral $I_{321}$ by replacing $x_1x_2+x_1x_3+x_2x_3$ by $U(x)$.

Plug \eqref{tensor-ex-3} back into \eqref{tensor-ex-2}, the final result is independent of $d_{\infty}$, as expected:
\begin{equation}
\begin{aligned}
&F\rightarrow \frac{1}{2}I_{323}(\eta^{\mu\rho}\eta^{\nu\sigma}+\eta^{\mu\sigma}\eta^{\nu\rho})+(\frac{1}{4}I_{321}+\frac{1}{2}I_{323})\eta^{\mu\nu}\eta^{\rho\sigma}.\ \\
\end{aligned}\label{tensor-ex-4}
\end{equation}

Now let us examine the behavior of these quantities in the $d_{\infty}\rightarrow \infty$ limit.
The $(D,d_{\infty})$ integrals behaves as
\begin{equation}
I_{n_1\cdots n_A}^{a_1\cdots a_A}=\mathcal{O}(d_{\infty}^{a}),\ 
a\equiv a_1+\cdots+a_A\ .
\end{equation}
It is convenient to rescale them by
\begin{equation}
\hat{I}_{n_1\cdots n_A}^{a_1\cdots a_A}=d_{\infty}^{-a}I_{n_1\cdots n_A}^{a_1\cdots a_A},
\end{equation}
then $\hat{I}_{n_1\cdots n_A}^{a_1\cdots a_A}\sim \mathcal{O}(d_{\infty}^0)$.
Only keep the leading $d_{\infty}$ terms, \eqref{tensor-ex-3} becomes
\begin{equation}
\begin{aligned}
&\hat{I}_{321}^{200}=\frac{1}{2}(I_{323}+2I_{332}+3I_{341})+\mathcal{O}(\frac{1}{d_{\infty}}),\ \\
&\hat{I}_{321}^{020}=\frac{1}{2}(I_{323}+3I_{422}+6I_{521})+\mathcal{O}(\frac{1}{d_{\infty}}),\ \\
&\hat{I}_{321}^{002}=\frac{1}{2}(I_{321}-2I_{332}+3I_{341}-3I_{422}
+6I_{521})+\mathcal{O}(\frac{1}{d_{\infty}}),\ \\
&\hat{I}_{321}^{110}=\frac{1}{4}I_{321}+\frac{1}{2}I_{323}+\mathcal{O}(\frac{1}{d_{\infty}}),\ \\
&\hat{I}_{321}^{011}=\frac{1}{4}I_{321}+3I_{521}+\mathcal{O}(\frac{1}{d_{\infty}}),\ \\
&\hat{I}_{321}^{101}=\frac{1}{4}I_{321}+\frac{3}{2}I_{341}+\mathcal{O}(\frac{1}{d_{\infty}}).\ \\
\end{aligned}\label{tensor-ex-5}
\end{equation}
Eq. \eqref{tensor-ex-2} becomes
\begin{equation}
\begin{aligned}
F\rightarrow &\frac{ 1}{4}
\left(\hat{I}_{321}^{200}+\hat{I}_{321}^{020}+\hat{I}_{321}^{002}
-2\hat{I}_{321}^{101}-2\hat{I}_{321}^{011}+  2 \hat{I}_{321}^{110}\right)
(\eta^{\mu\rho}\eta^{\nu\sigma}+\eta^{\mu\sigma}\eta^{\nu\rho})\\
&+\hat{I}_{321}^{110}\eta^{\mu\nu}\eta^{\rho\sigma}
+\mathcal{O}(\frac{1}{d_{\infty}})\ \\
=& \frac{1}{2}I_{323}(\eta^{\mu\rho}\eta^{\nu\sigma}+\eta^{\mu\sigma}\eta^{\nu\rho})+(\frac{1}{4}I_{321}+\frac{1}{2}I_{323})\eta^{\mu\nu}\eta^{\rho\sigma}
+\mathcal{O}(\frac{1}{d_{\infty}})\ \\
= & \frac{1}{2}I_{323}(\eta^{\mu\rho}\eta^{\nu\sigma}+\eta^{\mu\sigma}\eta^{\nu\rho})+(\frac{1}{4}I_{321}+\frac{1}{2}I_{323})\eta^{\mu\nu}\eta^{\rho\sigma}\ .\\
\end{aligned}\label{tensor-ex-6}
\end{equation}
In the last step, we have dropped the $\mathcal{O}(\frac{1}{d_{\infty}})$ terms, since they must vanish because the final result should not depend on $d_{\infty}$.

For general tensor structures, the PV reduction in the large $d_{\infty}$ limit is given by:
\begin{equation}
\begin{aligned}
&l_{1}^{\mu_1}\cdots l_{2a}^{\mu_{2a}}
=\Bigl[(\frac{\tilde{l}_1\cdot \tilde{l}_2 }{d_{\infty}}\eta^{\mu_1\mu_2})\cdots (\frac{\tilde{l}_{2a-1}\cdot \tilde{l}_{2a}}{d_{\infty}} \eta^{\mu_{2a-1}\mu_{2a}})+\text{permutations}\Bigr]+\mathcal{O}(\frac{1}{d_{\infty}}).\\
\end{aligned}\label{d-infinity-pv}
\end{equation}
Eq \eqref{d-infinity-pv} is completely symmetric and only contain $(2a-1)!!$ terms. It can be proved by contracting both sides of the equation with $ \tilde{\eta}^{\mu_1\mu_2}\cdots \tilde{\eta}^{\mu_{2a-1}\mu_{2a}}$, and notice that only the $(\tilde{l}_1\cdot \tilde{l}_2 \eta^{\mu_1\mu_2})\cdots (\tilde{l}_{2a-1}\cdot \tilde{l}_{2a} \eta^{\mu_{2a-1}\mu_{2a}})$ term contribute in the large $d_{\infty}$ limit. 

The dimensional shift formula \eqref{dim-shift-tilde1} also simplifies in the large $d_{\infty}$ limit. In order to have highest power of $d_{\infty}$, each $y$-derivative should act on the $U(x+y)^{-\frac{d_{\infty}}{2}}$ term:
\begin{equation}
\begin{aligned}
I_{n_1\cdots n_A}^{a_1\cdots a_A}
= &(\frac{d_{\infty}}{2})_{a}[U(x)]^{-\frac{D+2a}{2}}\prod_{i=1}^A\frac{x_i^{n_i-1}}{\Gamma(n_i)}\Bigl[\frac{\partial U(x+y)}{\partial y_i}\Bigr]^{a_i}\Bigr|_{y=0}
+\mathcal{O}(d_{\infty}^{a-1})\\
= &(\frac{d_{\infty}}{2})^{a}[U(x)]^{-\frac{D+2a}{2}}\prod_{i=1}^A\frac{x_i^{n_i-1}}{\Gamma(n_i)}\Bigl[\frac{\partial U(x)}{\partial x_i}\Bigr]^{a_i}
+\mathcal{O}(d_{\infty}^{a-1})\ .\\
\end{aligned}\label{dim-shift-large}
\end{equation}

As a simple example, let us consider the 1-loop $(D,d_{\infty})$ CV,
\begin{equation}
\begin{aligned}
I_{a+\frac{d}{2}}^a=&\frac{(\tilde{l}_1^2)^{a}}{(l_1^2)^{a+\frac{d}{2}}}
=(\frac{d_{\infty}}{2})^{a} x^{-\frac{D+2a}{2}}\frac{x^{a+\frac{d}{2}-1}}{\Gamma(a+\frac{d}{2})}+\mathcal{O}(d_{\infty}^{a-1})
=\frac{(\frac{d_{\infty}}{2})^{a}}{(\frac{d}{2})_{a}} I_{\frac{d}{2}}+\mathcal{O}(d_{\infty}^{a-1}),\\
\end{aligned}
\end{equation}
in which we used $U(x)=x$. The local divergence is given by
\begin{equation}
\begin{aligned}
\mathbf{L}I_{a+\frac{d}{2}}^a=&\frac{(\frac{d_{\infty}}{2})^a}{\Gamma(\frac{d}{2}+a)\epsilon}+\mathcal{O}(d_{\infty}^{a-1})\ .\\
\end{aligned}
\end{equation}

Both \eqref{d-infinity-pv} and \eqref{dim-shift-large} are much simpler compared with the original version of PV reduction and dimensional shift, because the majority of terms are suppressed by $\frac{1}{d_{\infty}}$ and can be neglected. However, as has been discussed before, at higher loops the bottleneck in the computation is the evaluation of higher dimensional scalar integrals, and it is preferable to compute the local divergence of $(D,d_{\infty})$ integrals directly.

\subsection{The local divergence of  $(D,d_{\infty})$ integrals}

As discussed in previous subsections, in 4 and higher loops, it is more efficient to compute the local divergence of $(D,d_{\infty})$ integrals directly using UV decomposition \eqref{uvdecom-ir-3}. This means we need to find efficient ways to evaluate sub UV divergences, IR divergences and the complete expression of $(D,d_{\infty})$ integrals in the large $d_{\infty}$ limit.

The $(D,d_{\infty})$ integrals contain $d_{\infty}$ dimensional Lorentz products $\hat{l}_i\cdot \hat{l}_j$ in the numerator. During the evaluation of the complete integral and IR divergence, the $\hat{l}_i\cdot \hat{l}_j$ terms can be reduced to $l_i\cdot l_j$ terms with the help of PV reduction. To see this, let us consider a tensor structure with rank-$2n$ vectors, $T^{\mu_1\cdots \mu_{2n}}=l_1^{\mu_1}\cdots l_{2n}^{\mu_{2n}}$. After PV reduction, the tensor structures have the basis $\{E_a^{\mu_1\cdots\mu_{2n}}|a=1,\cdots, (2n-1)!!\}$, in which $E_a$ can be generated from the following $E_1$ by  permuting $\mu_i$,
\begin{equation}
E_1^{\mu_1\cdots\mu_{2n}}=\eta^{\mu_1\mu_2}\cdots \eta^{\mu_{2n-1}\mu_{2n}}\ .
\end{equation}
We can define the metric $G_{ab}(D)=E_a\cdot E_b$, in which $\cdot$ means contracting all $\mu_i$ indices.

The $D$ dimensional PV reduction of $T^{\mu_1\cdots \mu_{2n}}$ is given by
\begin{equation}\label{pv-tilde-1}
T=\sum_{b,c} G^{bc}(D)(T\cdot E_c)E_b\ ,
\end{equation}
in which $G^{bc}$ is the inverse of $G_{ab}$.

If we contract both sides of \eqref{pv-tilde-1} with $\hat{E}_a\equiv E_a|_{\eta\rightarrow \hat{\eta}}$,
\begin{equation}\label{pv-tilde-2}
T\cdot \hat{E}_a=\sum_{b,c} G^{bc}(D)(T\cdot E_c)E_b
=\sum_{b,c} G_{ab}(d_{\infty})G^{bc}(D)(T\cdot E_c)\ .
\end{equation}
We observe that $T\cdot \hat{E}_a$ and $T\cdot E_a$ are products of  $\hat{l}_i\cdot \hat{l}_j$ and $l_i\cdot l_j$, respectively. Therefore, $(D,d_{\infty})$ integrals can be converted to D dimensional integrals using \eqref{pv-tilde-2}.

The number of elements in $\{E_a\}$ can be reduced by observing some of $l_i$ are the same.
For example, the tensor structures of 2-loop integrals are of the form $l_1^{\mu_1}\cdots l_1^{\mu_A}l_2^{\nu_1}\cdots l_2^{\nu_B}$, so one only need to consider $E_a^{\mu_1\cdots\mu_A\nu_1\cdots\nu_B}$ which is invariant under the permutations of $\mu_i$ and $\nu_i$. Eq. \eqref{pv-tilde-2} can be further simplified the taking the large $d_{\infty}$ limit in $G_{ab}(d_{\infty})$.

The large $d_{\infty}$ limit also helps in the computation of sub UV divergences. Suppose $\theta$ is a UV sub-integral of the integral $V(m)$, and let $l^h_i$ ($l^s_i$) be the loop (external) momenta of $\theta$, respectively. After the asymptotic expansion, $\mathcal{A}^0\theta$ contains terms like
\begin{equation}
N_{abc}=(\tilde{l}^h_i\cdot\tilde{l}^h_j)^a(\tilde{l}^h_i\cdot\tilde{l}^s_j)^b(l^h_i\cdot l^s_j)^c\ ,
\end{equation} 
in the numerator. In the large $d_{\infty}$ limit, $N_{abc}=\mathcal{O}(d_{\infty}^{a+b})$.

We know that $l_i^h,\ l_i^s\in \mathbb{R}^D$, and $\tilde{l}_i^h, \tilde{l}_i^s\in \mathbb{R}^{d_{\infty}}\subset \mathbb{R}^D$. However, there is an important difference between the soft momenta and the hard momenta. Both $\hat{l}^s_i$ and $l^s_i$ should be regarded as external momenta of the sub-integral $\theta$, so they are some constant vectors when we integrate over the hard loop momenta. Therefore it is always possible to choose $\mathbb{R}^{d_{\infty}}$ properly so that $\hat{l}^s_i, l^s_i\in \mathbb{R}^{d_{\infty}}$.
With this choice, $\tilde{l}^h_i\cdot\tilde{l}^s_j=l^h_i\cdot\tilde{l}^s_j$, and\footnote{Strictly speaking we should define a new $\mathbb{R}^{d^1_{\infty}}$ which satisfies $\mathbb{R}^{d_{\infty}}\subset \mathbb{R}^{d^1_{\infty}}$, and perform a $d^1_{\infty}$ dimensional PV reduction and determine the local divergence of $N_{abc}$. The $(\tilde{l}^h_i\cdot\tilde{l}^h_j)^a$ term lives in $\mathbb{R}^{d_{\infty}}$, and should be treated as a tensor during the $d^1_{\infty}$ dimensional PV reduction.}
\begin{equation}
N_{abc}=(\tilde{l}^h_i\cdot\tilde{l}^h_j)^a(l^h_i\cdot\tilde{l}^s_j)^b(l^h_i\cdot l^s_j)^c\ .
\end{equation} 

$N_{abc}$ can be regarded tensor structures of $l^h_i$ contracted with $ \tilde{l}^s_i$ and $l^s_i$. We can reduce the tensor structures using a $d_{\infty}$ dimensional PV reduction, and then contract the resulting $\eta^{\mu\nu}$ with $ \tilde{l}^s_i$ and $l^s_i$:
\begin{equation}
\begin{aligned}
N_{abc}\rightarrow &d_{\infty}^{-\frac{b+c}{2}}\sum_n(\tilde{l}^h_i\cdot\tilde{l}^h_j)^{a+\frac{b+c}{2}}(\tilde{l}^s_i\cdot\tilde{l}^s_j)^n(\tilde{l}^s_i\cdot l^s_j)^{b-2n}(l^s_i\cdot l^s_j)^{n+\frac{c-b}{2}}\\
=&d_{\infty}^{-\frac{b+c}{2}}\sum_n(\tilde{l}^h_i\cdot\tilde{l}^h_j)^{a+\frac{b+c}{2}}(\tilde{l}^s_i\cdot\tilde{l}^s_j)^{b-n}(l^s_i\cdot l^s_j)^{n+\frac{c-b}{2}}\ ,\\
\end{aligned}\label{Nabc-1}
\end{equation}
in which we used $\tilde{l}^s_i\cdot l^s_j=\tilde{l}^s_i\cdot\tilde{l}^s_j$.

It can be seen that in \eqref{Nabc-1} only the term with $n=0$ contributes in the large $d_{\infty}$ limit, because otherwise the term behaves like $\mathcal{O}(d_{\infty}^{a+b-n})$. This means all $\tilde{l}^s_i$ must be contracted to $l^s_i$ during the $d_{\infty}$ dimensional PV reduction, which also requires $b\le c$. So during the evaluation of sub UV divergences, a lot of terms can be dropped using the large $d_{\infty}$ limit.

\subsection{5-loop tensor integrals}
\label{5-loop-ex}
In this subsection we shall demonstrate the $d_{\infty}$ dimensional PV reduction by evaluating the local divergences of some 5-loop tensor integrals.
\begin{figure}[htb]
\centering
\includegraphics[scale=0.7]{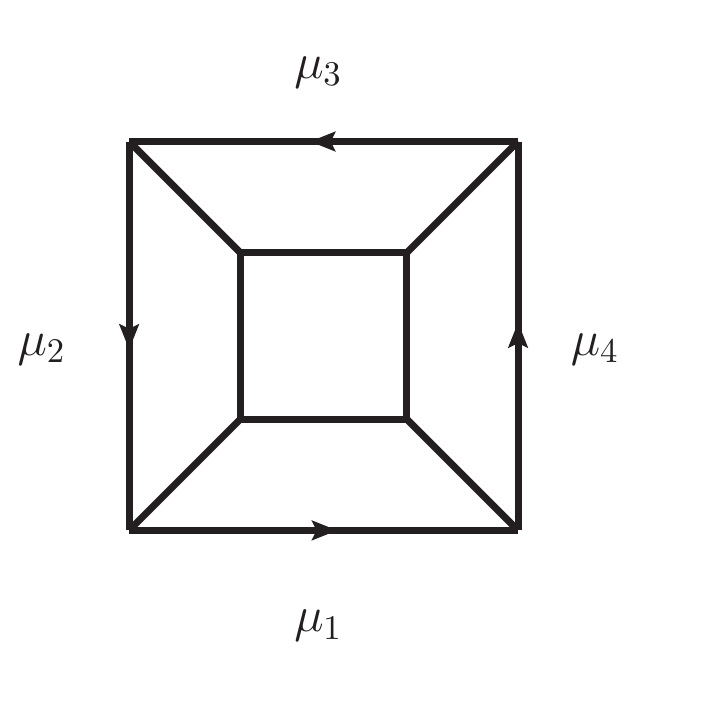}
\caption{A 5-loop tensor integral. The $\mu_i$ besides a propagator with momentum $l_a$ means there is a $l_a^{\mu_i}$ in the numerator.}
\label{fig:5looptensor}
\end{figure}

Let us consider the integral in Figure \ref{fig:5looptensor}. The integral can be parameterized as $l_1^{\mu_1}l_2^{\mu_2}l_3^{\mu_3}l_4^{\mu_4}$.
\begin{equation}
T=\frac{l_1^{\mu_1}l_2^{\mu_2}l_3^{\mu_3}l_4^{\mu_4}}{l_1^2l_2^2l_3^2l_4^2(l_1-l_2)^2(l_2-l_3)^2(l_3-l_4)^2(l_1-l_4)^2(l_1-l_5)^2(l_2-l_5)^2(l_3-l_5)^2(l_4-l_5)^2}\ .
\end{equation}

The $d_{\infty}$ dimensional PV reduction produces:
\begin{equation}
\begin{aligned}
&l_1^{\mu_1}l_2^{\mu_2}l_3^{\mu_3}l_4^{\mu_4}
\rightarrow \frac{1}{d_{\infty}^2}
\Bigl[\tilde{l}_1\cdot\tilde{l}_2\tilde{l}_3\cdot\tilde{l}_4\eta_1
+\tilde{l}_1\cdot\tilde{l}_3\tilde{l}_2\cdot\tilde{l}_4\eta_2
+\tilde{l}_1\cdot\tilde{l}_4\tilde{l}_2\cdot\tilde{l}_3\eta_3\Bigr]\ ,\\
&\eta_1=\eta^{\mu_1\mu_2}\eta^{\mu_3\mu_4},\ 
\eta_2=\eta^{\mu_1\mu_3}\eta^{\mu_2\mu_4},\ 
\eta_3=\eta^{\mu_1\mu_4}\eta^{\mu_2\mu_3}\ .\\
\end{aligned}
\end{equation}

Evaluating the local divergences of $(D,d_{\infty})$ integrals, we obtain
\begin{equation}
\begin{aligned}
\mathbf{L}T=
&\Bigl[-\frac{\zeta_{5}}{24\epsilon ^{2}}+\frac{-10\pi^{6}-2268\zeta_{3}^{2}+31500\zeta_{5}-36603\zeta_{7}}{90720\epsilon }\Bigr](\eta_{1}+\eta_{3})\\
&+\Bigl[-\frac{\zeta_{5}}{24\epsilon ^{2}}+\frac{-5\pi^{6}+7938\zeta_{3}^{2}+15750\zeta_{5}-33516\zeta_{7}}{45360\epsilon }\Bigr]\eta_{2}\ .\\
\end{aligned}
\end{equation}

If we modify the numerator but keep the propagator unchanged, we obtain
\begin{equation}
\begin{aligned}
l_1^{\mu_1}l_1^{\mu_2}l_2^{\mu_3}l_3^{\mu_4}\rightarrow
&\Bigl[-\frac{\zeta_{5}}{24\epsilon ^{2}}+\frac{-10\pi^{6}-6804\zeta_{3}^{2}+36540\zeta_{5}-18963\zeta_{7}}{90720\epsilon }\Bigr]\eta_{1}
\\
&+\Bigl[-\frac{\zeta_{5}}{24\epsilon ^{2}}+\frac{-40\pi^{6}-18144\zeta_{3}^{2}+85680\zeta_{5}-98343\zeta_{7}}{362880\epsilon }\Bigr](\eta_{2}+\eta_{3})\ ,\\
l_1^{\mu_1}l_1^{\mu_2}l_2^{\mu_3}l_4^{\mu_4}\rightarrow
&\Bigl[-\frac{\zeta_{5}}{24\epsilon ^{2}}+\frac{-20\pi^{6}-4536\zeta_{3}^{2}+73080\zeta_{5}-60417\zeta_{7}}{181440\epsilon }\Bigr]\eta_{1}\\
&+\Bigl[-\frac{\zeta_{5}}{24\epsilon ^{2}}+\frac{-10\pi^{6}-6804\zeta_{3}^{2}+21420\zeta_{5}-18963\zeta_{7}}{90720\epsilon }\Bigr](\eta_{2}+\eta_{3})\ ,
\\
l_1^{\mu_1}l_1^{\mu_2}l_1^{\mu_3}l_2^{\mu_4}
\rightarrow &\Bigl[-\frac{\zeta_{5}}{24\epsilon ^{2}}+\frac{-\pi^{6}-378\zeta_{3}^{2}+1638\zeta_{5}}{9072\epsilon }\Bigr](\eta_1+\eta_2+\eta_3)\ ,\\
l_1^{\mu_1}l_1^{\mu_2}l_1^{\mu_3}l_3^{\mu_4}\rightarrow
&\Bigl[-\frac{\zeta_{5}}{24\epsilon ^{2}}+\frac{-5\pi^{6}-4158\zeta_{3}^{2}+8190\zeta_{5}}{45360\epsilon }\Bigr](\eta_1+\eta_2+\eta_3)\ ,
\\
l_1^{\mu_1}l_1^{\mu_2}l_2^{\mu_3}l_2^{\mu_4}\rightarrow
&\Bigl[-\frac{\zeta_{5}}{24\epsilon ^{2}}+\frac{-\pi^{6}-378\zeta_{3}^{2}+4662\zeta_{5}}{9072\epsilon }\Bigr]\eta_{1}
\\
&+\Bigl[-\frac{\zeta_{5}}{24\epsilon ^{2}}+\frac{-\pi^{6}-378\zeta_{3}^{2}+126\zeta_{5}}{9072\epsilon }\Bigr](\eta_{2}+\eta_{3})\ ,\\
l_1^{\mu_1}l_1^{\mu_2}l_3^{\mu_3}l_3^{\mu_4}\rightarrow
&\Bigl[-\frac{\zeta_{5}}{24\epsilon ^{2}}+\frac{-5\pi^{6}-4158\zeta_{3}^{2}+23310\zeta_{5}}{45360\epsilon }\Bigr]\eta_{1}
\\
&+\Bigl[-\frac{\zeta_{5}}{24\epsilon ^{2}}+\frac{-5\pi^{6}-4158\zeta_{3}^{2}+630\zeta_{5}}{45360\epsilon }\Bigr](\eta_{2}+\eta_{3})\ ,\\
l_1^{\mu_1}l_1^{\mu_2}l_1^{\mu_3}l_1^{\mu_4}\rightarrow
&\Bigl[-\frac{\zeta_{5}}{12\epsilon ^{2}}+\frac{-\pi^{6}-378\zeta_{3}^{2}+1638\zeta_{5}}{4536\epsilon }\Bigr](\eta_1+\eta_2+\eta_3)\ .
\\
\end{aligned}
\end{equation}

\section{UV decomposition and renormalization}
\label{z-factors}
By now we have been focused on the UV decomposition of Feynman integrals. In this section, we discuss the UV divergences of physical quantities.
Anomalous dimensions and beta functions can be extracted from the UV divergences of correlation functions, scattering amplitudes and form factors, which can be regarded as the combination of several one-particle-irreducible (1PI) correlation functions.
We show that local divergence and sub-divergence can be naturally extended to correlation functions.
After the UV decomposition, the local divergence and various types of sub-divergences vanish separately. 

In Section \ref{phi3-decom}, we demonstrate the UV decomposition of correlation functions using the 6-d $\phi^3$ theory as an example. Then in Section \ref{subsection:uvdecom-general},  we will discuss the UV decomposition in more general theories.

\subsection{The UV decomposition in $\phi^3$ theory}
\label{phi3-decom}
The $\phi^3$ theory is one of the simplest quantum field theory. The beta functions and anomalous dimensions in this theory  have been computed to 5-loop \cite{Gracey:2015tta, Borinsky:2021jdb}, and they can be used to study the phase transitions in the Lee-Yang edge singularity problem \cite{Fisher:1978pf}. Here we re-examine the UV divergences of this theory using the UV decomposition method.

The Lagrangian of 6-d $\phi^3$ in Euclidean space is given by
\begin{equation}
L=\frac{1}{2}(\partial\phi)^2
+\frac{1}{2}(Z_{\phi}-1)(\partial\phi)^2+\frac{Z_gg}{3!}\phi^3\ ,
\end{equation}
in which we have split the Lagrangian into the free part and the interaction part.

\begin{figure}[htb]
\centering
\includegraphics[scale=0.5]{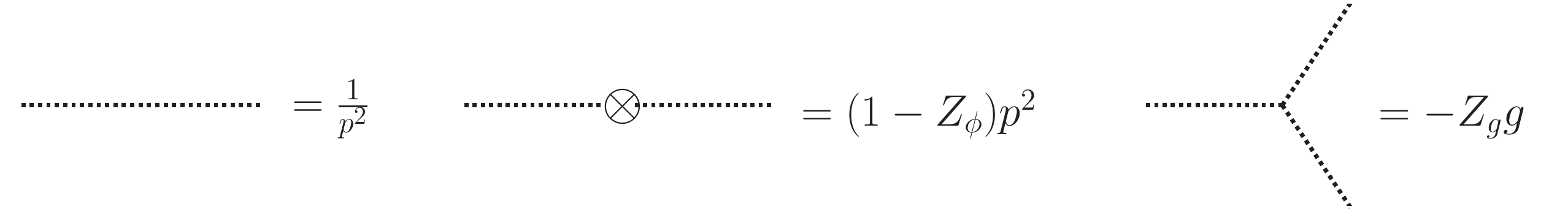}
\caption{Feynman rules of $\phi^3$ theory.}
\label{fig:phi3-feynman-rule}
\end{figure}
We will use the Feynman rules in Figure \ref{fig:phi3-feynman-rule}. 
The 1PI part of two-point correlation function can be written as
\begin{equation}
\begin{aligned}
\langle \phi\phi\rangle^{1PI}
=&\frac{1}{p^2}+\frac{1-Z_{\phi}}{p^2}+\frac{Z_g^2Z_{\phi}^{-2}g^2}{(p^2)^2}A_2^{(1)}
+\frac{Z_g^4Z_{\phi}^{-5}g^4}{(p^2)^2}A_2^{(2)}
+\frac{Z_g^6Z_{\phi}^{-8}g^6}{(p^2)^2}A_2^{(3)}+\cdots\ ,
\\
A_2^{(1)}=&\frac{1}{2l^2(l+p)^2}\ ,\\
A_2^{(2)}=&\frac{1}{2(l_1^2)^2(l_1-l_2)^2(l_2+p)^2}
+\frac{1}{2l_1^2l_2^2(l_1-l_2)^2(l_1+p)^2(l_2+p)^2}\ ,\\
A_2^{(3)}=&\frac{1}{8}I_{210201011}+\frac{1}{4}I_{311100011}
+I_{211101011}+\frac{1}{2}I_{221100101}\\
&+\frac{1}{4}I_{110110112}+\frac{1}{2}I_{110110112}+I_{111111101}
+\frac{1}{2}I_{211100111}\\
&+\frac{1}{4l_1^2l_2^2l_3^2(l_1+p)^2(l_3+p)^2(l_1-l_2)^2(l_2-l_3)^2(l_1-l_2+l_3+p_1)^2}\ ,\\
\end{aligned}\label{phi3n2-am}
\end{equation}
in which all 3-loop planar integrals are expressed using
\begin{equation}
I_{n_1\cdots n_9}=\frac{\prod_{i=1}^3(l_i^2)^{-n_i}[(l_i+p)^2]^{-n_{i+3}}}{
[(l_2-l_3)^2]^{n_7}[(l_1-l_3)^2]^{n_8}[(l_1-l_2)^2]^{n_9}}\ .
\end{equation}

In \eqref{phi3n2-am} we used an effective loop propagator in the loop integrand:
\begin{equation}
\frac{1}{l^2}+\frac{1}{l^2}(1-Z_{\phi})l^2\frac{1}{l^2}
+\frac{1}{l^2}(1-Z_{\phi})l^2\frac{1}{l^2}(1-Z_{\phi})l^2\frac{1}{l^2}+\cdots=\frac{Z_{\phi}^{-1}}{l^2}\ .
\end{equation}
But in the tree part\footnote{By "tree part" we mean the part of correlation function without loop integrals. For example, the $\frac{1}{p^2}+\frac{1-Z_{\phi}}{p^2}$ terms in \eqref{phi3n2-am}. Strictly speaking, $\frac{1-Z_{\phi}}{p^2}$ is a loop contribution since it contains counterterms.}, only the first two terms contribute, since the other terms are one-particle-reducible (1PR).

Let us examine the structure of local and sub-divergences in $\langle \phi\phi\rangle^{1PI}$. First, let us split the loop integrands into the "unrenormalized part" and the "counterterm part":
\begin{equation}
\begin{aligned}
\langle \phi\phi\rangle^{1PI}
=&\frac{1}{p^2}+\frac{g^2}{(p^2)^2}A_2^{(1)}
+\frac{g^4}{(p^2)^2}A_2^{(2)}
+\frac{g^6}{(p^2)^2}A_2^{(3)}+\cdots\ ,
\\
&+\frac{1-Z_{\phi}}{p^2}+\frac{(Z_g^2Z_{\phi}^{-2}-1)g^2}{(p^2)^2}A_2^{(1)}
+\frac{(Z_g^4Z_{\phi}^{-5}-1)g^4}{(p^2)^2}A_2^{(2)}\\
&+\frac{(Z_g^6Z_{\phi}^{-8}-1)g^6}{(p^2)^2}A_2^{(3)}+\cdots\ ,
\\
\end{aligned}\label{phi3n2-am2}
\end{equation}
Then we decompose the UV divergence of the "unrenormalized part":
\begin{equation}
\begin{aligned}
&A_2^{(1)}\sim -\frac{p^2}{12\epsilon }\ ,\\
&A_2^{(2)}\sim \Bigl(\frac{5}{144\epsilon ^{2}}-\frac{13}{864\epsilon }\Bigr)p^2
+\frac{5}{6\epsilon}A_2^{(1)}\ ,\\
&A_2^{(3)}\sim z_{\phi}^3p^2
+\frac{-285+194\epsilon }{432\epsilon ^{2}}A_2^{(1)}
+\frac{19}{12\epsilon }A_2^{(2)}\ ,\\
\end{aligned}
\end{equation}
in which 
\begin{equation}
z_{\phi}^3=-\frac{95}{5184\epsilon ^{3}}+\frac{341}{10368\epsilon ^{2}}+\frac{-5195+2592\zeta_{3}}{186624\epsilon }\ .
\end{equation}
The sub-divergences of $A_2^{(L)}$ can be neatly expressed by lower loop integrands times some  coefficients with $\epsilon$-poles, which must be canceled by the "counterterm part". Collecting all coefficients of $A_2^{(L)}$ in the "counterterm part" and the decomposition of the "unrenormalized part", we find
\begin{equation}
\begin{aligned}
\langle \phi\phi\rangle^{1PI}
\sim &\Bigl[2-Z_{\phi}-\frac{g^2}{12\epsilon }+g^4\Bigl(\frac{5}{144\epsilon ^{2}}-\frac{13}{864\epsilon }\Bigr)+g^6z_{\phi}^3\Bigr]\frac{1}{p^2}\\
&+\Bigl[Z_g^2Z_{\phi}^{-2}-1+\frac{5g^2}{6\epsilon}+g^4\frac{-285+194\epsilon }{432\epsilon ^{2}}\Bigr]\frac{g^2A_2^{(1)}}{(p^2)^2}\\
&+\Bigl[Z_g^4Z_{\phi}^{-5}-1+\frac{19g^2}{12\epsilon }\Bigr]\frac{g^4A_2^{(2)}}{(p^2)^2}+\cdots\ .
\\
\end{aligned}\label{phi3n2-am3}
\end{equation}

All three different types of UV divergences in \eqref{phi3n2-am3} must vanish separately, because each divergence must be canceled by the corresponding counterterm. This gives the following constraints to the Z-factors:
\begin{equation}
\begin{aligned}
&Z_{\phi}=1-\frac{g^2}{12\epsilon }+g^4\Bigl(\frac{5}{144\epsilon ^{2}}-\frac{13}{864\epsilon }\Bigr)+g^6z_{\phi}^3+\mathcal{O}(g^8)\ ,\\
&Z_g^2Z_{\phi}^{-2}=1-\frac{5g^2}{6\epsilon}+g^4\frac{285-194\epsilon }{432\epsilon ^{2}}+\mathcal{O}(g^6)\ ,\\
&Z_g^4Z_{\phi}^{-5}=1-\frac{19g^2}{12\epsilon }+\mathcal{O}(g^4)\ .\\
\end{aligned}\label{phi3-z2-constraint}
\end{equation}

Similarly, the decomposition of UV divergence of the three-point function gives
\begin{equation}
\begin{aligned}
& Z_g=1-\frac{g^2}{2\epsilon }+\frac{(30-23\epsilon )g^4}{96\epsilon ^{2}}
+g^6\Bigr(-\frac{5}{24\epsilon ^{3}}+\frac{83}{216\epsilon ^{2}}-\frac{1595+432\zeta_{3}}{5184\epsilon }\Bigr)
+\mathcal{O}(g^8)\ ,\\
&Z_g^3Z_{\phi}^{-3}=1-\frac{5g^2}{4\epsilon }
+\frac{(180-97\epsilon )g^4}{144\epsilon ^{2}}+\mathcal{O}(g^6)\ ,\\
&Z_g^5Z_{\phi}^{-6}=1-\frac{2g^2}{\epsilon }+\mathcal{O}(g^4)\ .\\
\end{aligned}\label{phi3-z3-constraint}
\end{equation}
It can be checked that the second and third line in \eqref{phi3-z2-constraint} and \eqref{phi3-z3-constraint} are consistent with the first line. Also, $Z_{\phi}$ and $Z_g$ are consistent with know results (see e.g.  \cite{Gracey:2015tta, Borinsky:2021jdb}).

\subsection{The UV decomposition in general quantum field theories}
\label{subsection:uvdecom-general}

Let us summarize the rules to compute local divergences of 1PI correlation functions:
\begin{enumerate}
\item In the tree part, $\mathbf{L}$ picks the terms with $\epsilon$-poles. For example,
\begin{equation}
\mathbf{L}Z_{\phi}=Z_{\phi}-1,\ \mathbf{L}Z_gg=(Z_g-1)g\ .
\end{equation}
\item In the loop part, $\mathbf{L}$ acts on the loop integrals, while Z factors are set to 1. For example, let $F(l_i)$ be the loop integral,
\begin{equation}
\mathbf{L}Z_g g F(l_i)=g\mathbf{L}F(l_i)\ .
\end{equation}

\end{enumerate}

Using these rules, the local divergences of  2 and 3 point correlation functions in $\phi^3$ theory are given by,
\begin{equation}
\begin{aligned}
\mathbf{L}\langle \phi\phi\rangle^{1PI}
= &\frac{1-Z_{\phi}}{p^2}+\frac{g^2}{(p^2)^2}\mathbf{L}A_2^{(1)}
+\frac{g^4}{(p^2)^2}\mathbf{L}A_2^{(2)}+\frac{g^6}{(p^2)^2}\mathbf{L}A_2^{(3)}
+\cdots\ ,\\
\mathbf{L}\langle \phi\phi\phi\rangle^{1PI}
= &-(Z_g-1)g+g^3\mathbf{L}A_3^{(1)}+g^5\mathbf{L}A_3^{(2)}
+g^7\mathbf{L}A_3^{(3)}+\cdots\ .\\
\end{aligned}
\end{equation}

As expected, the $Z$-factors are completely determined by the local divergence of the corresponding correlation functions. Actually, they can be determined by the local divergences of "unrenormalized" correlation functions:
\begin{equation}\label{phi3-z}
Z_{\phi}=1+\frac{1}{p^2}\sum_{i=1}^{\infty}g^{2L}\mathbf{L}A_2^{(L)},\ 
Z_g=1+\sum_{i=1}^{\infty}g^{2L}\mathbf{L}A_3^{(L)}.\ 
\end{equation}

Similar relations appear in generic theories with spin$\le\frac{1}{2}$. For example, in a $\lambda \phi^4$ model the $Z$ factors are
\begin{equation}\label{phi4-z}
Z_{\phi}=1+\frac{1}{p^2}\sum_{i=1}^{\infty}\lambda^{L}\mathbf{L}A_2^{(L)},\ 
Z_{\lambda}=1+\sum_{i=1}^{\infty}\lambda^{L}\mathbf{L}A_4^{(L)}.\ 
\end{equation}

Using UV decomposition, we have reproduced the beta functions in  $\phi^3$ and $\phi^4$ theories to 5-loop. In a recent work \cite{Jin:2022nqq}, we also computed the full $Q$-dependence of 5-loop anomalous dimensions of $\phi^Q$ operator in $O(N)$ $\phi^4$ theory, and the results are in agreement with the scaling dimensions obtained using semi-classical method \cite{Badel:2019oxl, Antipin:2020abu}.

In gauge and gravity theories, the correlation functions contain powers of $D$ or $\epsilon$, produced by the contraction of $\eta^{\mu\nu}$ in loops. Therefore the local divergence of the correlation functions cannot be directly extracted from the local divergence of the integrals. 
But relations similar to \eqref{phi3-z} exist in super-Yang-Mills and supergravity theories, in which the correlation functions in dimensional reduction scheme are free of explicit $\epsilon$-dependence.

In generic gauge theories, one may compute the Z factors by evaluating the total UV divergence of correlation functions \cite{Herzog:2017ohr}. 
Alternatively, R-operation can be applied before contracting Lorentz indices in Feynman rule to avoid $\epsilon$-terms, at the price of introducing new differentiated Feynman rules \cite{deVries:2019nsu}.
It would be desirable if the Z factors can still be determined solely from the local divergences, as in \eqref{phi3-z}, and we will discuss this possibility in a future work.

\section{Discussions}

The method only apply to integrals with quadratic propagators. It would be interesting to study the UV decomposition of integrals with linear propagators, which would appear in CSW\cite{Cachazo:2013hca,Cachazo:2013iea}, partial fraction, SCET \cite{Becher:2014oda,Broggio:2015dga,Gardi:2009zv} , Q-cut \cite{Baadsgaard:2015twa,Huang:2015cwh}, HQET \cite{Hussain:1994zr,Grozin:1992yq}, light cone gauge, etc.

Unitarity based methods (see e.g. \cite{Bern:1994zx, Bern:2008qj, Boels:2008ef,Bern:2012uc,Cachazo:2013hca,Cachazo:2013iea, Bern:2015ooa,Yang:2016ear}) are very efficient in the computation of  multiloop scattering amplitudes and form factors. However, unitarity cut fails to capture some bubble-type integrals which attach to the external legs. In massless theories, these integrals integrate to zero and do not contribute the the amplitude, but they may have non-zero UV divergences, and their contribution must be included in order to find the correct UV divergence using our approach. It is desirable to develop a compensation method to solve this problem.

In this paper we mainly worked Euclidean space. The local divergences of integrals in Minkowski space are the same as the Euclidean counterparts except for some extra $i$ factors from Wick rotation. More details can be found in Appendix \ref{wick rotation}.

\acknowledgments

We would like to thank Bo Feng, Song He, Rijun Huang, Zuotang Liang, Hui Luo, Mingxing Luo, Radu Roiban, Matthias Wilhelm, Gang Yang for helpful discussions, and Roman Lee for help in the computation of vacuum master integrals using DRA method, and Rijun Huang, Frenz Herzog, Ben Zuijl and Gang Yang for reading a preliminary version of the draft and giving valuable comments. 

\appendix

\section{The local divergences of disconnected integrals}
\label{appendix:disconnected}
In this appendix, we will prove the following statement which will be useful to understand the UV decomposition of generic integral:
\begin{enumerate}[\textbf{DC1}]
\item If the UV decomposition formula \eqref{uvdecom-2} holds for any integral with loop number $\mathbb{L}\le L_0$, then a $L_0$-loop disconnected integral with two components $A$ and $B$ satisfies 
\begin{equation}\label{disconnected-1}
\mathbf{L}(AB)\sim -\mathbf{L}(A)\mathbf{L}(B)\ .
\end{equation}
\end{enumerate}
We will prove \eqref{disconnected-1} by induction. We will assume \eqref{disconnected-1} holds for any disconnected integral $A_1B_1$ with $\mathbb{L}(A_1)\le \mathbb{L}(A),\ \mathbb{L}(B_1)< \mathbb{L}(B)$, or $\mathbb{L}(A_1)< \mathbb{L}(A),\ \mathbb{L}(B_1)\le \mathbb{L}(B)$.
Using the definition of UV and IR sub-integrals, it can be shown that
\begin{equation}
\begin{aligned}
\Theta(AB)=&\Bigl\{\theta\cup\eta\Bigr|\theta\in\Theta(A),\eta\in\Theta(B)\Bigr\}\\
=&\Bigl\{\theta\cup\eta\Bigr|\theta\in\Theta'(A),\eta\in\Theta'(B)\Bigr\}\cup\Theta(A)\cup \Theta(B)\ ,\\
\end{aligned}\label{dis-sub}
\end{equation}
in which $\Theta'(A)$ is the set of non-empty UV sub-integrals of $A$.

 Using \eqref{dis-sub}, the UV divergence of $AB$ has the following decomposition,
\begin{equation}
\begin{aligned}
&AB
\sim\sum_{\theta\in\Theta'(A)}\sum_{\eta\in\Theta'(B)}\mathcal{V}_{\theta\cup\eta}(AB)
+\sum_{\eta\in\Theta(B)}\mathcal{V}_{\eta}(AB)
+\sum_{\theta\in\Theta(A)}\mathcal{V}_{\theta}(AB)\ .\\
\end{aligned}\label{uv disconnected1}
\end{equation}
If $(\theta,\eta)\ne(A,B)$, the first term on the r.h.s. of  \eqref{uv disconnected1} can be written as
\begin{equation}
\begin{aligned}
&\mathcal{V}_{\theta\cup\eta}(AB)
=(A\setminus\theta)(B\setminus\eta)\mathbf{L}(\theta \rho)
=-(A\setminus\theta)(B\setminus\eta)\mathbf{L}\theta \mathbf{L}\rho
=-\mathcal{V}_{\theta}(A)\mathcal{V}_{\eta}(B)\ .\\
\end{aligned}
\end{equation}
In the derivation we used 
$\mathbf{L}(\theta \rho)=-\mathbf{L}\theta \mathbf{L}\rho$, which is true by the induction assumption. We also used the fact that the sub-divergence corresponding to an empty set is zero.
Then we have
\begin{equation}
\begin{aligned}
&\sum_{\theta\in\Theta'(A)}\sum_{\eta\in\Theta'(B)}\mathcal{V}_{\theta\cup\eta}(AB)
=\mathbf{L}(AB)-\sum_{\theta\in\Theta'(A)\eta\in\Theta'(B)}^{(\theta,\eta)\ne(A,B)}\mathcal{V}_{\theta}(A)\mathcal{V}_{\eta}(B)\\
=&\mathbf{L}(AB)+\mathbf{L}(A)\mathbf{L}(B)
-\sum_{\theta\in\Theta'(A)\eta\in\Theta'(B)}\mathcal{V}_{\theta}(A)\mathcal{V}_{\eta}(B)\\
=&\mathbf{L}(AB)+\mathbf{L}(A)\mathbf{L}(B)
-\sum_{\theta\in\Theta(A)}\mathcal{V}_{\theta}(A)
\sum_{\eta\in\Theta(B)}\mathcal{V}_{\eta}(B)\ .\\
\end{aligned}
\end{equation}

The second term on the r.h.s. of \eqref{uv disconnected1} can be simplified using
\begin{equation}
\mathcal{V}_{\eta}(AB)=A(B\setminus \eta)\mathbf{L}\eta
=A\mathcal{V}_{\eta}(B)\ .
\end{equation}
Then \eqref{uv disconnected1} becomes
\begin{equation}
\begin{aligned}
AB
\sim&\mathbf{L}(AB)+\mathbf{L}(A)\mathbf{L}(B)
-\sum_{\theta\in\Theta(A)}\mathcal{V}_{\theta}(A)
\sum_{\eta\in\Theta(B)}\mathcal{V}_{\eta}(B)
+A\sum_{\eta\in\Theta(B)}\mathcal{V}_{\eta}(B)\\
&+B\sum_{\theta\in\Theta(A)}\mathcal{V}_{\theta}(A)\\
=&\mathbf{L}(AB)+\mathbf{L}(A)\mathbf{L}(B)+AB-\Bigl[A-\sum_{\theta\in\Theta(A)}\mathcal{V}_{\theta}(A)\Bigr]\Bigl[B-\sum_{\eta\in\Theta(B)}\mathcal{V}_{\eta}(B)\Bigr]\\
\sim&AB+\mathbf{L}(A)\mathbf{L}(B)
+\mathbf{L}(AB)\ .\\
\end{aligned}\label{uv disconnected2}
\end{equation}
In the last step, we dropped $\Bigl[A-\sum_{\theta\in\Theta(A)}\mathcal{V}_{\theta}(A)\Bigr]\Bigl[B-\sum_{\eta\in\Theta(B)}\mathcal{V}_{\eta}(B)\Bigr]$ because it is a product of two UV finite terms. Eq. \eqref{uv disconnected2} verified $\mathbf{L}(AB)\sim -\mathbf{L}(A)\mathbf{L}(B)$ and finished the proof of \eqref{disconnected-1}.

If the integral has more disconnected components,
\begin{equation}\label{dis-fn-1}
\mathbf{L}(F_1\cdots F_n)= (-1)^{n-1}\mathbf{L}(F_1)\cdots\mathbf{L}(F_n)\ .
\end{equation}
The formula will be free of the $(-1)^{n-1}$ factor if it is rewritten in terms of the counterterm function $Z$ in \eqref{BPHZ-R-1}, since $Z(F_i)=-\mathbf{L}F_i$:
\begin{equation}
Z(F_1\cdots F_n)= Z(F_1)\cdots Z(F_n)\ .
\end{equation}


As an example, we compute the local divergence of the following disconnected CV,
\begin{equation}
\begin{aligned}
&\mathbf{L}\frac{(-2l_1\cdot l_2)^{2i-d}}{(l_1^2)^{i}(l_2^2)^{i}}
\sim -(-2)^{2i-d}\mathbf{L}\frac{l_1^{\mu_1}\cdots l_1^{\mu_{2i-d}}}{(l_1^2)^{i}}
\mathbf{L}\frac{l_{2\mu_1}\cdots l_{2\mu_{2i-d}}}{(l_2^2)^{i}}\\
=& -\frac{1}{\Gamma^2(i)\epsilon^2}\eta_s^{\mu_1 \cdots \mu_{2i-d}}(\eta_s)_{\mu_1 \cdots \mu_{2i-d}}
=-\frac{1}{\Gamma^2(i)\epsilon^2}\frac{(2i-d)!}{(i-\frac{d}{2})!}
(\frac{d}{2}-\epsilon)_{i-\frac{d}{2}}\\
\sim&\frac{(2i-d)!}{\Gamma(i)\Gamma(\frac{d}{2})(i-\frac{d}{2})!}
\Bigl[-\frac{1}{\epsilon^2}+\frac{1}{\epsilon}(H_{i-1}-H_{\frac{d}{2}-1})\Bigr]\ ,\\
\end{aligned}\label{n3le0-3}
\end{equation}
in which we used \eqref{fn}, \eqref{disconnected-1} and
\begin{equation}
\begin{aligned}
& \eta_s^{\mu_1 \cdots \mu_{2a}}(\eta_s)_{\mu_1 \cdots \mu_{2a}}
=\frac{(2a)!(\frac{D}{2})_a}{a!}\ .\\
\end{aligned}\label{etas-contraction-1}
\end{equation}
In \eqref{etas-contraction-1}, $(\frac{D}{2})_a\equiv \frac{D}{2}(\frac{D}{2}+1)\cdots (\frac{D}{2}+a-1)$ is the Pochhammer symbol.

The local divergence of disconnected integrals are free of $\frac{1}{\epsilon}$ terms if there are no Lorentz contractions among different components. For example, if $i=\frac{d}{2}$ in \eqref{n3le0-3},
\begin{equation}
\begin{aligned}
&\mathbf{L}\frac{1}{(l_1^2)^{\frac{d}{2}}(l_2^2)^{\frac{d}{2}}}
\sim -\frac{1}{\epsilon^2\Gamma^2(\frac{d}{2})}\ .\\
\end{aligned}\label{n3le0-4}
\end{equation}
Notice that only $\mathcal{O}(\frac{1}{\epsilon})$ order terms contribute to beta functions and anomalous dimensions, so disconnected integrals without Lorentz contractions have no contribution to beta functions or anomalous dimensions, assuming that the amplitude has no $\epsilon$ dependent prefactors multiplied to the integral. For example in $\phi^4$ theory, the disconnected integral $\frac{g^3}{l_1^2(l_1+p)^2l_2^2(l_2+p)^2}$ would not contribute the the beta functions(anomalous dimensions).  This is not true in gauge and gravity theories, where extra $\epsilon$ dependence may appear from the contraction of metric, gamma matrices, etc.

\section{UV divergence in odd and fractional dimension space}
\label{appendix:fractional}
As discussed in Section \ref{subsection:asymptotic}, the local divergences of generic integrals can be determined by local divergences of CV. If there is no CV in a certain loop, then there is no UV divergence in this loop.

Let $V$ be a $\mathbb{L}$-loop CV in d-dimension. $V$ can be written as
\begin{equation}
V=\frac{\prod_{j=1}^Kl_j^{\mu_j}}{\prod_{i=1}^Nl_i^2}\ .
\end{equation}
In order for $\mathbf{L}V\ne 0$, $K$ must be even. Otherwise $V\rightarrow -V$ under the transformation $l_i\rightarrow -l_i$. Let us define $n=N-\frac{K}{2}$, then $n$ is an integer. The critical condition is given by $\mathbb{L}=\frac{2n}{d}$, then there are UV divergences in $\mathbb{L}$ loop with 
\begin{equation}
\mathbb{L}\in \Omega_d\equiv\frac{2}{d}\mathbb{Z}_+\cap \mathbb{Z}_+,\ 
\end{equation}
in which $\mathbb{Z}_+$ represents the set of positive integers.
Obviously, $\Omega_d=\mathbb{Z}_+$ if $d$ is an even integer, and $\Omega_d=2\mathbb{Z}_+$ if $d$ is an odd integer. 

Sometimes one may also consider fractional dimensions. $d$ can be written as $\frac{p}{q}$, in which $p$ and $q$ are coprime integers. It can be checked that
\begin{equation}
\begin{aligned}
\Omega_{\frac{p}{q}}=&
\left\{
\begin{aligned}
&2q\mathbb{Z}_+,&\ 
&\text{if $p$ is odd,}\\
&q\mathbb{Z}_+,&\ 
&\text{if $p$ is even.}\\
\end{aligned}
\right. \\
\end{aligned}
\end{equation}
For example in \cite{Bern:2017ucb, Bern:2018jmv} the 5-loop UV divergence of 4-point amplitude in $\mathcal{N}=8$ Supergravity are considered in $\frac{22}{5}$ and $\frac{24}{5}$ dimensions.

The UV divergences in even dimension have most complicated structures, because local and sub-divergences appear in any loop. In this work we assume $d$ is an even integer unless otherwise specified.

\section{The 3 loop master integral with non-adjacent masses}
\label{appendix-non-adj}

In this appendix we give details in the evaluation of the 3 loop vacuum integral in Figure \ref{fig:3 loop vacuum}(b) in Section \ref{IR regulation by adding two masses} following the DRA approach.
The master integral $I_{111111}$ has poles at $D=4$, and it is more convenient to compute $I_{211111}$, which only has one simple pole at $D=\frac{8}{3}$ in the basic stripe $(2,4]$,
\begin{equation}\label{f211111}
F(D)=I_{211111}=\frac{3(4-D)}{4}I_{111111}.
\end{equation}

Dimensional shift gives
\begin{equation}
\begin{aligned}
F(D-2)=&\frac{(D-6)(D-3)(D-2)}{4}F(D)+R_+(D-2)+R_-(D)\ ,\\
R_+(D-2)=&-\frac{3(-6+D)\pi^{2}\csc(\frac{D\pi}{2})^{2}\Gamma(4-D)}{\Gamma(-1+\frac{D}{2})}\\
&+\frac{9(-6+D)\pi^{\frac{7}{2}}\csc(\frac{D\pi}{2})^{2}\csc(\frac{3D\pi}{2})}{2^{D-3}\Gamma(\frac{-3+D}{2})\Gamma(-3+D)}\ ,\\
R_-(D)=&\frac{3(-6+D)(-24+5D)\pi^{\frac{3}{2}}\csc(\frac{D\pi}{2})\Gamma(6-\frac{3D}{2})\Gamma(4-D)}{4^{5-D}\Gamma(\frac{11}{2}-D)}\ .\\
\end{aligned}\label{f211111shift}
\end{equation}

The special solution is
\begin{equation}
\begin{aligned}
F_1(D)=&\sum_{k=0}^{\infty}\Bigl(-\frac{3\left(-\frac{1}{2}\right)^{k}\pi^{3}\csc(\frac{D\pi}{2})^{2}\csc(D\pi)}{(-3+D)(-2+D+2k)\Gamma(-4+D)\Gamma(\frac{D}{2})}\\
&+\frac{9\pi^{\frac{7}{2}}\csc(\frac{D\pi}{2})^{2}\csc(\frac{3D\pi}{2})\Gamma(-1+\frac{D}{2}+k)}{2^{D+3k}(-3+D)\Gamma(-4+D)\Gamma(\frac{D}{2})\Gamma(\frac{-1+D}{2}+k)}\\
&+\frac{3(24-5D+10k)\pi^{\frac{3}{2}}\csc(\frac{D\pi}{2})\Gamma(2-D)\Gamma(3-\frac{D}{2})\Gamma(6-\frac{3D}{2}+3k)}{2^{8-2D+3k}\Gamma(3-\frac{D}{2}+k)\Gamma(\frac{11}{2}-D+2k)}\Bigr)\ .\\
\end{aligned}
\end{equation}

We choose the summing factor as
\begin{equation}
\Sigma^{-1}(D)=\frac{2^{\frac{D}{2}}\Gamma(3-\frac{D}{2})\Gamma(2-D)}{\sin\frac{\pi}{2}(D-\frac{8}{3})}\ .
\end{equation}

When $Im(D)\rightarrow \pm\infty$,
\begin{equation}
\begin{aligned}
&F(D)\sim |\Gamma(7-\frac{3D}{2})|\sim |Im D|^{\frac{13-3Re D}{2}}e^{-\frac{3\pi |Im D|}{4}}\ ,\\
&F_1(D)\sim e^{-\frac{5\pi |Im D|}{4}},\ \Sigma(D)=e^{\frac{5\pi |Im D|}{4}}\ .\\
\end{aligned}
\end{equation}
So we have $\frac{\Sigma(D)F(D)}{e^{\pi|Im D|}}\rightarrow 0$ and $\frac{\Sigma(D)F_1(D)}{e^{\pi|Im D|}}\rightarrow 0$.

$\Sigma(D)F_1(D)$ has simple poles at $D=\frac{10}{3}+2\mathbb{Z}$, and has $\frac{1}{\epsilon^2}$ pole at $D=4+2\mathbb{Z}$.
Notice $\Sigma(D) F(D)$ has no poles in the basic stripe, and the pole structure of $\omega(D)=\Sigma(D)F(D)-\Sigma(D)F_1(D)$ can be determined from $\Sigma(D)F_1(D)$. We choose the following form for $\omega(D)$,
\begin{equation}
\begin{aligned}
&\omega(D)=a+b_1\cot\frac{\pi}{2}(D-\frac{10}{3})
+b_2\cot\frac{\pi}{2}(D-4)+b_3\cot^2\frac{\pi}{2}(D-4)\ .\\
\end{aligned}
\end{equation}

At $D=\frac{10}{3}-2\epsilon$,
\begin{equation}
\begin{aligned}
&\omega+\Sigma F_1
\sim \frac{1}{\epsilon}(-\frac{b_1}{\pi}+\frac{\pi}{4})
+\mathcal{O}(\epsilon^0)\ .\\
\end{aligned}
\end{equation}

At $D=4-2\epsilon$,
\begin{equation}
\begin{aligned}
&\omega+\Sigma F_1
\sim \frac{1}{\epsilon^2}(\frac{b_3}{\pi^2}-\sqrt{3})+\frac{1}{\epsilon}(-\frac{b_2}{\pi}+\pi)
+\mathcal{O}(\epsilon^0)\ .\\
\end{aligned}
\end{equation}

The constant $a$ can be fixed at $D=3$, where $\Sigma(D)F(D)=\Sigma(D)F_1(D)=0$, and
\begin{equation}
\omega(3)=a-\sqrt{3}b_1
\end{equation}
So we have

\begin{equation}
\begin{aligned}
&\omega(D)=\pi^2\left[\frac{\sqrt{3}}{4}+\frac{1}{4}\cot\frac{\pi}{2}(D-\frac{10}{3})
+\cot\frac{\pi}{2}(D-4)+\sqrt{3}\cot^2\frac{\pi}{2}(D-4)\right]\ .\\
\end{aligned}
\end{equation}

To verify the result, we evaluated $F(D)$ at $D=4-2\epsilon$, and found that
\begin{equation}
\begin{aligned}
\frac{I_{111111}}{\Gamma^3(1-\frac{D}{2})}\Bigr|_{D=4-2\epsilon}
=&-2\zeta_3\epsilon^2+\left(\frac{7\pi^4}{60}
+\frac{2}{3}\ln^2 2(\pi^2-\ln^2 2)-16\text{Li}_4(\frac{1}{2})\right)\epsilon^3\\
&+28.6007184522938416617755059822969702148\epsilon^4+\cdots\\
\end{aligned}\label{f111111}
\end{equation}
which is in agreement with the numerical result in \cite{Schroder:2005va}.

\section{An example of local divergence from IR subtraction}
\label{appendix:example IR subtraction}

As an example of IR subtraction we recompute the local divergence of $I_{411411}$ in subsection \ref{IR regulation by adding two masses} using IR subtraction. We regulate the overall IR divergence by adding a single mass to the $l_1$-propagator, and the total divergence of the integral is 
\begin{equation}
\begin{aligned}
&I^m_{411411}=\frac{1}{(l_1^2+m^2)^4l_2^2l_3^2[(l_2-l_3)^2]^4(l_3-l_1)^2(l_1-l_2)^2}\\
\sim& -\frac{1}{1296\epsilon^{3}}+\frac{1+36\ln m}{7776\epsilon^{2}}+\frac{-1637-225\pi^{2}-180\ln m-3240\ln^2m+1296\zeta_3}{233280\epsilon}\ .\\
\end{aligned}
\end{equation}

The integral has a single IR divergence\footnote{In the computation of \eqref{IR 411411} it is helpful to first redefine $l_3\rightarrow l_2-l_3$, and shift the IR divergence to the region $l_3\rightarrow 0$.} corresponding to $\gamma_4=\{L(l_2-l_3)\}$. We can 
\begin{equation}
\begin{aligned}
&\mathcal{IR}_{\gamma_4}I^m_{411411}
=-\frac{1}{6\epsilon}\frac{1}{(l_1^2+m^2)^4(l_2^2)^2[(l_1-l_2)^2]^2}\\
\sim&-\frac{1}{432\epsilon^{3}}+\frac{1+24\ln~m}{2592\epsilon^{2}}+\frac{11-18\pi^{2}-24\ln m-288\ln^2m}{15552\epsilon}\ .\\
\end{aligned}\label{IR 411411}
\end{equation}

The UV divergence structures of $I^{m}_{411411}$ and  $I^{mm}_{411411}$ are the same. Still there is a 1 loop UV divergence,
\begin{equation}
\begin{aligned}
\mathcal{UV}_{\gamma_1}I^m_{411411}
=&\frac{1}{6\epsilon}\mathcal{U}\frac{1}{(l_1^2+m^2)^4[(l_2-l_3)^2]^4}
=\frac{1}{36\epsilon^2}\frac{1}{(l_1^2+m^2)^4}\\
\sim &\frac{1}{216\epsilon^3}-\frac{\ln~m}{108\epsilon^2}+\frac{\pi^2+24\ln^2m}{2592\epsilon}\ ,\\
\end{aligned}
\end{equation}
and two 2 loop UV divergence,
\begin{equation}
\begin{aligned}
\mathcal{UV}_{\gamma_2}I^{m}_{411411}
=&\frac{1}{(l_1^2+m^2)^4}\mathbf{L}\frac{1}{(l_2^2)^2(l_3^2)^2[(l_2-l_3)^2]^4}\\
\sim &-\frac{1}{432\epsilon^{3}}+\frac{\frac{-1}{2592}+\frac{\ln m}{216}}{\epsilon^{2}}+\frac{-\pi^{2}+4\ln m-24\ln^2m}{5184\epsilon}\ ,\\
\mathcal{UV}_{\gamma_3}I^{m}_{411411}
=&\mathcal{U}\frac{1}{[(l_2-l_3)^2]^4}\mathbf{L}\frac{1}{(l_1^2)^4(l_2^2)^2[(l_1-l_2)^2]^2}
\sim -\frac{1}{432\epsilon^{3}}-\frac{1}{2592\epsilon^{2}}\ .\\
\end{aligned}
\end{equation}
In the last line, we used the fact that $\mathcal{U}$ is the same as $\mathbf{L}$ for 1 loop scalar vacuum integrals.
\begin{equation}
\mathcal{U}\frac{1}{[(l_2-l_3)^2]^4}=\frac{1}{6\epsilon}\ .
\end{equation}

Combining these results, we find the same local divergence as in \eqref{ld411411}:
\begin{equation}
(1-\mathcal{IR}_{\gamma_4}-\mathcal{UV}_{\gamma_1}-\mathcal{UV}_{\gamma_2}
-\mathcal{UV}_{\gamma_3})I^m_{411411}
\sim\frac{1}{648\epsilon^{3}}+\frac{1}{1944\epsilon^{2}}+\frac{-901+648\zeta_3}{116640\epsilon}\ .
\end{equation}

\section{Wick rotation of local divergence}
\label{wick rotation}

We discussed the computation of local divergence  in Euclidean space. In order to apply our method to theories in Minkowski space, we need to do a Wick rotation. The asymptotic expansion $\mathcal{A}$ is not sensitive to the signature of the metric. After the expansion
\begin{equation}
\mathcal{A}(F)\sim \frac{\prod_{i=1}^A l_{a_i}^{\mu_i}}{\prod_{j=1}^NL_j}\ ,
\end{equation}
is a massless tensor vacuum integral.

 We can pair each $l_{a_i}^{\mu_i}$ with an auxiliary momentum $p_{i\mu_i}$, and after the Wick rotation,
\begin{equation}
 \frac{\prod_{i=1}^A l_{a_i}\cdot p_i}{\prod_{j=1}^NL_j}
\rightarrow  (-1)^{A-N}\frac{\prod_{i=1}^A l_{Ea_i}\cdot p_{E i}}{\prod_{j=1}^NL_{E j}}\ .
\end{equation}
Then the local divergence of the tensor integral can be computed in Euclidean space. Each term of the result is a product of $\frac{A}{2}$ metric tensors,
\begin{equation}
 \frac{\prod_{i=1}^A l_{a_i}\cdot p_i}{\prod_{j=1}^NL_j}
\sim   (-1)^{A-N}\prod_{i=1}^{\frac{A}{2}}  p_{E \alpha_i}\cdot p_{E \beta_i}
\sim   (-1)^{\frac{A}{2}-N}\prod_{i=1}^{\frac{A}{2}}  p_{\alpha_i}\cdot p_{\beta_i}\ .
\end{equation}

We know that the local divergence vanishes unless $A+L d=2N$, so the factor $(-1)^{\frac{A}{2}-N}=i^{Ld}$. Combining with the extra $i$ factor associated with each loop integration, we have in all an $i^{L(d+1)}$ factor. So we can do the computation pretending we are working in Euclidean space, and add a $i^{L(d+1)}$ factor in the end. In 4-d, the factor is $i^L$.





\bibliographystyle{JHEP}
\bibliography{/Users/jin/Documents/tex/PSUThesis/Biblio-Database}{}

\newpage

\printnomenclature

\newpage

\end{document}